\documentclass[preprint]{elsarticle}
\usepackage{amssymb}
\usepackage[final]{epsfig}
\usepackage{color}
\usepackage{graphicx}
\usepackage{lineno}
\newcommand{\beq}{\begin{equation}}
\newcommand{\eeq}{\end{equation}}
\newcommand{\bdi}{\begin{displaymath}}
\newcommand{\edi}{\end{displaymath}}
\newcommand{\no}{\nonumber}
\newcommand{\bea}{\begin{eqnarray}}
\newcommand{\eea}{\end{eqnarray}}
\newcommand{\vep}{\varepsilon}

\newcommand{\grad}{\vec{\nabla}}
\newcommand{\vx}{\vec{x}}

\newcommand{\text}{}

\newcommand{\LL}[1]{{\bf
 L} \left[ #1 \right]   }

\begin{document}


\begin{frontmatter}

\title{Electric fields, weighting fields, signals and charge diffusion in detectors including resistive materials}

\author[]{W. Riegler, CERN 
}
\address{CERN EP, CH-1211 Geneve 23}

\begin{abstract}

In this report we discuss static and time dependent electric fields in detector geometries with an arbitrary number of parallel layers of a given permittivity and weak conductivity. We derive the Green's functions i.e. the field of a point charge, as well as the weighting fields for readout pads and readout strips in these geometries. The effect of 'bulk' resistivity on electric fields and signals is investigated. The spreading of charge on thin resistive layers is also discussed in detail, and the conditions for allowing the effect to be described by the diffusion equation is discussed. We apply the results to derive fields and induced signals in Resistive Plate Chambers, Micromega detectors including resistive layers for charge spreading and discharge protection as well as detectors using resistive charge division readout like the MicroCAT detector. We also discuss in detail how resistive layers affect signal shapes and increase crosstalk between readout electrodes.

\end{abstract}

\end{frontmatter} 
\clearpage
\tableofcontents
\clearpage
\section{Introduction}

This report discusses electric fields and signals in detectors that represent parallel plate geometries with segmented readout like GEMs \cite{gem1}, Micromegas \cite{micromega1}, RPCs \cite{rpc1}, liquid nobel gas calorimeters \cite{larg1} and to some extent in silicon strip and pixel detectors. In all of these detectors, the  charges generated inside the sensor volume act as a source of the signal. Through the quasi-static approximation of Maxwell's equations \cite{fano}\cite{melcher}\cite{schnizer1} the derived solutions also apply to detectors that contain elements of finite resistivity.
\\
\\
The field of a point charge in such geometries, i.e. the Green's function of the problem, is first of all needed for  calculation of the mutual influence of the charge carriers on each other. These are the so called space-charge effects that play e.g. a very prominent role in RPCs. This solution is also employed to calculate charge-up effects and related field distortions in detectors with finite resistivity.
\\
\\
The movement of the charges in these detectors induces signals on the metallic readout electrodes that are connected to the readout electronics. Theorems employing so called weighting fields exist for calculation of these signals for the case of grounded electrodes \cite{ramo}\cite{shockley}, electrodes connected with a linear impedance network \cite{radeka1}\cite{driftchamberbook}, and detectors where the sensitive volume is in addition filled with materials of finite resistivity \cite{riegler1}\cite{riegler2}. The above mentioned Green's function is used to calculate the weighting fields of strips and pads (pixels) for general parallel plate geometries.
\\
\\
The report is an extension of the techniques developed in \cite{schnizer2}\cite{riegler2} and specific attention is given to the numerical evaluation of the formulas in view of practical application.  Section 2 discusses the Green's function of an infinitely extended two layer geometry to develop the general concepts and also investigates solutions of this geometry for finite extension i.e. including boundary conditions on circles and rectangles. Section 3 extends these solutions to geometries that employ an arbitrary number of parallel layers of different permittivity and conductivity. Section 4 applies the solutions to Resistive Plate Chambers. Section 5 and 6 then discusses in some detail the dispersion of charge on thin resistive layers. Section 7 investigates the potentials on thin resistive layers for uniformly distributed currents. Section 8 is finally treating the effect of resistive layers on the signals induced on readout electrodes. Section 9 discusses some of the mathematical tools used to evaluate the integrals used in this report.
\\
\\
To conclude the introduction we recall the application of the quasi-static approximation of Maxwell's equations: Knowing the solution of the Poisson equation for a charge distribution $\rho (\vx)$ embedded in a geometry of a given permittivity $\vep (\vx)$, we find the time dependent solution (in the Laplace domain with parameter $s$) for an 'externally impressed' charge density $\rho_e(\vx,s)$ and a geometry that in addition includes a finite (weak) conductivity $\sigma (\vx)$ by replacing $\vep (\vx)$ with  $\vep (\vx)+\sigma(\vx)/s$ and $\rho(\vx)$ with $\rho_e(\vx,s)$. For detector applications the volume resistivity $\rho(\vx)=1/\sigma(\vx)$ is traditionally used.
\\
\\
As an example we look at the potential of a point charge $Q$ in a medium of constant permittivity $\vep$, which is given by
\beq \label{green_infinite}
    \phi(r) = \frac{Q}{4\vep \pi\, r}
\eeq
In case the medium has a conductivity $\sigma$ and we place the 'external' charge $Q$ at $t=0$, i.e. $Q(t)=Q\Theta(t)$ and therefore $Q(s)=Q_0/s$, we replace $\vep$ by $\vep+\sigma/s$ and $Q$ by $Q/s$ and perform the inverse Laplace transform, which gives
\beq
    \phi(r,s) = \frac{Q}{4\pi (s\vep + \sigma)r} \quad \rightarrow \quad 
    \phi (r, t)= 
    \frac{Q}{4\pi\vep\, r} \, \,e^{\frac{-t}{\tau}} \qquad  \tau = \vep/\sigma=\rho \vep
\eeq
The potential is equal to the electrostatic one, but 'destroyed' with the time constant $\tau$. Finally we recall that the current signal induced by two moving charges $q$ and $-q$ on a grounded electrode in a detector containing resistive elements is given by 
\beq
  I(t) = -\frac{q}{V_w}\int_0^t \vec{E}_w(\vec{x}_1(t'),t-t') \vec{\dot x}_1(t')dt' + \frac{q}{V_w}\int_0^t \vec{E}_w(\vec{x}_2(t'),t-t') \vec{\dot x}_2(t')dt' 
\eeq
The time dependent weighting field $\vec{E}_w(\vx,t)$ represents the electric field in the detector volume in case all electrodes are grounded and a delta potential $V(t)=V_w\delta(t)$ is applied to the electrode in question.  $\vec{x}_1(t)$ and  $\vec{x}_2(t)$ are the trajectories of the charges $q$ and $-q$ where $\vx_1(0)=\vx_2(0)$. It is important to consider the movement of two opposite charges originating from the same point, since otherwise the unphysical effect of creating a net charge at a given space point will be added to the result. As in the previous example, this field can be derived using the quasistatic approximation by simply applying the 'constant' potential $V_w=\LL{V_x\delta(t)}$ to the geometry in the Laplace domain.
\section{Electric fields and weighting fields in a 2-layer geometry}

In this section we discuss electric fields in a geometry consisting of two layers of different permittivity, that are bound by grounded planes. The results will be extended to the general case of $N$ layers in the following section. 

\subsection{Potential of a point charge centred at the origin}

\begin{figure}[ht]
 \begin{center}
  \includegraphics[width=7cm]{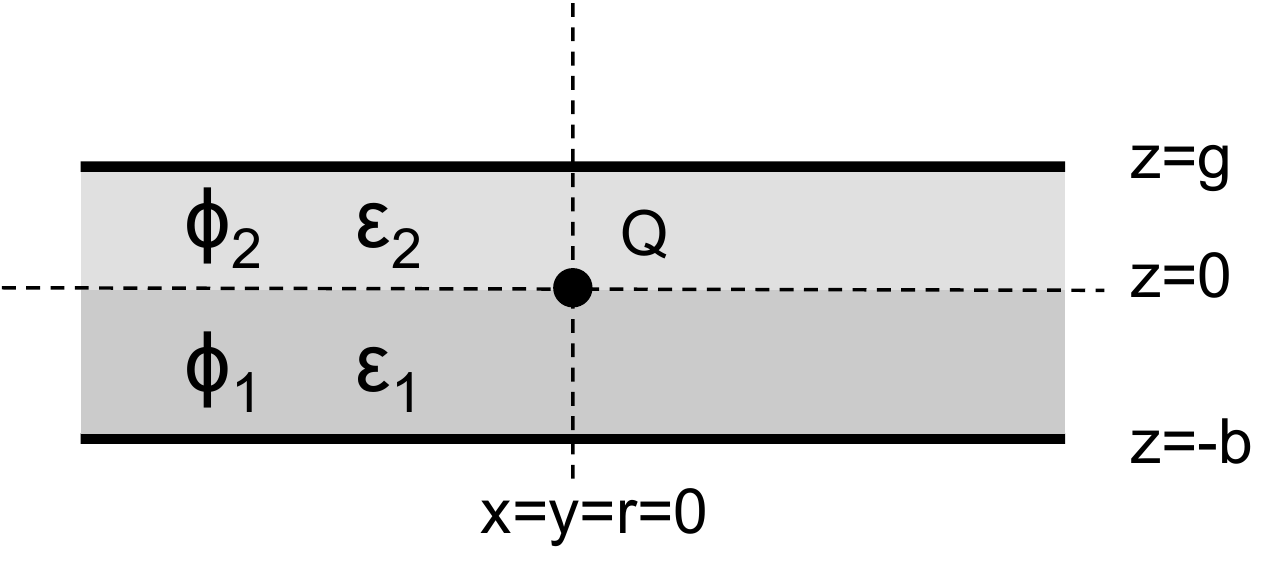}
  \caption{A point charge $Q$ on the boundary between two dielectric layers.}
  \label{two_layer_sketch}
  \end{center}
\end{figure}

We first investigate the electric field of a point charge in a two layer geometry as shown in Fig. \ref{two_layer_sketch}. We assume two layers of thickness $b$ and $g$ with constant dielectric permittivity of $\vep_1, \vep_2$, surrounded by grounded metal plates. A point charge $Q$ is placed on the boundary between the two layers at $r=0, z=0$. The problem has rotational symmetry and we therefore use cylindrical coordinates. In both layers there are no charges present, so the potential $\phi$ must be a solution of the Laplace equation. Separation of the Laplace equation in cylinder coordinates and assuming rotational symmetry yields the solutions $J_0(kr)$\cite{besselJ} and $Y_0(kr)$\cite{besselY} for the radial part and $e^{kz}$ and $e^{-kz}$ for the axial part. Because $Y_0(kr)$ diverges at $r=0$, but the electric field must be finite for $r=0$ (at $z \neq 0$), the coefficients of $Y_0(kr)$ are zero and the general solution in the two layers is
\bdi \label{pot1}
  \phi_1(r,z) = \frac{1}{2\pi} \int_0^\infty J_0(kr)
  \left[
   A_1(k)e^{kz} + B_1(k) e^{-kz}
  \right] dk
  \qquad -b<z<0
\edi
\bdi  \label{pot2}
  \phi_2(r,z) = \frac{1}{2\pi} \int_0^\infty J_0(kr)
  \left[
   A_2(k)e^{kz} + B_2(k) e^{-kz}
  \right] dk
   \qquad \phantom{-}0<z<g
\edi
The coefficients $A(k)$ and $B(k)$ must be determined by boundary conditions.
The grounded plates at $z=-b$ and $z=g$ define the conditions $\phi_1(-b,r)=0$ and $\phi_2(g,r)=0$, which gives
\bea
   A_1 e^{-kb}+B_1 e^{kb}  & = & 0 \label{cond1} \\
   A_2 e^{kg}+B_2 e^{-kg}  & = & 0 \label{cond2}
\eea
On the boundary between the two dielectric layers we assume a surface charge density $q(r)$. From Gauss' Law $\grad \vep(\vx) \vec{E}(\vx) = \rho(\vx)$ for a medium of inhomogeneous permittivity we derive that 'passing through an infinitely thin sheet of charge' with a surface charge density $q(r)$\,[C/cm$^2$], the potential is continuous  i.e. $\phi_1(r,0)=\phi_2(r,0)$ which gives
\beq \label{cond3}
  A_1+B_1=A_2+B_2
\eeq
and the $\vep E$ component perpendicular to the sheet 'jumps' by $q(r)$
\beq
  \vep_1 \frac{\partial \phi_1(r,z)}{\partial z}\vert_{z=0} - \vep_2 \frac{\partial \phi_2(r,z)}{\partial z}\vert_{z=0}
  =q(r)
\eeq
The surface charge density corresponding to the point charge $Q$ at $r=0$ is $q(r)=Q \delta(r)/2\pi r$, so this last equation reads as
\bdi
   \frac{1}{2\pi} \int_0^\infty J_0(kr) k  \left[
    \vep_1 (A_1-B_1) -
    \vep_2 (A_2-B_2)\right] dk
    =\frac{Q}{2\pi r} \delta (r)
\edi
Multiplying both sides of the equation with $r J_0(k'r)$, integrating them over $r$ from $0$ to $\infty$ and using the relation $\int_0^\infty rJ_0(kr)J_0(k'r)dr=\delta(k-k')/k$ \cite{jackson} we have
\beq \label{cond4}
  \vep_1(A_1-B_1)- \vep_2(A_2-B_2)=Q
\eeq
The four Eqs. \ref{cond1}, \ref{cond2}, \ref{cond3}, \ref{cond4} determine $A_1, B_1, A_2, B_2$ to
\beq \label{two_layer_coefficients}
  A_1 = \frac{2Q \sinh (kg)}{D(k)}\, e^{bk} \quad
  B_1 = -\frac{2Q \sinh (kg)}{D(k)}\, e^{-bk} \quad
  A_2 = -\frac{2Q \sinh (kb)}{D(k)}\, e^{-gk} \quad
  B_2 = \frac{2Q \sinh (kb)}{D(k)}\, e^{gk} \quad
\eeq
\beq \label{two_layer_determinant}
   D(k) =  4[\vep_1 \cosh (bk)\,\sinh (gk)+
       \vep_2 \sinh (bk) \, \cosh (gk)]
\eeq
%
%
%
%
%
%
%
%
%
The solutions then read as
\beq \label{two_layer_potential1}
  \phi_1(r,z) =\frac{Q}{2\pi} \int_0^\infty J_0(kr) \frac{4\sinh (gk) \sinh(k(b+z))}{D(k)} dk \qquad -b<z<0
\eeq
\beq \label{two_layer_potential2}
  \phi_2(r,z) =
 \frac{Q}{2\pi}\int_0^\infty J_0(kr)  \frac{4\sinh (bk) \sinh(k(g-z))}{D(k)} dk \qquad \phantom{-}0<z<g
\eeq

\subsection{Evaluation and divergence removal \label{section_divergence_removal}}

The integrals in the expressions Eq. \ref{two_layer_potential1} and Eq. \ref{two_layer_potential2} cannot be expressed in closed form, so for evaluation we have to either use numerical integration or find appropriate techniques to express the result as an infinite series. We first focus on the numerical integration method and investigate the behaviour of the integrand with respect to $k$ in order to find an appropriate upper integration limit. For large values of $k$ the integrand behaves as 
\bdi
   \frac{4\sinh (gk) \sinh(k(b+z))}{D(k)} \, \rightarrow \, \frac{e^{kz}}{\vep_1+\vep_2} \qquad
%
%
%
  \frac{4\sinh (bk) \sinh(k(g-z))}{D(k)} \, \rightarrow \, \frac{e^{-kz}}{\vep_1+\vep_2}
\edi
The integrand behaves like $\exp(-k\vert z\vert)$, so by using an upper integration limit for $k$ of a multiple of $1/z$ will give an accurate result. We see however that at $z=0$, the expressions become constant. The Bessel function $J_0(kr)$ behaves as $\sqrt{2/\pi kr}\cos(kr-\pi/4)$ for large values of $r$,
so the $1/\sqrt{k}$ factor together with the oscillatory behavior results in convergence, although very slow. The expression for $E_z$ however has an additional factor of $k$ so the integrand behaves like $\sqrt{k}$ and therefore diverges. To cure this problem we employ the following idea \cite{schnizer2}: In case we move the two metal plates to infinity i.e. $b,g\rightarrow  \infty$ we find
\bdi
    \lim_{b\rightarrow \infty} \frac{4\sinh (g k) \sinh(k(b+z))}{D(k)} = \frac{e^{kz}}{\vep_1+\vep_2} \qquad 
    \lim_{g\rightarrow \infty}\frac{4\sinh (bk) \sinh(k(g-z))}{D(k)} = \frac{e^{-kz}}{\vep_1+\vep_2}
\edi
Having moved the grounded plates infinitely far away, this geometry corresponds to a point charge $Q$ sitting at the
boundary of two infinite half-spaces of permittivity $\vep_1$ and $\vep_2$, for which we know to be potential to be
$Q/(2\pi(\vep_1+\vep_2)1/\sqrt{r^2+z^2})$. We therefore have the identity
\beq \label{inverse_square_identity}
   \frac{1}{\sqrt{r^2+z^2}} = \int_0^\infty J_0(kr) e^{-k\vert z \vert} dk
\eeq
We can now write the integrand of $\phi_1$ as

\bdi
     \frac{e^{kz}}{\vep_1+\vep_2} + \frac{4\sinh (gk) \sinh(k(b+z))}{D(k)} - \frac{e^{kz}}{\vep_1+\vep_2} = \frac{e^{kz}}{\vep_1+\vep_2} + f_1(k,z)
\edi
and arrive with Eq. \ref{inverse_square_identity} at
\beq
  \phi_1(r,z)  =  \frac{Q}{2\pi (\vep_1+\vep_2)}\, \frac{1}{\sqrt{r^2+z^2}} 
    + \frac{Q}{2\pi} \int_0^\infty J_0(kr) f_1(k,z) dk
\eeq
and similarly for $\phi_2(r,z)$. This expression corresponds to the potential of a point charge $Q$ on the boundary of two infinite half-spaces of permittivity $\vep_1$ and $\vep_2$ together with a 'correction' term that accounts for the presence of the grounded plates. The behaviour of $f_1(k,z)$ for large $k$ and is now given by
\beq
  f_1(k,z) \rightarrow -\frac{e^{-k(2b+z)}}{\vep_1+\vep_2}  - \frac{2\vep_2\,e^{-k(2g-z)} }{\vep_1+\vep_2} 
\eeq
The potential $\phi_1$ is defined for $-b<z<0$, so the expression decays exponentially in the entire volume, and even in the plane of the point charge at $z=0$, an upper integration limit of a multiple of $k=1/(b+g)$ will give an accurate evaluation. We can now repeat this procedure i.e. subtract the above expression from $f_1$ and perform the explicit integration of the two exponential terms and have
\bea
    \phi_1(r,z) &=&  \frac{Q}{2\pi (\vep_1+\vep_2)}\, \frac{1}{\sqrt{r^2+z^2}}  -\frac{Q}{2\pi (\vep_1+\vep_2)}\, \frac{1}{\sqrt{r^2+(2b+z)^2}} \\ \no
  &&  -\frac{2Q\vep_2}{2\pi (\vep_1+\vep_2)}\, \frac{1}{\sqrt{r^2+(2g-z)^2}} + \frac{Q}{2\pi} \int_0^\infty J_0(kr) f_2(k,z) dk
\eea
The two new terms correspond to two 'mirror' charges of values $-Q$ at $z=-2b$ and $-2\vep_2Q$ at $z=2g$. The process can be repeated ad infinitum and the potential is expressed as a sum of mirror charges with an integral term that contributes less and less, the more mirror charges we use. Explicitly we can write this in the following form
\beq 
  D(k) =\,e^{(b+g)k}(\vep_1+\vep_2)[1-p(b,g,k)]
\eeq
\beq
   p(b,g,k) = e^{-2(b+g)k}-\frac{\vep_1-\vep_2}{\vep_1+\vep_2}e^{-2bk}+\frac{\vep_1-\vep_2}{\vep_1+\vep_2}e^{-2gk} 
\eeq
We can verify that $0<p(b,g,k)<1 \, \forall (b,g,k, \vep_1, \vep_2)>0$ so we have
\beq 
       \frac{1}{D(k)} = \frac{e^{-(b+g)k}}{\vep_1+\vep_2}\,\frac{1}{1-p(b,g,k)} = \frac{e^{-(b+g)k}}{\vep_1+\vep_2}\sum_{n=0}^\infty p(b,g,k)^n
\eeq
Inserting this expression in Eq. \ref{two_layer_potential1} we find that the integrand consists of an infinite number of terms of the form $J_0(kr)e^{-\alpha k}$ so they can all be explicitly integrated with Eq. \ref{inverse_square_identity} and the potential is expressed as an infinite number of mirror charges. As an example we investigate the field for $\vep_1 = \vep_2 = \vep_0$ and have $p(b,g,k)=e^{-2(b+g)k}$
\beq
  \phi_1(r,z) = \frac{Q}{4\pi\vep_0} \sum_{n=0}^\infty J_0(kr) e^{-(b+g)k}
  \left(e^{gk}-e^{-gk}\right)
  \left(e^{(b+z)k}-e^{-(b+z)k}\right)\,
  e^{-2n(b+g)k}dk
\eeq
Evaluating all expressions with Eq. \ref{inverse_square_identity} and replacing $g \rightarrow d-z_0, b \rightarrow z_0, z \rightarrow z-z_0$ gives the correct expression for the potential of a point charge in an empty condenser as presented in \cite{riegler3}.
\\
\\
To conclude this section we evaluate the potential $\phi_1$ of Eq. \ref{two_layer_potential1}  by using the method of residuals as discussed in  Section \ref{integral_app}. The integrand has an infinite number of poles at $k_m=im\pi/(b+g)$ so the resulting expression is (Section \ref{integral_1})
\beq
  \phi_1(r,z)=\frac{Q}{2\pi \vep_0} \int_0^\infty J_0(kr)
  \frac{\sinh(gk) \sinh(k(b+z))}{\sinh(k(b+g))} dk = 
\eeq
\beq  
   =\frac{Q}{\pi \vep_0 (b+g)} \sum_{n=1}^\infty -(-1)^n 
  \sin \left( \frac{n\pi g}{b+g} \right)
  \sin \left(  \frac{n \pi (b+z)}{b+g}\right)
  K_0\left(  \frac{n\pi r}{b+g} \right)
\eeq
The expression diverges for $r=0$, so the numerical evaluation does not work at $r=0$ and convergence will be slow close to $r=0$. The expression is however well suited to evaluate the fields at large $r$ since 
the modified Bessel functions $K_0(x)$ behave as $e^{-x}/\sqrt{x}$ for large values of $x$.  So for $r>b+g$ the potential behaves as $1/\sqrt{r} \exp (-\pi r/(b+g)$), because the higher order terms $K_0(n\pi r/(b+g))\approx 1/\sqrt{r} \exp (-n\pi r/(b+g))$ all decay more rapidly with $r$.
\\
\\
In case the two layers have different permittivities we have to find the zeros of $D(k)$ in Eq. \ref{two_layer_determinant}. Since the zeros all lie only along the imaginary axis we write $k=iy$ and the equation reads as 
\beq
     \vep_1 \tan (gy) = -\vep_2 \tan (by)
\eeq
This is a transcendent equation and we can find the zeros only with numerical methods, however by plotting the two sides of the equation on top of each other we see that the first zero has to satisfy the condition
\beq
    \frac{1}{2b} < y_1 <\frac{1}{2g} \quad \mbox{for} \quad b>g \qquad \qquad 
    \frac{1}{2g} < y_1 <\frac{1}{2b} \quad \mbox{for} \quad b<g
\eeq
By evaluating the residual at $k_1$ we find a term $K_0(y_1 r)$ so we learn that for large values of $r$ the potential behaves as $e^{-y_1r}\sqrt{r}$.

\subsection{Potential of a point charge centred at $r_0, \varphi_0$}

In case the point charge in Fig. \ref{two_layer_sketch} is not centred at the origin but at a position $r_0, \varphi_0$  (Fig. \ref{N_layer_sketch}a), we have to replace $r$ by the distance $P$ between the charge and the observer point $r, \varphi$, which is given by
\beq
  P=\sqrt{r^2+r_0^2-2 r r_0 \cos (\varphi-\varphi_0)}
\eeq
Using the identity \cite{jackson}
\beq
  J_0(kP) = \sum_{m=-\infty}^\infty e^{im(\varphi - \varphi_0)}J_m(kr)J_m(kr_0)
\eeq
the solution becomes
\beq \label{point_charge_r0_phi0}
   \phi_1(r, \varphi, z) = \frac{1}{2 \pi} \int_0^\infty \sum_{m=-\infty}^\infty\,
    e^{im(\varphi - \varphi_0)}J_m(kr)J_m(kr_0)
    \left[
   A_1(k)e^{kz} + B_1(k) e^{-kz}
  \right] dk
\eeq
and equally for $\phi_2$.

\subsection{Potential of a point charge in a geometry grounded on a circle}

In case the geometry from Fig. \ref{two_layer_sketch} is not extended to infinity but grounded on a circular boundary at $r=c$ (Fig. \ref{N_layer_sketch}b), the condition that $\phi(r=c, \varphi)=0$ implies that $J_m(kc)=0$, and therefore only the values of $kc=j_{ml}$, where $j_{ml}$ is the $l^{th}$ zero of $J_m(x)$, are permitted. The solution of the problem is therefore written as
 \beq
   \phi_1(r, z) = \sum_{l=1}^\infty  \sum_{m=-\infty}^\infty\,
    e^{im(\varphi - \varphi_0)}J_m(k_{ml}r)J_m(k_{ml}r_0)
    \left[
   C_1(k_{ml})e^{k_{ml}z} + D_1(k_{ml}) e^{-k_{ml}z}
  \right] dk
\eeq
with $k_{ml}=j_{ml}/c$, and similar for $C_2, D_2$ of $\phi_2(r, z)$.  Three conditions for $C_1, D_1, C_2, D_2$ are equal to the ones for $A_1, B_1, A_2, B_2$ from Eqs. \ref{cond1}, \ref{cond2}, \ref{cond3} and for the fourth condition we use the relation and
\beq
   \vep_1\frac{\partial \phi_1}{\partial z}-\vep_2\frac{\partial \phi_2}{\partial z} = \frac{Q}{r}\delta(r-r_0)\delta(\varphi-\varphi_0)
\eeq
Multiplying the equation with $J_m(k_{ml'}r)$ and $e^{-im'\varphi}$ and employing the relations
\beq
   \int_0^{2\pi} e^{-im \varphi} e^{im'\varphi} d\varphi = 2\pi \delta_{mm'} \quad
  \int_0^crJ_m(k_{ml}r)J_m(k_{ml'}r)dr = \frac{c^2}{2}[J_{m+1}(j_{ml})]^2\delta_{ll'}
\eeq
yields
\beq
    \vep_1(C_1-D_1)-\vep_2(C_2-D_2) = \frac{Q}{c\pi}\, \frac{1}{j_{ml}[J_{m+1}(j_{ml})]^2}
\eeq
Comparing this to Eq. \ref{cond4} we find that the coefficients $C, D$ are related to $A, B$ by
\beq
   C_1(k_{ml}) = \frac{A_1(k_{ml})}{c\pi\,j_{ml}[J_{m+1}(j_{ml})]^2} \qquad 
   D_1(k_{ml}) = \frac{B_1(k_{ml})}{c\pi\,j_{ml}[J_{m+1}(j_{ml})]^2} 
\eeq
and similar for $C_2, D_2$.

\subsection{Potential of a point charge in a geometry grounded on a rectangle}

For the case where the geometry is grounded at $x=0,a$ and $y=0,b$ (Fig. \ref{N_layer_sketch}c) and the charge is placed at position $x_0, y_0$ we have to solve the Laplace equation in Cartesian coordinates, and the most general solution that satisfies these boundary conditions is 
\beq \label{grounded_rectangle}
  \phi_1(x,y,z) = \sum_{l=1}^\infty \sum_{m=1}^\infty 
  \sin \left( \frac{l\pi x}{a} \right) \sin \left( \frac{m \pi y}{b} \right)
   \left[
   E_1(k_{lm})e^{k_{lm}z} + F_1(k_{lm}) e^{-k_{lm}z}
  \right] 
\eeq
with $k_{lm}=\pi\sqrt{l^2/a^2+m^2/b^2}$. As before, three conditions for $E$ and $F$  are equivalent to the ones for $A$ and $B$, and the fourth conditions is  
\beq
   \vep_1\frac{\partial \phi_1}{\partial z}-\vep_2\frac{\partial \phi_2}{\partial z} = Q\delta(x-x_0)\delta(y-y_0)
\eeq
multiplying the equation with $\sin(l'\pi x/a)$ and $\sin(m'\pi y/b)$ and exploiting the orthogonality of the expressions gives
\beq
   \vep_1(E_1-F_1)-\vep_2(E_2-F_2) = \frac{Q \sin\frac{lx_0}{a} \sin\frac{ly_0}{b} }{ab k_{lm}}
\eeq
so the coefficients are related to $A$ and $B$ by
\beq 
    E_1(k_{ml}) =  \frac{A_1(k_{lm}) \sin\frac{lx_0}{a} \sin\frac{ly_0}{b} }{ab k_{lm}} \qquad 
    F_1(k_{ml}) =  \frac{B_1(k_{lm}) \sin\frac{lx_0}{a} \sin\frac{ly_0}{b} }{ab k_{lm}}
\eeq
We can find the solution for an infinitely extended geometry in Cartesian coordinates by shifting the coordinate system by $a/2$ and $b/2$ such that the origin is in the center of the rectangle and taking the limit of $a$ and $b$ to infinity. Writing $k_x=l\pi/a$ and $k_y=m\pi/b$ and replacing the sum by an integral $\sum_l \rightarrow \int dl = a/\pi \int dk_x$ and $\sum_m \rightarrow \int dm = b/\pi \int dk_y$ we find the solution as
\beq \label{two_layer_cartesian}
  \phi_1(x, y, z) = \frac{1}{\pi^2}\int_0^\infty \int_0^\infty \cos[k_x(x-x_0)]\cos[k_y(y-y_0)]
   \frac{1}{k}\left[
   A_1(k)e^{kz} + B_1(k) e^{-kz}
  \right] dk_x dk_y
\eeq
In case the geometry is grounded at $x=0, a$ but insulated at $y=0, b$ (Fig. \ref{N_layer_sketch}d), we have to employ the condition that $\partial \phi_n/\partial y=0$ at $y=0,b$ and derive a solution similar to  Eq. \ref{grounded_rectangle}, that will be quoted later.

\subsection{Weighting fields}

\begin{figure}[ht]
 \begin{center}
  \includegraphics[width=7cm]{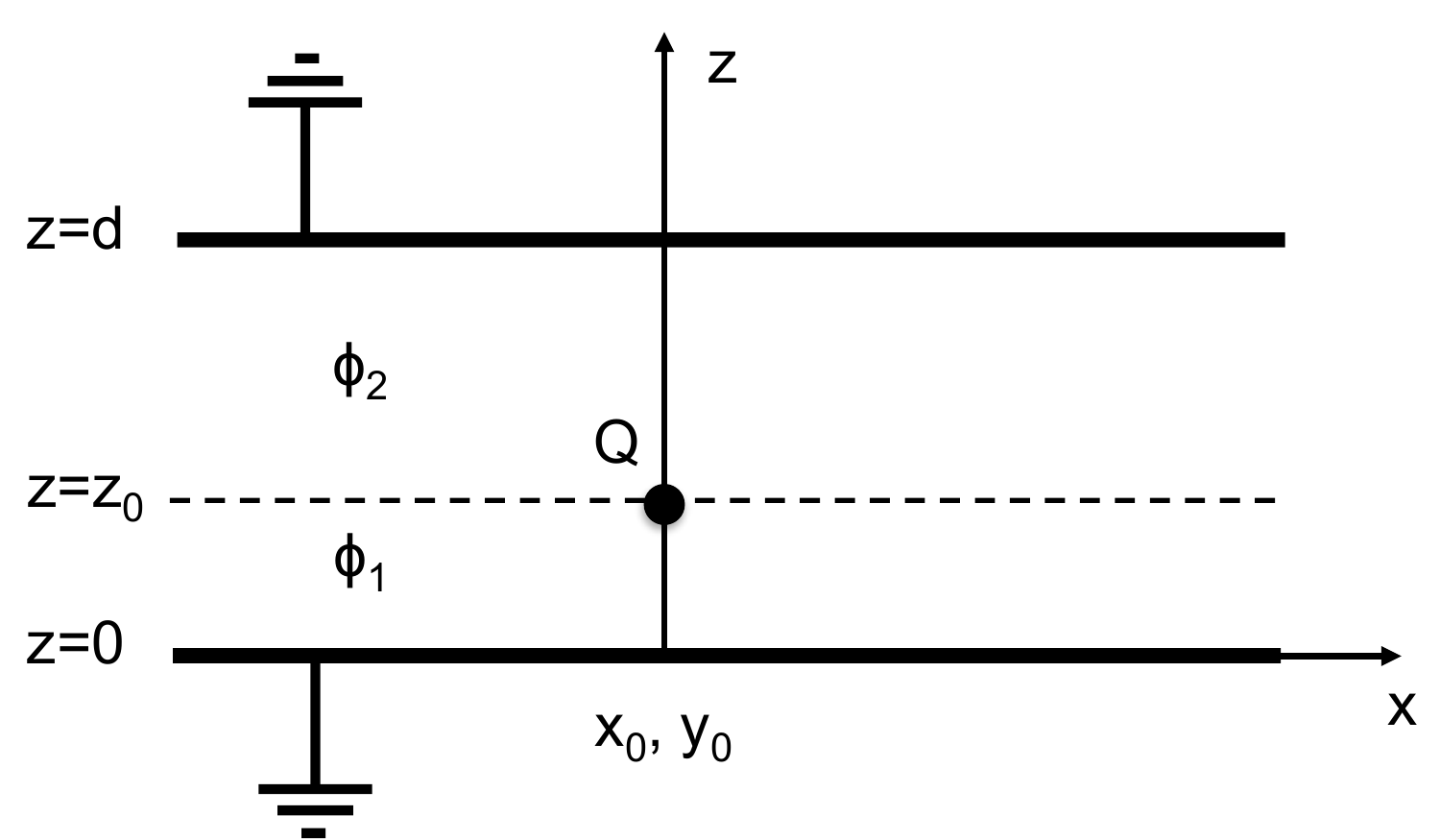}
  \includegraphics[width=7cm]{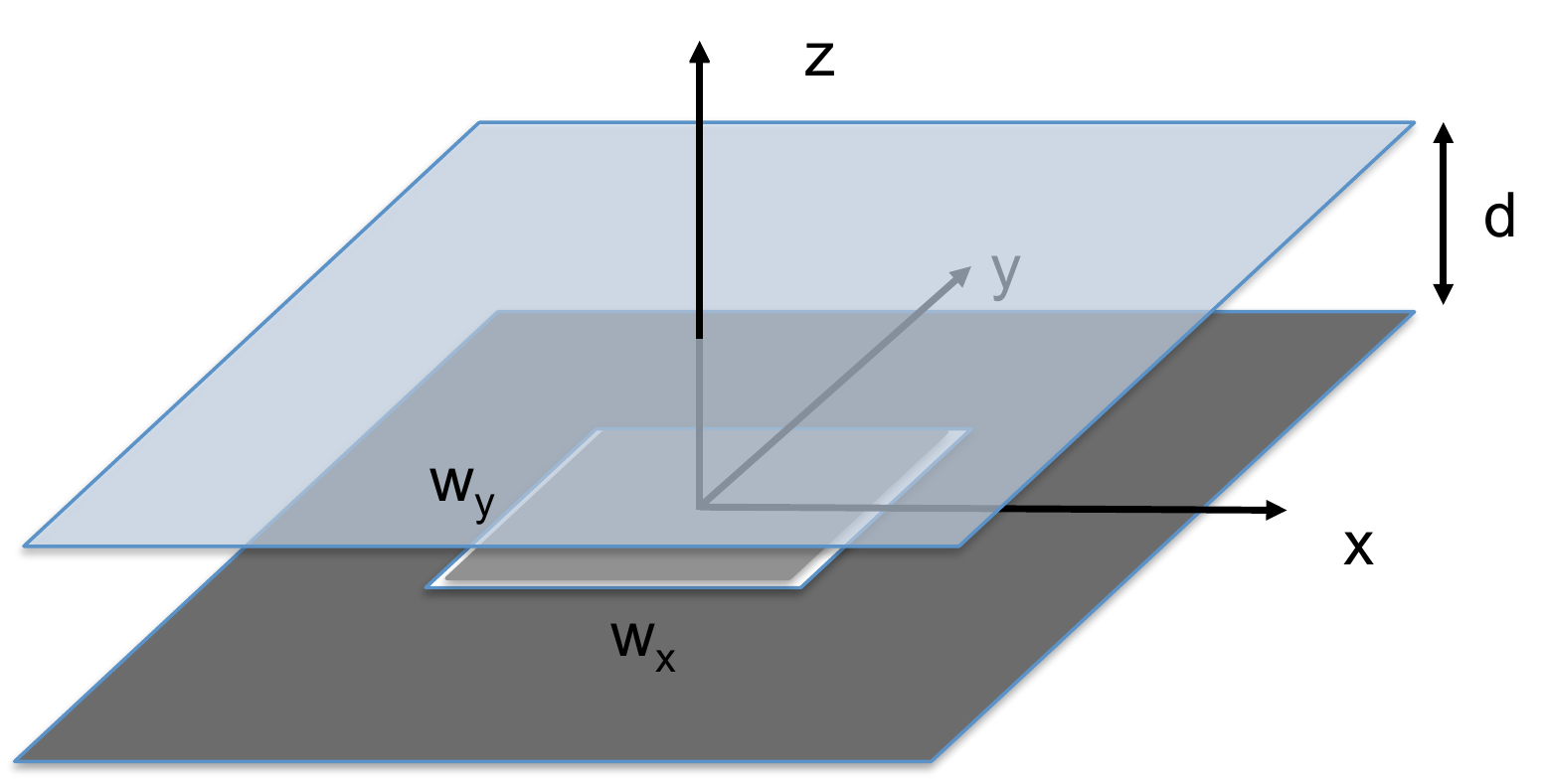}  
  \caption{Point charge in an empty condenser (left) and a rectangular readout pad (right).}
  \label{weighting_field_example}
  \end{center}
\end{figure}

In this section we want to calculate the weighting field of a rectangular pad centred at $x=y=0$ with a width of $w_x, w_y$ for the geometry of Fig. \ref{weighting_field_example}, which is infinitely extended and where the permittivity of both layers is equal to $\vep_0$. We use the solution in Cartesian coordinates (Eq. \ref{two_layer_cartesian}) and shift the coordinate system such that there is a grounded plate at $z=0$ and $z=g$ and the point charge is placed at $x_0, y_0, z_0$. Using the coefficients from Eq. \ref{two_layer_coefficients} and replacing $g=d-z_0, b=z_0$ we have 
\beq
  \phi_1(x,y,z,x_0,y_0,z_0) = \frac{Q}{\pi^2 \vep_0}
  \int_0^\infty \int_0^\infty \cos\left[ k_x(x-x_0)\right]\cos\left[ k_y(y-y_0)\right]\frac{\sinh(kz)\sinh [k(d-z_0)]}{k\,\sinh (kd)}dk_x dk_y
\eeq 
and $\phi_2$ is given by the same expression with $z$ and $z_0$ exchanged. The charge induced on the rectangular pad is related to the electric field on the surface by 
\beq
  Q_{ind}(x_0, y_0, z_0)=\int_{-w_x/2}^{w_x/2} \int_{-w_y/2}^{w_y/2} -\vep_0 \frac{\partial \phi_1}{\partial z}\vert_{z=0} dx dy
\eeq
Through the reciprocity theorem we know that $Q_{ind}=-Q/V_w \phi_w(x_0,y_0,z_0)$ where $\phi_w$ is the potential at $x_0, y_0, z_0$ in case the charge is 
removed and the pad is put to potential $V_w$. We therefore have 
\beq
  \phi_w(x, y, z) = \frac{4 V_w}{\pi^2} \int_0^\infty  \int_0^\infty
  \cos (k_x x) \sin (k_x \frac{w_x}{2})  \cos (k_y y) \sin (k_y \frac{w_y}{2})
   \frac{\sinh k(d-z)}{k_x k_y\sinh(kd)} dk_x dk_y 
\eeq
and
\beq
  \vec{E}_w = -\grad \phi_w
\eeq
Details of this expression are given in \cite{riegler3}.

\section{Electric fields and weighting fields in a N-layer geometry \label{point_charge_N_layer_section}}

In this section we generalize the results form the previous section to a geometry with an arbitrary number of layers. 

\subsection{Potential for N point charges}

\begin{figure}[ht]
 \begin{center}
  \includegraphics[width=6cm]{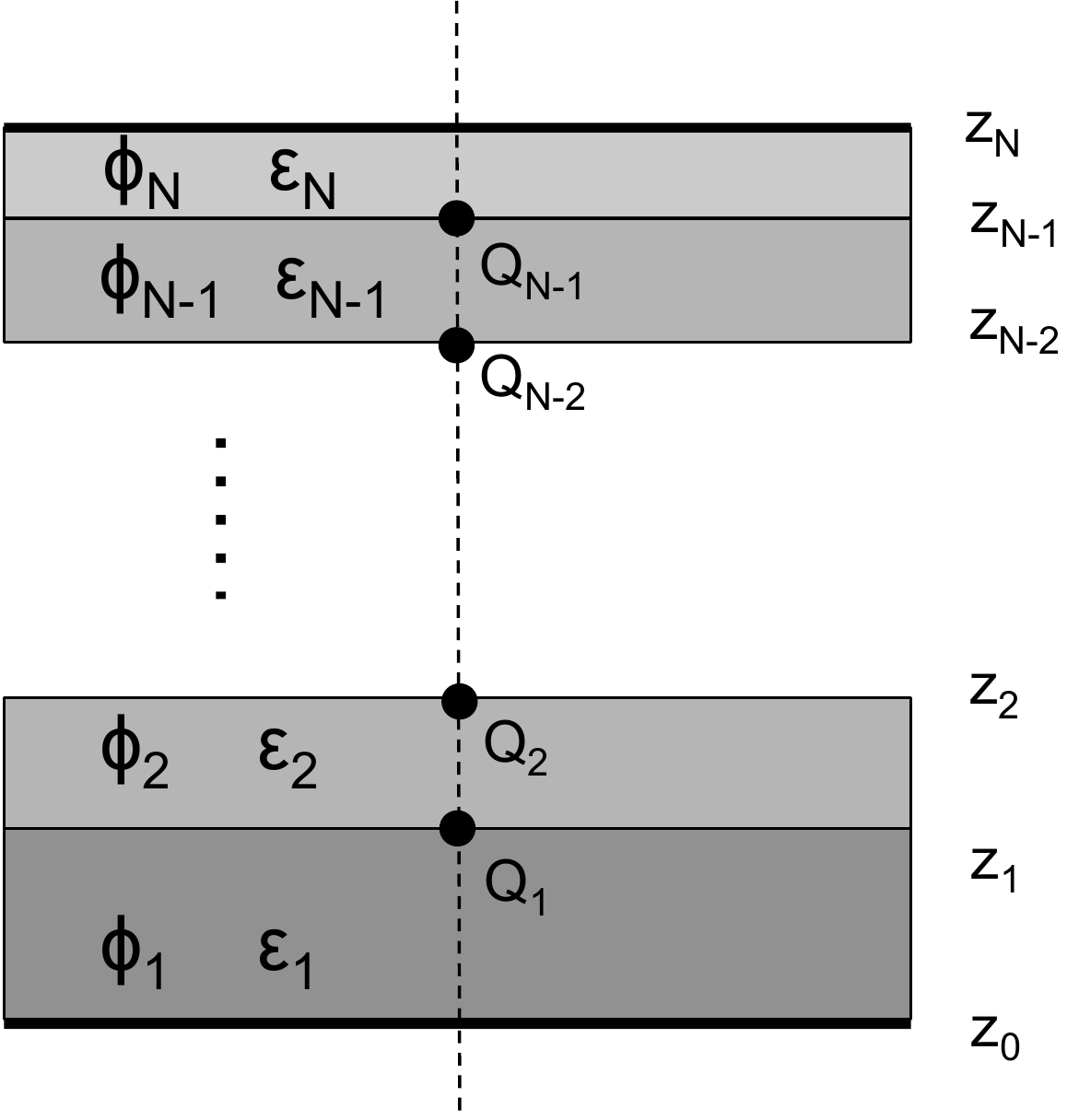}
    \includegraphics[width=8cm]{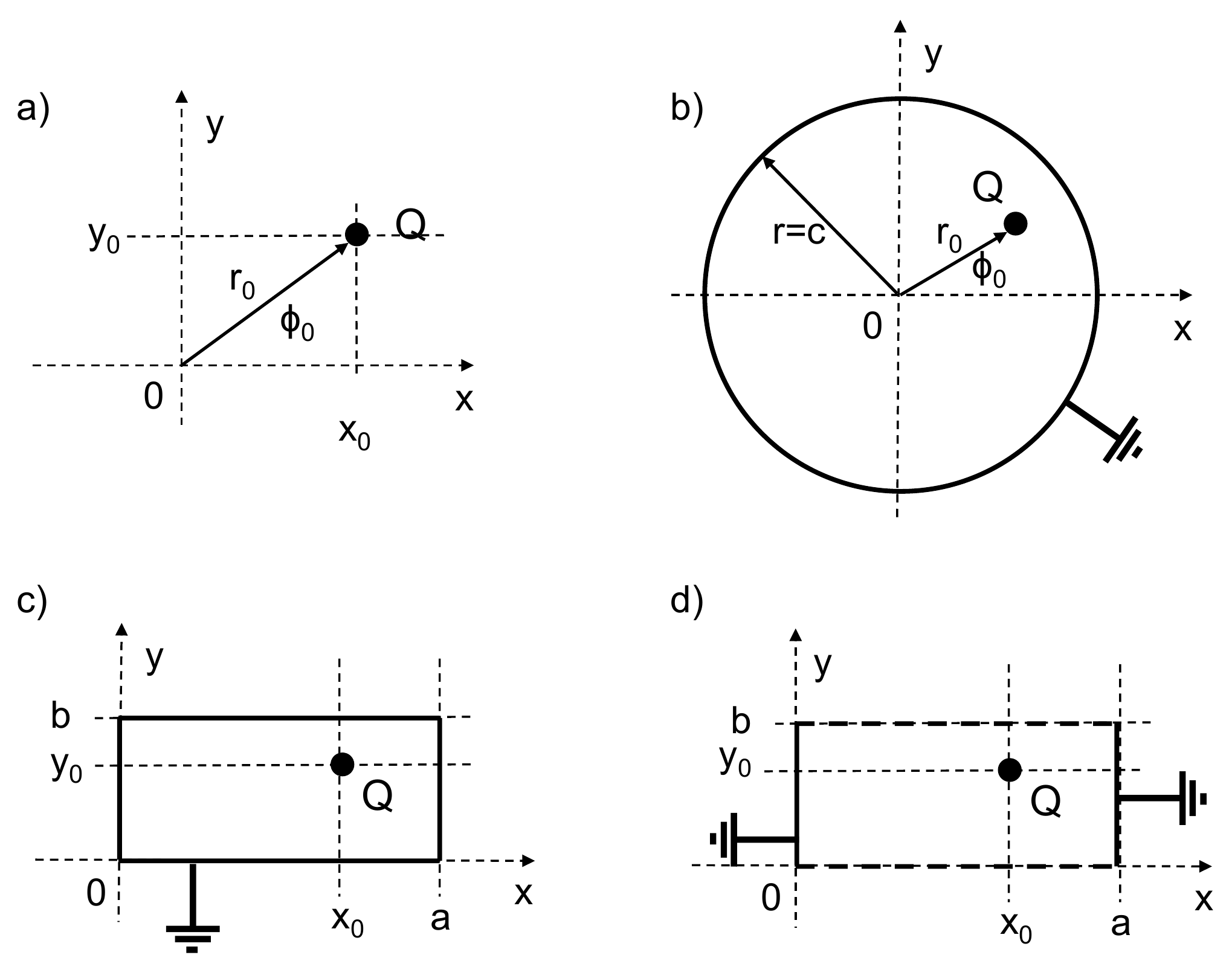}
  \caption{Left: A geometry of $N$ dielectric layers enclosed by grounded metal plates. On the boundary between two layers at $r=0$ there are point charges $Q_n$. Right: Different boundary conditions in the $x$-$y$ plane.}
  \label{N_layer_sketch}
  \end{center}
\end{figure}
We consider the geometry shown in Fig. \ref{N_layer_sketch}, for which the solutions can be written in analogy to the previous section. We assume $N$ dielectric layers ranging from $z_{n-1}<z<z_n$ of constant permittivity $\vep_n$. On the boundaries at $z=z_n$ there are charges $Q_n$. At $z=z_0$ and $z=z_N$ there are grounded metal plates. We define a characteristic function $f_n(k,z)$ for each layer as
\beq
  f_n(k,z) = A_n e^{kz}+B_n e^{-kz} \qquad n=1 ... N
\eeq
and can write the solutions for different boundaries:
\\
\\
For a geometry that extends to infinity in $x$-$y$ direction with the charges $Q_n$ at positions $r_0, \varphi_0$ (Fig. \ref{N_layer_sketch}a), the solution for the potential in layer $n$ in cylindrical coordinates is given by 
\beq \label{sol1}
  \phi_n(r, \varphi, z) = \frac{1}{2\pi}\int_0^\infty \sum_{m=-\infty}^\infty\, e^{im(\varphi-\varphi_0)}
  J_m(kr)J_m(kr_0) f_n(k,z) dk
\eeq
In case the charges are placed at $r_0=0$ the solutions are
\beq \label{sol1b}
     \phi_n(r,z) = \frac{1}{2\pi}\int_0^\infty J_0(kr) f_n(k,z) dk
\eeq
\\
\\
The solution for an infinitely extended geometry with the charges at position $x_0, y_0$ in Cartesian coordinates is given by ($k=\sqrt{k_x^2+k_y^2}$)
\bea \label{sol5}
  \phi_n(x,y,z) & = &   \frac{1}{\pi^2} \int_0^\infty \int_0^\infty
  \cos [k_x (x-x_0)] \cos [k_y (y-y_0)] 
     \frac{f_n(k, z)}{k}    dk_x dk_y
\eea
\\
\\
The solution for a geometry that is grounded on a boundary at radius $r=c$ (Fig. \ref{N_layer_sketch}b) with the charges at $r_0, \varphi_0$ is given by ($k_{ml}=j_{ml}/c$ where $j_{ml}$ is the $l^{th}$ zero of $J_m(x)$). 
\bea \label{sol2}
  \phi_n(r,z) =   \frac{1}{c\pi} \sum_{l=1}^\infty \sum_{m=-\infty}^\infty e^{im(\varphi-\varphi_0)}
   \frac{J_m(k_{ml} r) J_m(k_{ml}r_0)}{j_{ml} [J_{m+1}(j_{ml})]^2} \, f_n(k_{ml}, z) 
\eea
\\
\\
For the case where the geometry is grounded on a rectangle at $x=0, a$ and $y=0, b$  (Fig. \ref{N_layer_sketch}c) the solution is ($k_{lm} = \pi \, \sqrt{\frac{l^2}{a^2}+\frac{m^2}{b^2}}$)
\beq \label{sol3}
   \phi_n(x,y,z)  =  \frac{4}{ab} \sum_{l=1}^\infty \sum_{m=1}^\infty  
   \sin \left( {l\pi\frac{x}{a}}       \right)  
   \sin \left( {l\pi\frac{x_0}{a}}   \right)  
   \sin \left( {m\pi\frac{y}{b}}     \right) 
   \sin \left( {m\pi\frac{y_0}{b}} \right) 
    \, \frac{ f_n(k_{lm}, z)}{k_{lm}} 
\eeq
\\
\\
If the boundary is grounded at $x=0, a$ and insulated at $y=0, b$ (Fig. \ref{N_layer_sketch}d) the solution is ($k_{lm} = \pi \, \sqrt{\frac{l^2}{a^2}+\frac{m^2}{b^2}}$)
\bea \label{sol4}
   \phi_n(x,y,z)  =  \frac{4}{ab} \sum_{l=1}^\infty \sum_{m=0}^\infty  
   \sin \left({l\pi\frac{x}{a}}  \right)  
   \sin \left({l\pi\frac{x_0}{a}} \right)  
   \cos \left({m\pi\frac{y}{b}} \right) 
   \cos \left({m\pi\frac{y_0}{b}} \right) 
   \, \left(1-\frac{\delta_{0m}}{2} \right) \frac{f_n(k_{lm}, z)}{k_{lm}} 
\eea
\\
\\
The $2N$ coefficients $A_n(k)$ and $B_n(k)$ are defined by the two conditions at the grounded plates and at the $2(N-1)$ conditions at the $N-1$ dielectric interfaces
\beq \label{N_layer_equations1}
  A_1 e^{kz_0} + B_1 e^{-kz_0}  =  0  \qquad
  A_N e^{kz_N} + B_N e^{-kz_N}  =  0 
\eeq
\bdi
  A_n e^{kz_n} + B_n e^{-kz_n} 
  = A_{n+1} e^{kz_n} + B_{n+1} e^{-kz_n} 
\edi
\bdi
   \vep_n A_ne^{kz_n} - \vep_n B_n e^{-kz_n} 
  = \vep_{n+1}A_{n+1} e^{kz_n} - \vep_{n+1} B_{n+1}
  e^{-kz_n} + Q_n   
\edi
with $n=1...N-1$.  For solving these equations with symbolic equation manipulation programs it is useful to write them in matrix-form with a $2N\times 2N$ matrix $M$. Using the Kronecker  delta $\delta(n,m)$ we have for $m=1...2N$ 
\bea
  M_{1,m} &= & \delta (m,1)\,e^{kz_0}+  \delta (m,2) \,e^{-kz_0}   \\
  M_{2N,m} & = &\delta (m,2N-1)\,e^{kz_N} +  \delta (m,2N)\,e^{-kz_N} \no
\eea
and for $n=1 ... N-1$ and $m=1,2N$ 
\bea
  M_{2n,m}&=& \delta (m,2 n-1) \, e^{kz_n} 
          + \delta (m,2 n) \, e^{-kz_n}  \no \\
          &-& \delta (m,2 n+1) \, e^{kz_n} 
          - \delta (m,2 n+2) \, e^{-kz_n}  \\[5mm]
  M_{2n+1,m} & = &  \vep_n \delta (m,2 n-1) \, e^{kz_n} 
              -  \vep_n \delta (m,2 n) \, e^{-kz_n} \no \\
              &-&  \vep_{n+1} \delta (m,2 n+1) \, e^{kz_n} 
              +  \vep_{n+1} \delta (m,2 n+2) \, e^{-kz_n} \no
\eea
In addition we define the vectors $\vec a$ and $\vec b$ as
\beq
  \vec{a} = (A_1, B_1, A_2, B_2, ..., A_N, B_N)^T
\eeq
\beq
  \vec{b} = (0,0,Q_1,0,Q_2,0, ..., Q_{N-2}, 0, Q_{N-1}, 0)^T
\eeq
or in Matrixform
\beq
   a_{2n-1}=A_n \qquad a_{2n} = B_n \qquad n=1...N
\eeq
\beq
  b_{2n+1} = Q_n \qquad n=1...N-1
\eeq
The equation to solve is then
\beq \label{N_layer_equations2}
   M \vec{a} = \vec{b}  \qquad \rightarrow \qquad \vec{a} = M^{-1} \vec{b}
\eeq
\begin{figure}[ht]
 \begin{center}
 a)
  \includegraphics[width=7cm]{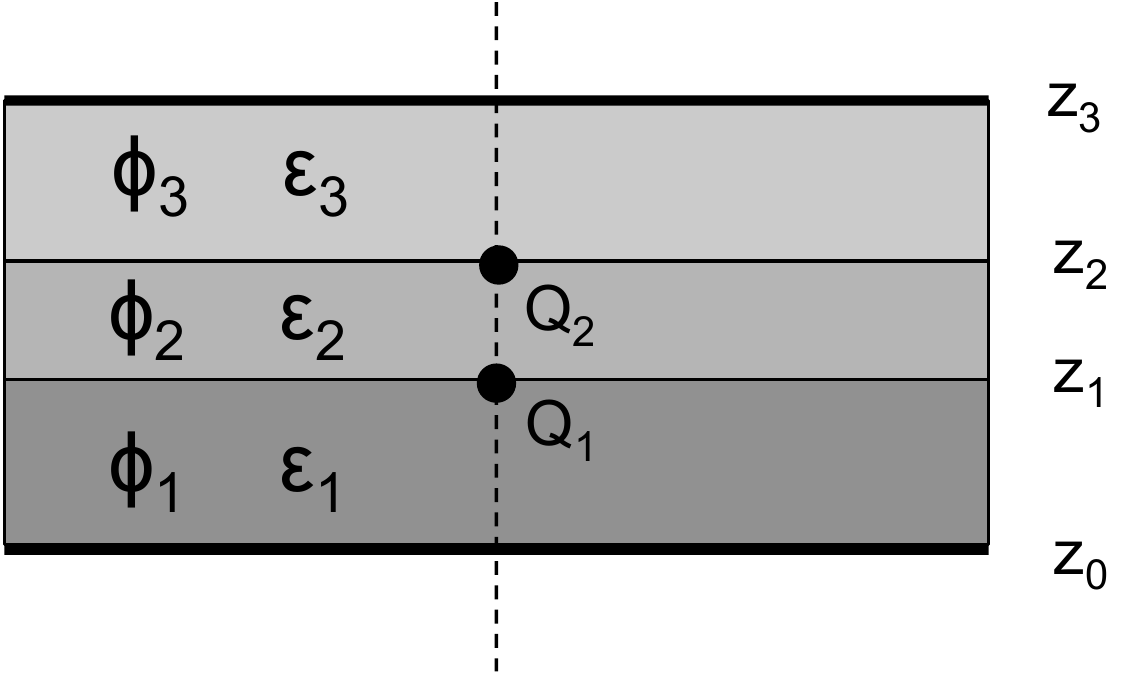}
  b)
   \includegraphics[width=7cm]{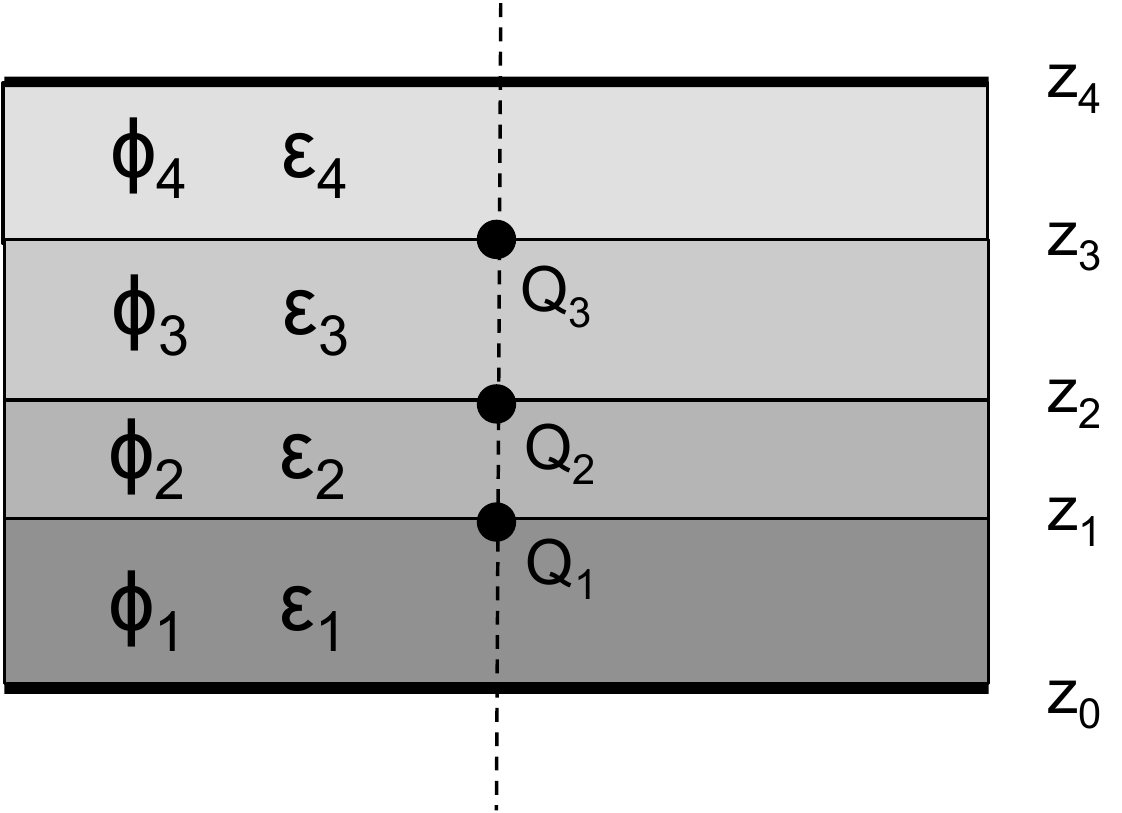}
  \caption{a) A geometry with three dielectric layers b) A geometry with four dielectric layers.}
  \label{34layer_sketch}
  \end{center}
\end{figure}
For later use we write down explicitly the matrix equation for the 3-layer geometry (Fig. \ref{34layer_sketch}a)
\beq  \label{matrix_3_layers}
      M = \left(
      \begin{array}{llllll}
      e^{kz_0} & e^{-kz_0} & 0 & 0 & 0 & 0\\
      e^{kz_1} & e^{-kz_1} &   -e^{kz_1} & -e^{-kz_1} & 0 & 0\\
      \vep_1e^{kz_1} & -\vep_1e^{-kz_1} & -\vep_2e^{kz_1} & \vep_2e^{-kz_1} & 0 & 0 \\
      0& 0 & e^{kz_2} & e^{-kz_2} &   -e^{kz_2} & -e^{-kz_2} \\
      0 & 0 &\vep_2e^{kz_2} & -\vep_2e^{-kz_2} & -\vep_3e^{kz_2} & \vep_3e^{-kz_2}  \\
     0 & 0 & 0 & 0 & e^{kz_3} & e^{-kz_3} \\
      \end{array}
   \right)
\eeq
\bdi
    \vec{a} = (A_1, B_1, A_2, B_2, A_3, B_3)^T \qquad  \vec{b} = (0, 0, Q_1, 0, Q_2, 0)^T
\edi
and 4-layer geometry (Fig. \ref{34layer_sketch}b)
\beq  \label{matrix_4_layers}
   M=\left(
      \begin{array}{llllllll}
      e^{kz_0} & e^{-kz_0} & 0 & 0 & 0 & 0 & 0 & 0\\
      e^{kz_1} & e^{-kz_1} &   -e^{kz_1} & -e^{-kz_1} & 0 & 0 & 0& 0 \\
      \vep_1e^{kz_1} & -\vep_1e^{-kz_1} & -\vep_2e^{kz_1} & \vep_2e^{-kz_1} & 0 & 0 & 0 & 0 \\
      0& 0 & e^{kz_2} & e^{-kz_2} &   -e^{kz_2} & -e^{-kz_2} & 0 & 0 \\
      0 & 0 &\vep_2e^{kz_2} & -\vep_2e^{-kz_2} & -\vep_3e^{kz_2} & \vep_3e^{-kz_2} & 0 & 0 \\
      0 & 0 & 0 & 0 & e^{kz_3} & e^{-kz_3} &   -e^{kz_3} & -e^{-kz_3}  \\
      0 & 0 & 0 & 0 &\vep_3e^{kz_3} & -\vep_3e^{-kz_3} & -\vep_4e^{kz_3} & \vep_4e^{-kz_3}  \\
      0 & 0 & 0 & 0 & 0 & 0 & e^{kz_4} & e^{-kz_4} \\
      \end{array}
   \right)
\eeq
\bdi
   \vec{a} = (A_1, B_1, A_2, B_2, A_3, B_3, A_4, B_4)^T  \qquad  \vec{b} = (0, 0, Q_1, 0, Q_2, 0, Q_3, 0)^T
\edi
\\ \\
We investigate the structure of the matrix $M$ to draw some conclusions on the general solutions. An inverse matrix can be written in the form $1/$det$(M)$ times powers $M^n$ and traces of tr$(M)$. Since all elements of $M$ have exponential factors of the form $e^{- k \alpha }$ and since the determinant, powers and traces of $M$ are all just sums and products of the matrix elements of $M$ we know that the characteristic functions $f_n$ for each layer are of the form
\beq
        f_n(k, z) = \frac{u_n(k)e^{kz}+v_n(k)e^{-kz}}{D(k)} \qquad   D(k)=\det (M)
\eeq
where $u_n(k), v_n(k)$ and $D(k)$ are expressions that just consist of sums of exponentials terms $e^{-k \alpha}$.  Inspecting the matrix shows that, except for $k=0$, the columns can never form linear dependent set of vectors for any value of $k>0$, so we know that $D(k)$ does not have any zeroes for $k > 0$.  This in turn means that $D(k)$ is either always positive or always negative for any value of $k \in \mathbb{R} >0 $. For evaluation of the integrals with the method of residues, the complex zeroes of $D(k)$ are therefore the relevant quantities. Specifically it can be shown that $D(k)$ is of the form 
\beq
    D(k) = (-1)^N\,e^{k (z_N-z_0)}
    \left(
    \prod_{m=1}^{N-1} (\vep_{m}+\vep_{m+1})
    \right)
    \left(
    1 + \sum_{m=1}^{2^N-1}  \gamma_m\,e^{-k\delta_m}       
    \right)
\eeq
with
\beq
    -1<\gamma_m(\vep_1 ... \vep_N)<1
    \qquad \delta_m(z_0 .... z_N)>0 \quad \mbox{for} \quad m =1 ... 2^N-1
\eeq
This allows $f_n$ to be written in the form
\beq
      f_n(k,z) =\frac{\sum_m \alpha_{nm}\,e^{-k\beta_{nm}}}{
    1 + \sum_{m=1}^{2^N-1}  \gamma_m\,e^{-k\delta_m}}
    \qquad \beta_{nm} (z_0 ... z_N, z) > 0 \qquad  \alpha_{nm} (\vep_1 ... \vep_N, Q_1 ... Q_{N-1})
\eeq
We can now expand the denominator around an appropriate value according to
\beq
  \frac{1}{1+x}= \sum_{n=0}^\infty (-1)^n \frac{(x-x_0)^n}{(1+x_0)^{n+1}} \qquad
  -1 < x < 2x_0+1
\eeq
Since we know that
\beq
     -1 < \sum_{m=1}^{2^N-1}  \gamma_m\,e^{-k\delta_m} < 2^N-1 \qquad \forall k>0
\eeq
we can use expand around $x_0=2^{N-1}-1$ and therefore express $f_n(k, z)$ an an infinite sum of expression of the form $e^{-k \alpha}$. Using Eq. \ref{inverse_square_identity} and we can therefore express the solution as an infinite sum of 'mirror charges'. This concludes the proof that the potential of a point charge in a general parallel layer geometry can be expressed by an infinite sum of 'free space point charge potentials'.
%
%
%
%
%
%
%
%
%
%
%
%
%
\\
\\
As pointed out in the previous section, the integrals of Eq. \ref{sol1}, \ref{sol1b} and \ref{sol5} are difficult to evaluate in the $z$-planes of the charges $Q_n$ i.e. at $z_n$. Using the technique of extracting the slowly converging parts from the integrand we can arrive at expressions that are easier to evaluate. We use the example for the infinitely extended geometry in cylindrical coordinates with the charges centred at $r_0=0$ and have for $n=1,N$ the expressions 
\bea
  \phi_n(r,z)&=&\frac{Q_{n-1}}{2\pi (\vep_{n-1}+\vep_n)} \, \frac{1}{\sqrt{r^2+(z-z_{n-1})^2}} 
   +  \frac{Q_{n}}{2\pi (\vep_n+\vep_{n+1})} \, \frac{1}{\sqrt{r^2+(z-z_n)^2}} \no \\
  & + &  \frac{1}{2\pi}\int_0^\infty J_0(kr)  \left( A_n e^{kz} + B_n e^{-kz} 
  -\frac{Q_{n-1}}{\vep_{n-1}+\vep_n}e^{-k(z-z_{n-1})} 
   -   \frac{Q_n}{\vep_n+\vep_{n+1}}e^{-k(z_n-z)} \no
 \right) dk 
\eea

\subsection{Inclusion of resistivity}

\begin{figure}[ht]
 \begin{center}
 a)
  \includegraphics[width=6cm]{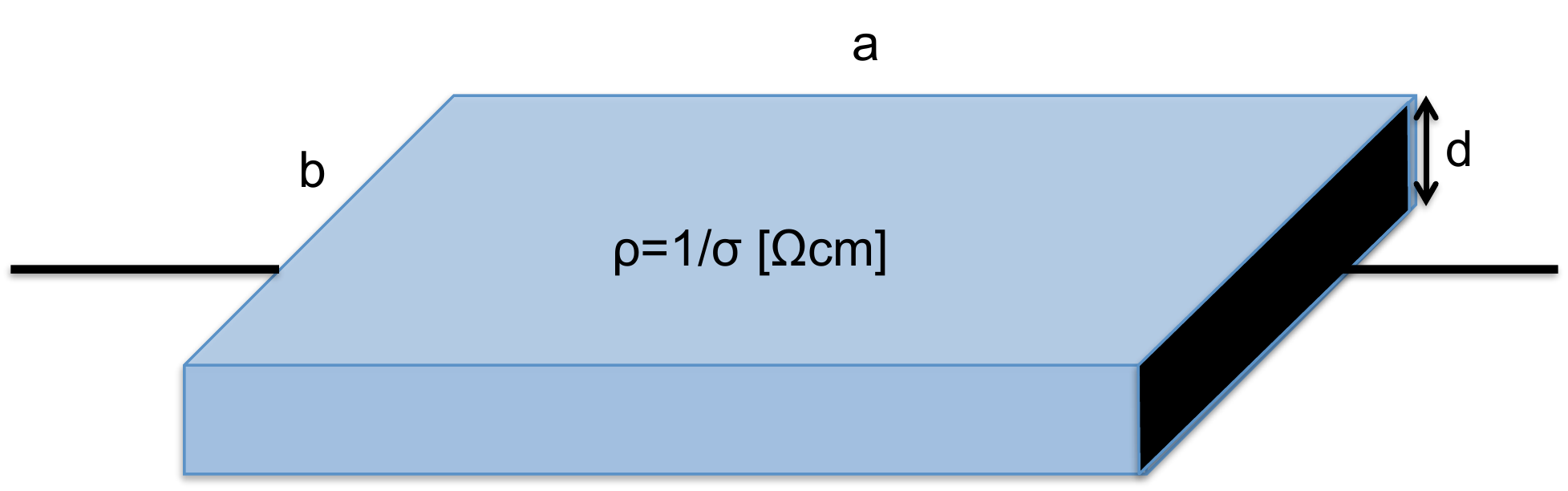}
  \qquad
   b) \includegraphics[width=6cm]{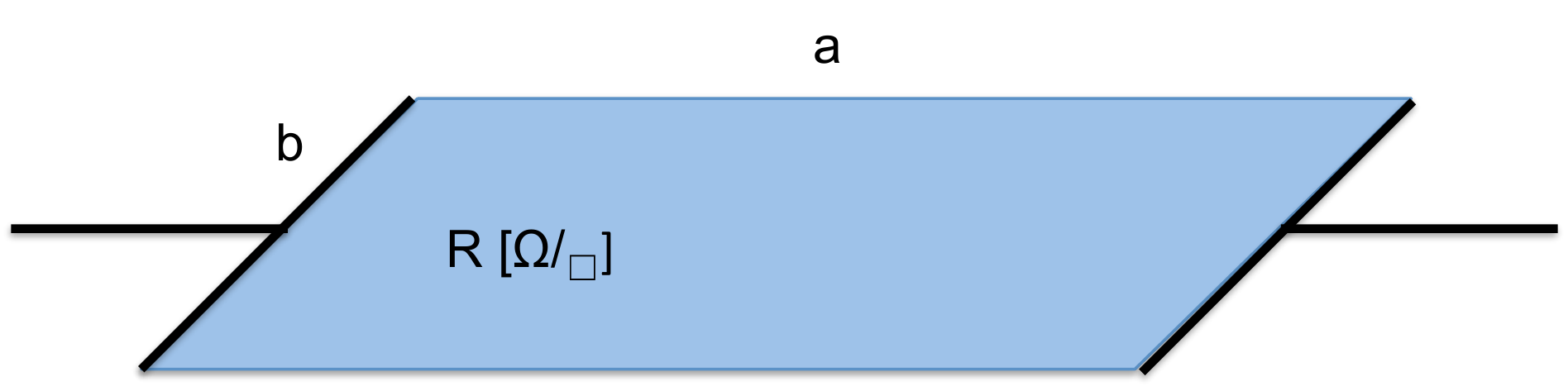}
  \caption{a) A block of material with volume resistivity $\rho\,[\Omega cm$]. b) A thin sheet of material with surface resisitivity $R\,[\Omega$/square].}
  \label{resistivity_explanations}
  \end{center}
\end{figure}
Using the quasi-static approximation of Maxwell's equations  \cite{schnizer1}, all results from the previous section can also be applied to geometries where the layers have finite conductivity. The conductivity of a material is defined by the volume resistivity $\rho\,[\Omega$cm]. Typical values of materials used for RPCs are $10^{10}\,\Omega$cm for Bakelite and $10^{12}\,\Omega$cm for glass. The conductivity is defined by $\sigma(\vx)=1/\rho(\vx)$. The current density $\vec {j}_1(\vx)\,$[A/cm$^2$] inside the resistive layer is related to the electric field inside the layer by  $\vec{j}_1(\vx)=1/\rho \vec{E}(\vx)$. The resistance represented by the material block in Fig. \ref{resistivity_explanations}a is therefore given by $\rho \,a/(bd)$.
\\ \\
If we make the resistive layer very thin, the current can only flow in '2 dimensions' and the current density $\vec{j_2}(x,y)\,$[A/cm] is related to the electric field inside the layer by  
\beq
     \vec{j}_2(x,y)=\vec{j}_1(x,y) d =  \frac{d}{\rho}  \vec{E}(x,y) = \frac{1}{R}  \vec{E}(x,y) \qquad R=\frac{\rho}{d}
\eeq
The resistance represented by the resistive sheet in Fig. \ref{resistivity_explanations}b is given by $R \,a/b$.
We can therefore conclude that for layers that have finite conductivity $\sigma_n=1/\rho_n$, where $\rho_n$ represents the volume resistivity of the layer, we find the fields in the Laplace domain by replacing $\vep_n$ by $\vep_n+1/(\rho_n s)$ in all expressions.  In case we want a specific layer $m$ i.e. $z_{m-1} < z < z_m$ to represent a thin sheet of a given surface resistivity $R\,[\Omega$/square], we have to replace $\vep_m$ of this layer by 
\beq
   \vep_m \quad \rightarrow \quad \vep_m + \frac{1}{(z_m-z_{m-1})Rs} 
 \eeq
In case we want to make this layer infinitely thin we have to perform the limit $\lim_{z_m\rightarrow z_{m-1}} \phi_n$ for all expressions.
\\ \\
If we use the static solutions for charges $Q_1, Q_2, ..., Q_N$, replace $\vep_m$ by $\vep_m+1/(\rho_m s)$ 
and perform the inverse Laplace transforms of the expressions, we find the time dependent fields for the case where charges $Q_1\delta(t), Q_2 \delta(t), ..., Q_N\delta(t)$ are placed on the boundaries of the layers, since we have the Laplace transform $\LL{Q\delta(t)}=Q$. \\ In case we want the solutions for the situation where charges $Q_n$ are placed at $t=0$, i.e. $Q(t)=Q\,\Theta(t)$, with $\Theta(t)$ being the Heaviside step function, we have to replace the $Q_n$ in the static solutions by $Q_n/s$, since we have $\LL{Q\Theta(t)}=Q/s$. \\
In case there are currents $I_n$ placed on the resistive layers we have $Q_n(t)=I_n\,t$ and therefore $\LL{Q_n(t)}=I_n/s^2$, so we have to replace the $Q_n$ of the static solutions by $I_n/s^2$ before performing the inverse Laplace transform.
\\ \\ 
Finally we note that in many occasions we are interested in the potentials and fields at $t=0$ and for long times $t \rightarrow \infty$. These expressions can be directly calculated in the Laplace domain and there is no need to perform the inverse Laplace transform, since the following relations hold:
\beq
  F(s) = \LL{f(t)} \qquad 
  f( t \rightarrow \infty ) = \lim_{s \rightarrow 0} sF(s) \qquad 
  f( t \rightarrow 0 ) = \lim_{s \rightarrow \infty} sF(s) 
\eeq

\subsection{Weighting fields \label{weighting_field_section}}

\begin{figure}[ht]
 \begin{center}
a)
\includegraphics[width=4.5cm]{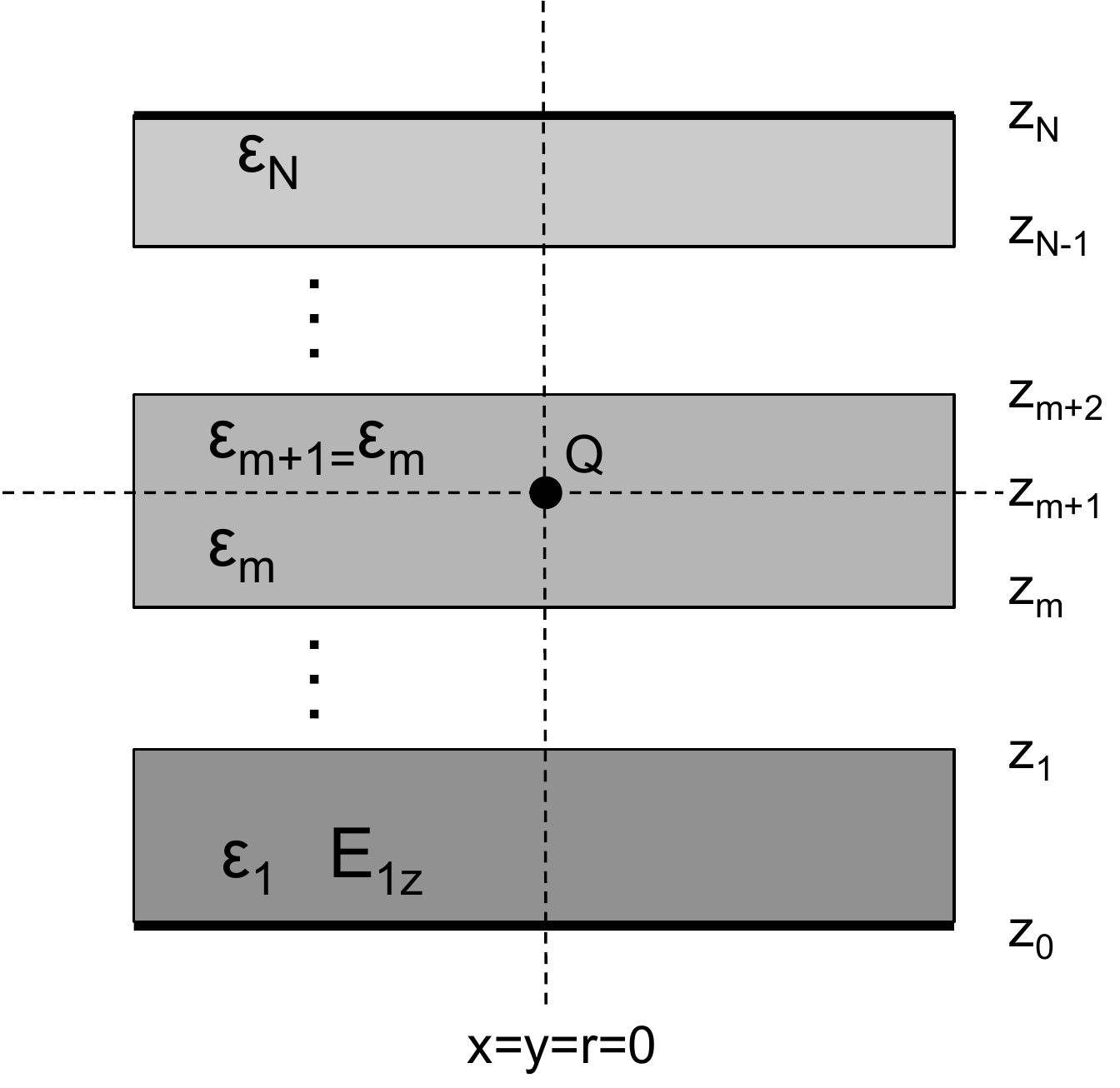}
b)
\includegraphics[width=4.5cm]{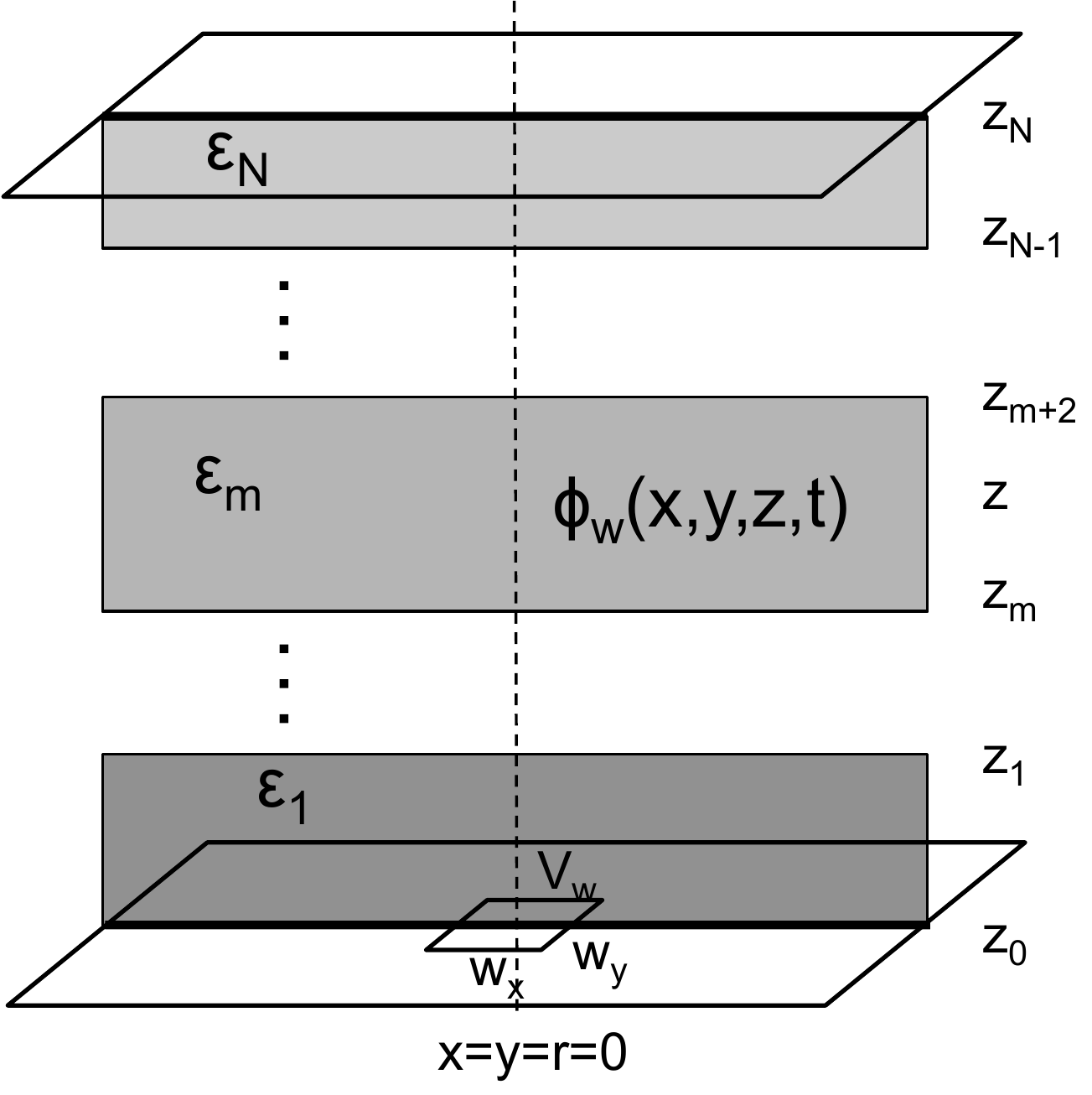}
c)
\includegraphics[width=4.5cm]{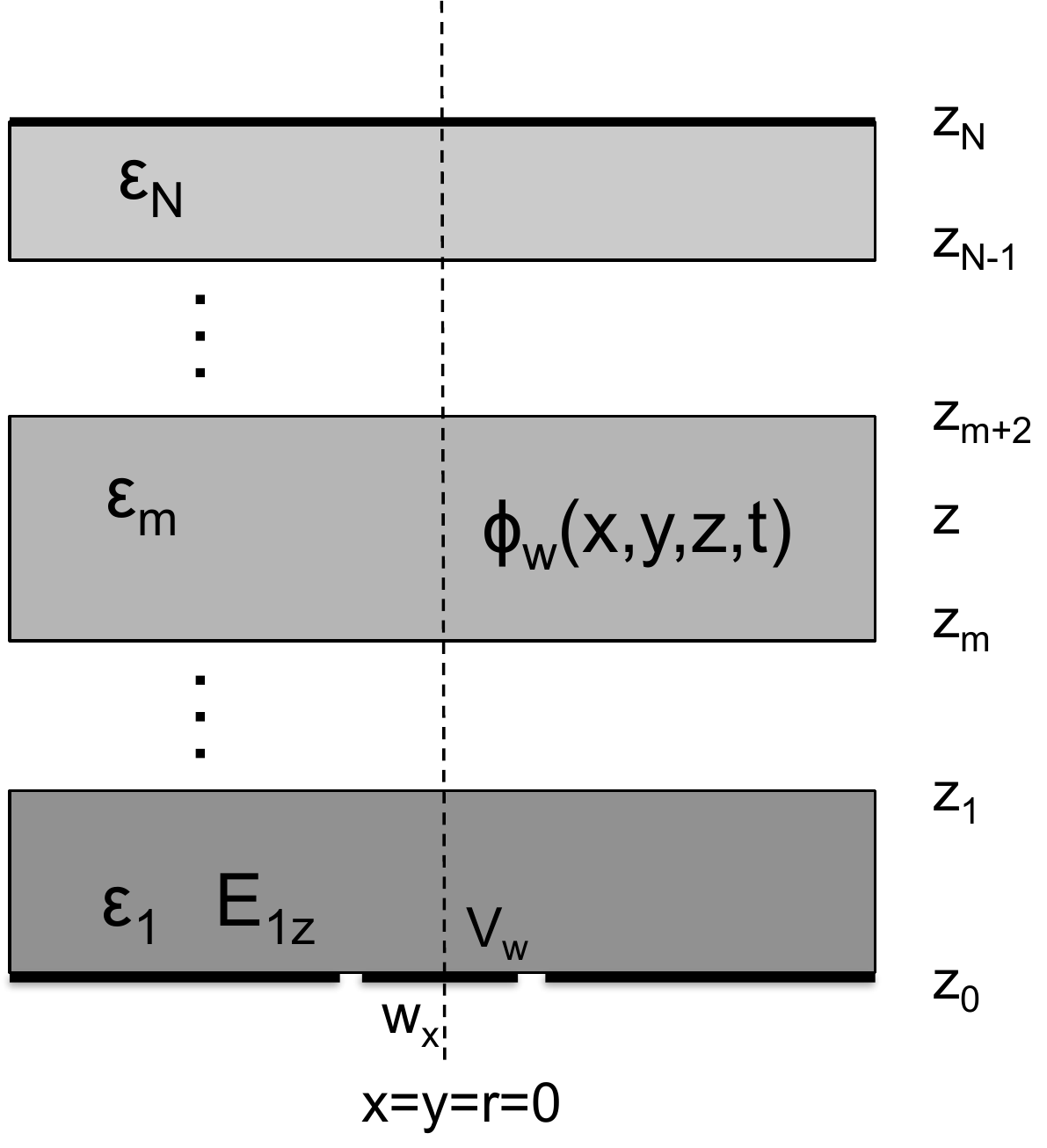}
  \caption{a) Point charge in a N layer geometry. b) Potential $\phi_w$ due to a rectangular pad at potential of $V_w$.  c) Potential $\phi_w$ due to an infinitely extended strip at potential $V_w$. }
  \label{N_layer_weighting}
  \end{center}
\end{figure}

Before moving to explicit geometries we investigate the general formulas that allow the calculation of signals that are induced on one of the grounded electrodes by the movement of charges in the different layers. We assume the geometry of Fig. \ref{N_layer_weighting} where a point charge is placed between $z_m$ and $z_{m+2}$.
\\ \\
We calculate coefficients $A_n$ and $B_n$ for the geometry where the layers below and above the point charge have the same permittivity $\vep_m$. The electric field on the surface of the grounded plate at $z=z_0$ is related to the induced surface charge density $q$ by $q = \vep_1 E_{z_0}$ so we find the charge $Q_{ind}$ induced on an area $A$ of the metal surface to be 
\beq
   q(x, y) = -\vep_1 \, \frac{\partial \phi_1}{\partial z}\,\vert_{z=z_0}  \qquad 
   Q_{ind} = \int\int_A  q(x, y) \, dx dy
\eeq 
By the reciprocity theorem we have $Q_{ind} = -Q/V_w\,\phi_w (x_0, y_0, z_{m+1})$ where $\phi$ is the potential at position $x_0, y_0, z_{m+1}$ in case the point charge $Q$ is removed and the area $A$ on the grounded plate is set to potential $V_w$ while the rest stays grounded, so we have
%
%
%
%
%
%
%
%
%
\beq
   \phi^w_n(x_0, y_0, z_{n+1}) = \vep_1 \frac{V_w}{Q}\int \int_A  \frac{\partial \phi_1}{\partial z}\,\vert_{z=z_0} dx dy 
   \qquad
   E_w = -\grad \phi_w
\eeq
%
%
%
%
%
%
%
%
%
%
%
%
and therefore
\bea \label{weighting_field_formula1}
    \phi^w_n(x, y, z) & = &
     \vep_1 \frac{V_w}{Q}\frac{4}{\pi^2}\int_0^\infty \int_0^\infty        
     \frac{\cos (k_x x) \sin (k_x w_x/2) \cos (k_y y) \sin (k_y w_y/2)}{k_x k_y} \no \\
     &&  \times \left[ A_1(k,z_{n+1}=z) e^{kz_0}-B_1(k, z_{n+1}=z) e^{-kz_0} \right] dk_x dk_y
\eea
For the case of an infinitely long strip, i.e. $w_y \rightarrow \infty$ we change variables to $s_y=k_y w_y/2$, let $w_y \rightarrow \infty$ and use $\int_0^\infty \sin(s_y)/s_yds_y = \pi/2$ which gives
\beq \label{weighting_field_formula2}
     \phi^w_n(x,z) = \vep_1 \frac{V_w}{Q}\frac{2}{\pi}\int_0^\infty 
       \frac{\cos (k x) \sin (k w_x/2)}{k} 
        \times \left[ A_1(k, z_{n+1}=z) e^{k z_0}-B_1(k, z_{n+1}=z) e^{-kz_0} \right] dk
\eeq
In case also $w_x$ goes to infinity we have the weighting potential of the entire electrode which becomes
\beq \label{weighting_field_formula3}
     \phi^w_n(z) = \vep_1 \frac{V_w}{Q} \left[ A_1(k=0, z_{n+1}=z) -B_1(k=0, z_{n+1}=z)  \right]
\eeq
In this case the weighting field and potential can be evaluated to 
\beq
    E^w_n = \frac{V_w}{\vep_n} \left( \sum_{m=1}^N \frac{z_m-z_{m-1}}{\vep_m} \right)^{-1} \qquad
    z_{n-1} < z < z_n
\eeq
\beq
    \phi^w_n(z) = V_w-\sum_{m=1}^{n-1} (z_m-z_{m-1})E^w_m - (z-z_{n-1})E^w_n \qquad
    z_{n-1}<z<z_n
\eeq

\clearpage

\section{Geometry representing a Resistive Plate Chamber}

In this section we present the explicit formulas for some common RPC geometries.

\subsection{Single layer RPC}

\begin{figure}[ht]
 \begin{center}
 a)
  \includegraphics[width=7cm]{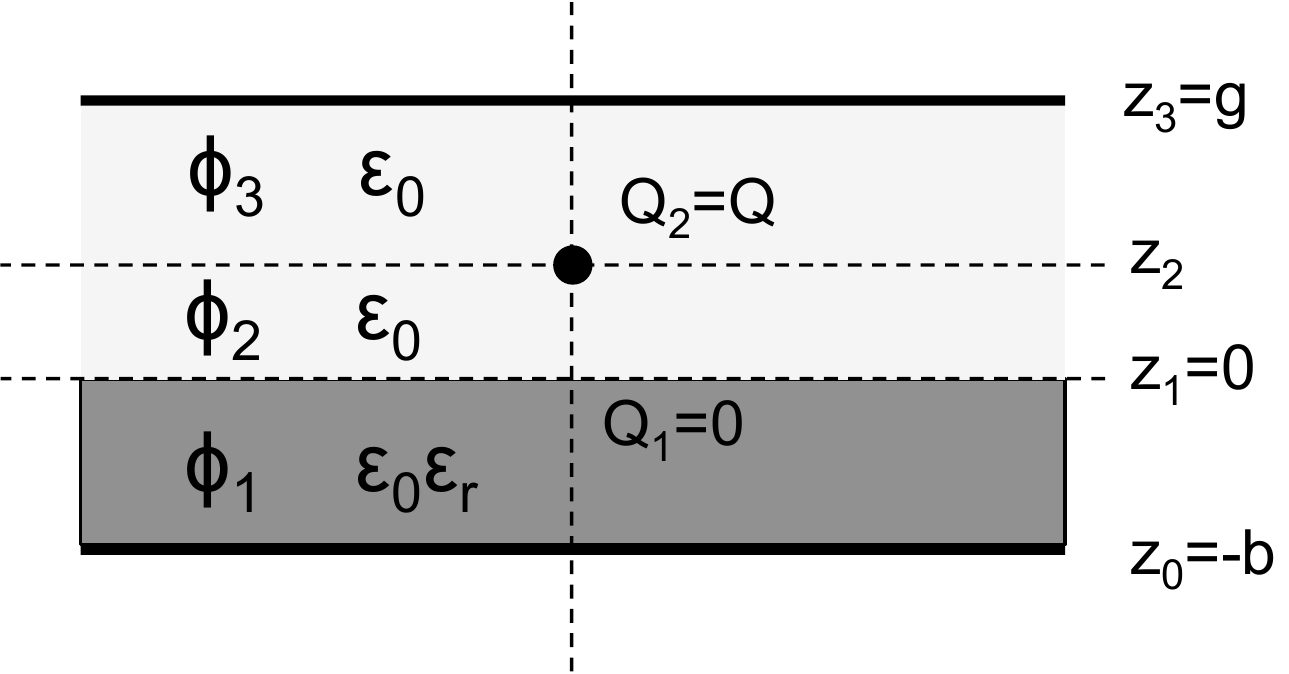}
  b)
  \includegraphics[width=7cm]{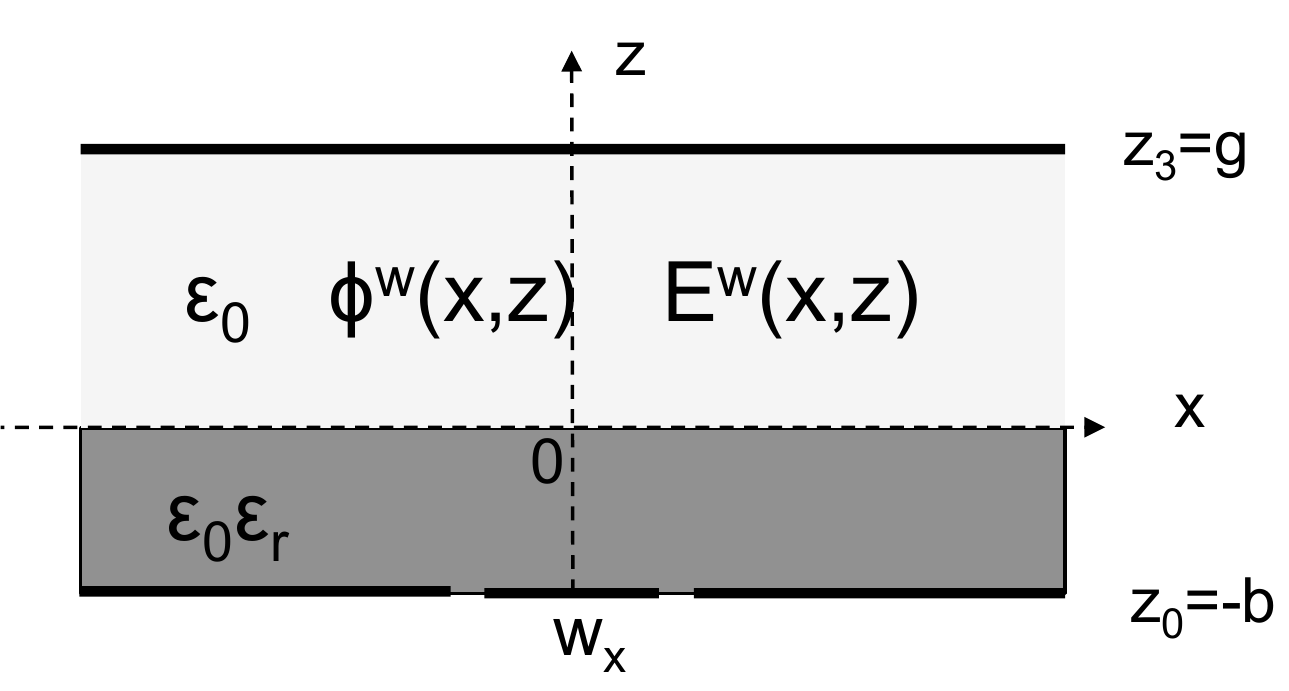}
  \caption{a) A geometry with three layers and one point charge representing e.g. a single gap RPC. b) Weighting field for a strip electrode of width$w_x$ and infinite extension in $y$-direction.}
  \label{rpc_figures}
  \end{center}
\end{figure}

As a first application of the formalism developed in the previous sections we investigate a geometry with 3 layers, shown in Fig. \ref{rpc_figures}a, that represents e.g. a single gap RPC with one resistive layer. 
To find the coefficients $A_1, B_1, A_2, B_2, A_3, B_3$ for this geometry we have to solve the equations Eq. \ref{N_layer_equations1}, \ref{N_layer_equations2} that are explicitly written in Eq. \ref{matrix_3_layers}. We set
\beq
  z_0 = -b \qquad z_1=0 \qquad z_3=g \qquad \vep_1=\vep_r\vep_0 \qquad \vep_2=\vep_3=\vep_0 \qquad 
  Q_1=0 \qquad Q_2 = Q
\eeq
and get for the the characteristic functions $f_n(k,z)$ 
\bea \label{f_rpc_geometry}
  f_1(k,z) & = & Q \sinh (k (b+z)) \sinh (k (g-\text{z_2}))/(\vep_0 D(k))\\
 f_2(k,z) & = &  Q \sinh (k (g-\text{z_2})) [\sinh (b k) \cosh (k z)+\text{\vep_r} \cosh (b k) \sinh (k z)]/(\vep_0 D(k))\\
 f_3(k,z)  & = &  Q \sinh (k (g-z)) [ \sinh (b k) \cosh (k \text{z_2})+\text{\vep_r} \cosh (b k) \sinh (k \text{z_2})]/(\vep_0 D(k))
\eea
with
\bdi
   D(k)=\sinh (b k) \cosh (g k)+ \text{\vep_r} \cosh (b k) \sinh (g k)
\edi
This solution can now be used to calculate the potential and electric field due to charges inside the gas gap of the RPC, which is essential for studies of space-charge effects in these detectors.
\beq
   \phi_2(r,z) = \frac{1}{2\pi} \int_0^\infty J_0(kr) f_2(k,z)dk \qquad
   \phi_3(r,z) = \frac{1}{2\pi} \int_0^\infty J_0(kr) f_3(k,z)dk 
\eeq
As pointed out earlier the numerical evaluation of the integral is difficult when $z$ is close to the 'plane' at $z=z_2$ where the charge is sitting. By using the trick described in Section \ref{section_divergence_removal} the expression can be written as
\beq
     \phi_2(r,z) = \frac{Q}{4\pi\vep_0 \sqrt{r^2+(z_2-z)^2}}+\frac{1}{2\pi} \int_0^\infty J_0(kr) 
     \left[f_2(k,z)- \frac{Q}{2\vep_0}\,e^{-k(z_2-z)} \right] dk
\eeq
\beq
     \phi_3(r,z) = \frac{Q}{4\pi\vep_0 \sqrt{r^2+(z-z_2)^2}}+\frac{1}{2\pi} \int_0^\infty J_0(kr) 
     \left[f_3(k,z)- \frac{Q}{2\vep_0}\,e^{-k(z-z_2)}  \right] dk
\eeq
The expressions represent a point charge $Q$ in free space together with a term that accounts for the presence of the dielectric layer and the grounded plates, which is more suited for numerical evaluation. As shown in Section \ref{section_divergence_removal} and in \cite{riegler3} one can continue to use further 'mirror charges' to reduce the contribution integral term to arbitrarily small values.
\\
\\
To find the weighting potential for a readout pad or readout strip we evaluate the Eq. \ref{weighting_field_formula1},  \ref{weighting_field_formula2}, \ref{weighting_field_formula3} and have 
\beq \label{rpc_weighting_pixel}
 \phi^w(x,y,z) =   \frac{4 \vep_r V_w}{\pi^2 }\int_0^\infty \int_0^\infty  
  \frac{\cos (k_x x) \sin (k_x w_x/2) \cos (k_y y) \sin (k_y w_y/2) \sinh(k(g-z))}{k_x k_y D(k)} dk_x dk_y
\eeq
\beq \label{rpc_weighting_strip}
     \phi^w(x, z)=  \frac{2 \vep_r V_w}{\pi} \int_0^\infty \frac{\cos (k x) \sin (k w_x/2) \sinh(k(g-z))}{k D(k)}  dk
\eeq
\beq  \label{rpc_weighting_inf}
     \phi^w(z)=  \frac{\vep_r V_w (g-z)}{b+\vep_rg} \qquad E^w_z =\frac{\vep_r V_w}{b+\vep_r g}
\eeq
\begin{figure}[ht]
 \begin{center}
 a)
  \includegraphics[width=7cm]{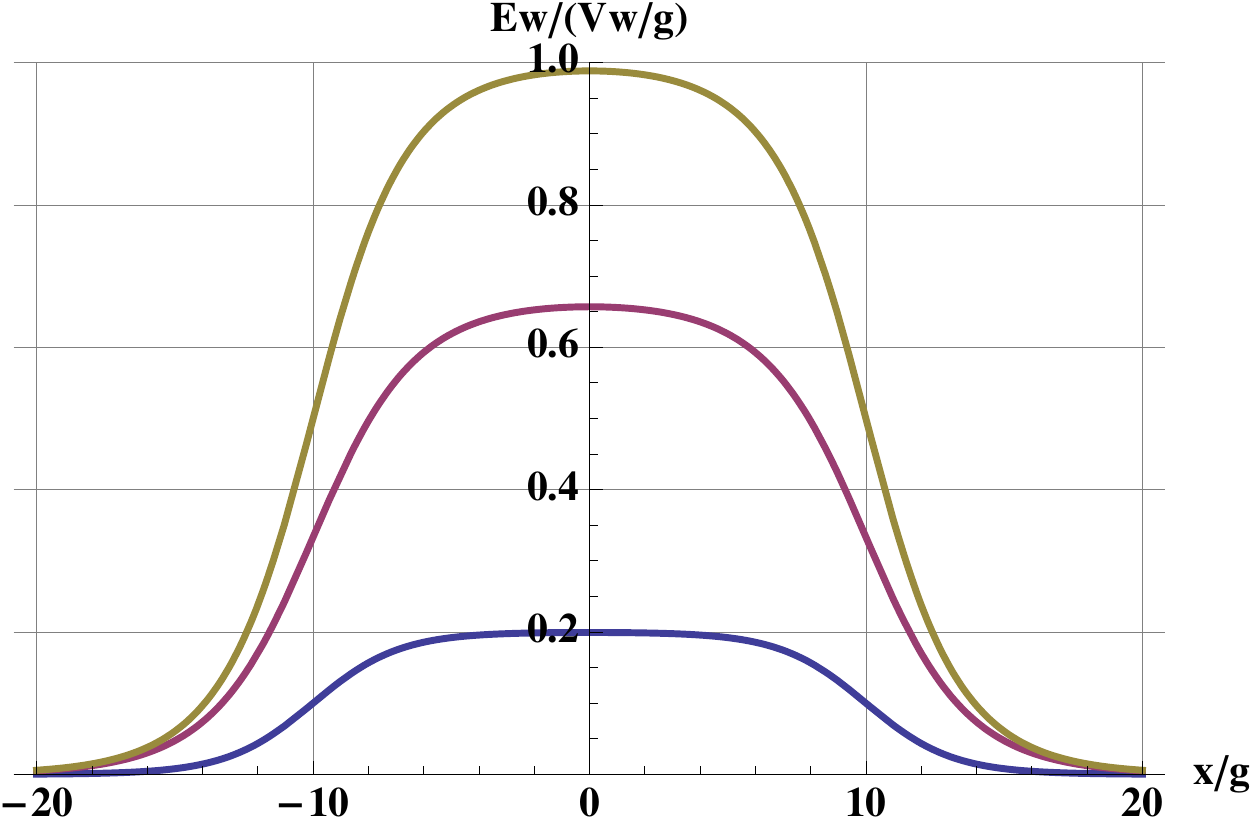}
  b)
  \includegraphics[width=7cm]{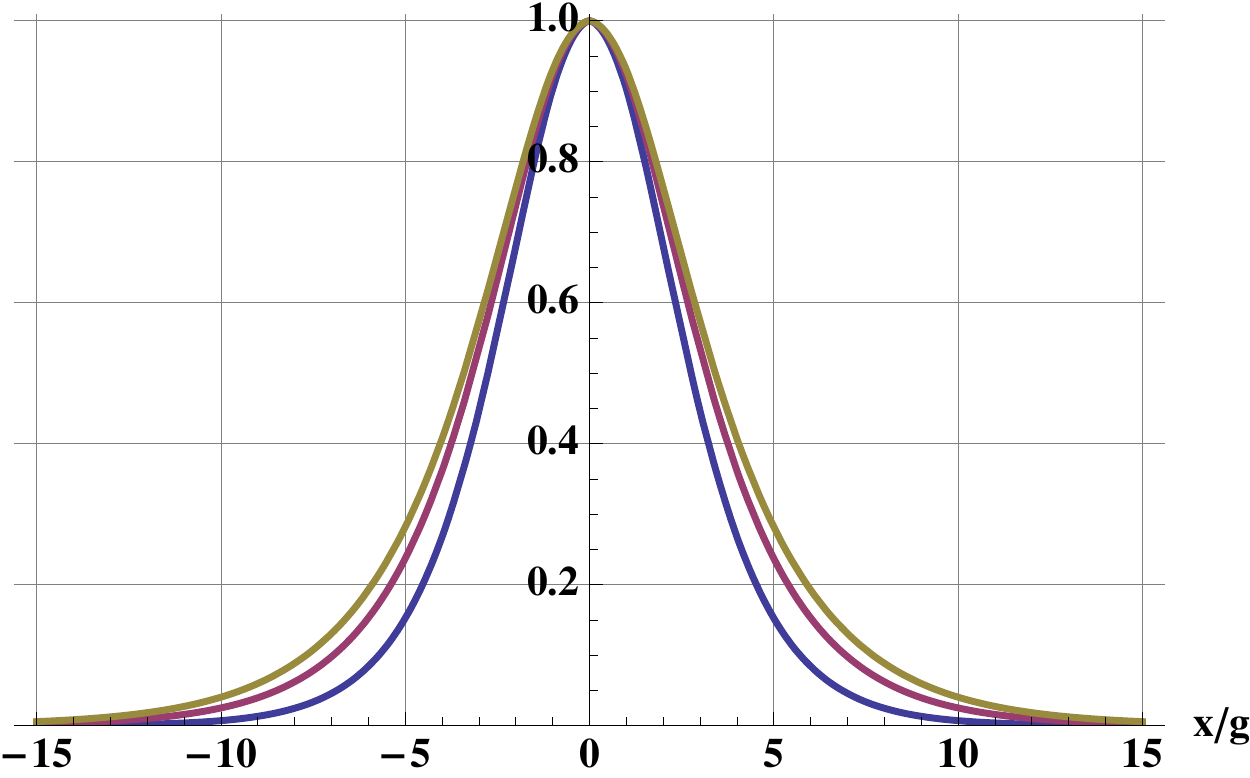}
  \caption{a) Weighting field $E_z$ at position $z=g/2$ for $b=4g$ and $w_x=20g$. The three curves represent $\vep_r =1$ (bottom), $\vep_r =8$ (middle) and $\vep_r =\infty$ (top). b) Normalized weighting field for the same geometry with $w_x=g$ for $\vep_r = 1 $(inner), $\vep_r=8$ (middle) and $\vep_r=\infty$ (outer). }
  \label{rpc_plots}
  \end{center}
\end{figure}
Fig. \ref{rpc_figures}b) represents the geometry with a readout strip of width $w_x$. We first assume the geometry to represent a single layer RPC with a gas gap of $g=0.25$\,mm and a resistive layer of dielectric permittivity $\vep_r$ and thickness $b=1$\,mm. We assume a very wide readout strip width $w_x=5$\,mm and we find for the $z$-component of the weighting field in the center of the gas gap ($z=0.125$\,mm) the numbers shown in Fig.  \ref{rpc_plots}a).  The three curves represent dielectric permittivities of $\vep_r=1$ (bottom), 8 (middle), $\infty$ (top).  The strip extends between $-10 < x/g < 10$ and the value at $x/g=10$ is therefore half of the peak as required by symmetry for a wide readout strip. The value in the center of the strip is close to the one from Eq. \ref{rpc_weighting_inf} for the 'infinitely wide' strip and it is clear from this expression that a higher dielectric permittivity of the resistive plate will increase the weighting field and therefore the induced signal. For precise position measurements one has to use narrow strips, and Fig. \ref{rpc_plots}a) shows the weighting field for a strip of width $w_x=g=0.25$\,mm and $\vep_r = 1, 8, \infty$. The curves are normalized to the peak of the weighting field, so we see that the higher permittivity will result in a slightly wider pad response function. The value of $\vep_r=8$ which is typical for glass and bakelite used in RPCs gives a shape that is already close to the one for an arbitrarily large permittivity.
\\
\\
The effect of typical resistivities of $10^{10}-10^{12}\,\Omega$cm used in RPCs results in very long time constants and has no impact on the fast RPC signal shape. The impact of the resistivity on the electric fields in the gas gap will be discussed in the next section.
\subsection{Effect of resistivity}

\begin{figure}[ht]
 \begin{center}
 a)
  \includegraphics[width=7cm]{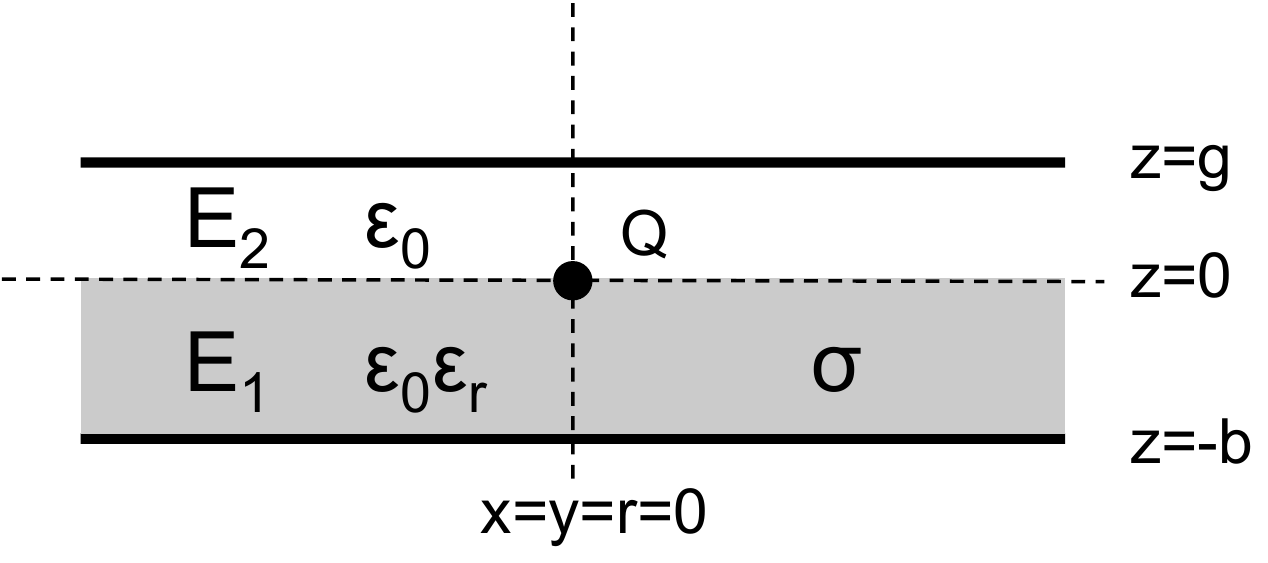}
  b)
  \includegraphics[width=7cm]{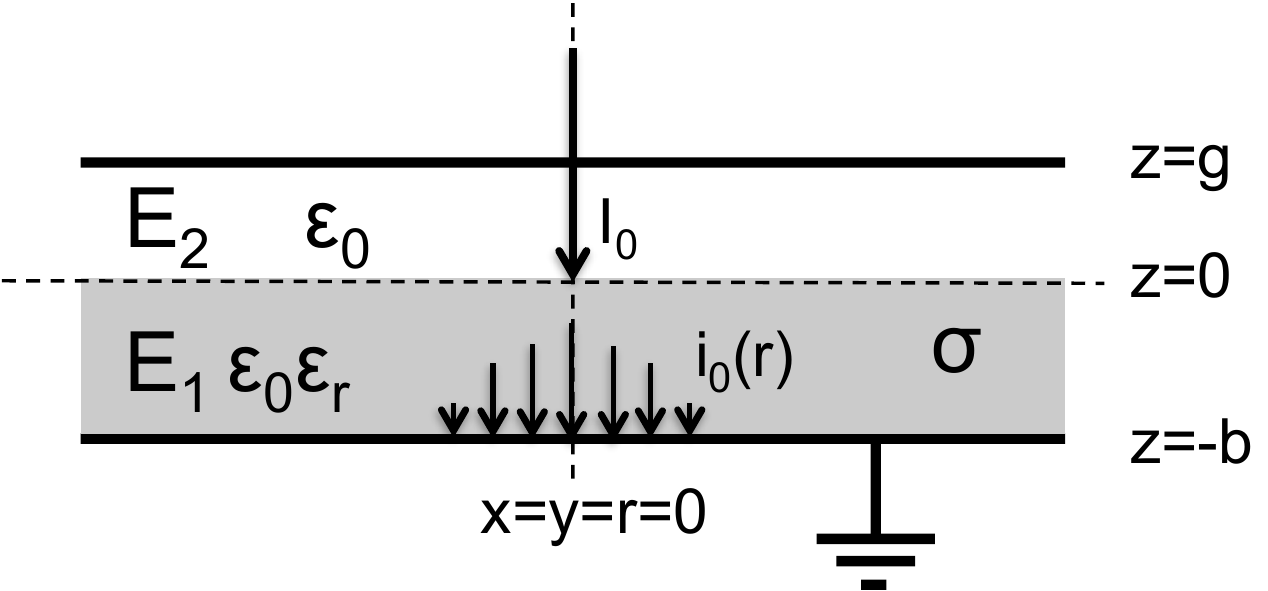}
  \caption{a) A point charge $Q$ placed on the resistive layer at $t=0$. b) A current $I_0$ 'impressed' on the resistive plate at $r=0$.}
  \label{rpc_current}
  \end{center}
\end{figure}
In this section we want to investigate the effect of resistivity in the single gap RPC using the quasi static approximation as outlined in the introduction. We assume layer 1 of the geometry shown in Fig. \ref{rpc_current}a) to have  finite conductivity $\sigma = 1/\rho$. We first recall a few time constants related to this conductivity. 
\begin{figure}[ht]
 \begin{center}
 a)
  \includegraphics[width=7cm]{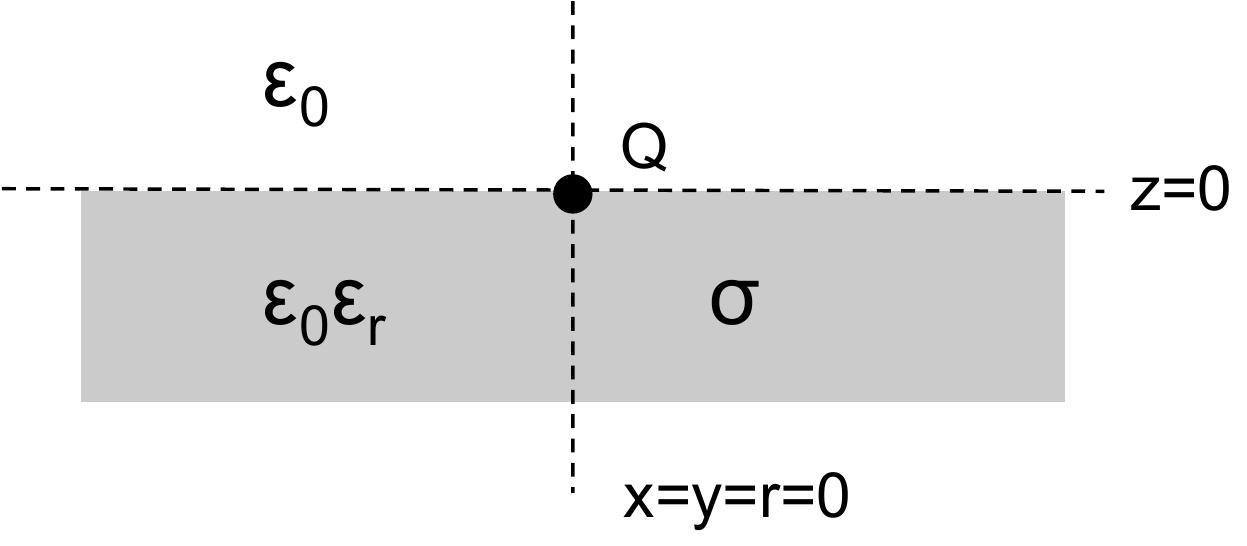}
  b)
  \includegraphics[width=7cm]{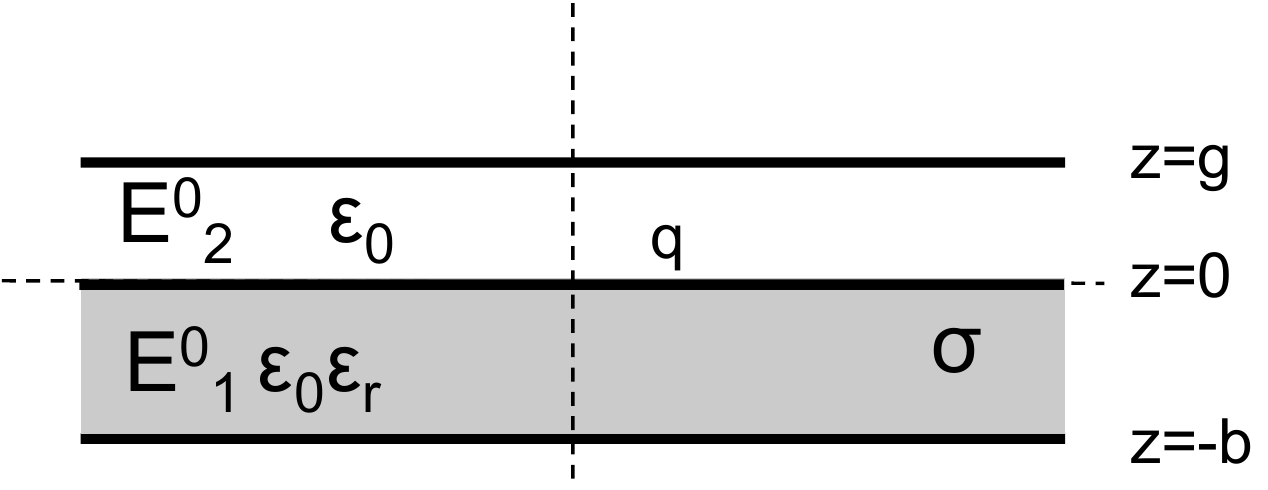}
  \caption{a) A point charge $Q$ on the surface of an infinite half-space with conductivity $\sigma$ and permittivity $\vep_r\vep_0$. b) An infinitely extended sheet of charge of density $q$ on the surface of the resistive layer inside the RPC geometry. }
  \label{rpc_time_constants}
  \end{center}
\end{figure}
In Fig. \ref{rpc_time_constants}a) we have a point charge $Q$ on the boundary of two infinite half spaces with permittivity $\vep_1$ and $\vep_2$. The potential is given by $\phi(r)=Q/(2\pi(\vep_1+\vep_2)r)$ \cite{jackson}. We put the point charge $Q$ at $t=0$ i.e. $Q(t) = Q \Theta(t)$ which reads as $Q(s) = Q/s$ in the Laplace domain. The potential in the Laplace domain is then given by writing $\vep_1=\vep_0\vep_r+\sigma/s$ and $\vep_2=\vep_0$ and we have
\beq 
   \phi(r,s) = \frac{Q}{sr}\frac{1}{2\pi\vep_0(\vep_r+\sigma/(s\vep_0)+1)} \qquad
   \phi(r,t) = \frac{Q}{r2\pi\vep_0(1+\vep_r)}\,e^{-t/\tau_1} \qquad \tau_1 = \frac{\vep_0(1+\vep_r)}{\sigma}
\eeq
The charge is therefore 'destroyed' with a time constant $\tau_1$. Next we look at the geometry in Fig. \ref{rpc_time_constants}b) where a layer of charge with density $q$ is placed on the boundary between the two layers. Using Gauss' law we can calculate the electric field in the two layers to be
\beq \label{uniform_surface_charge}
       E_1^0(s) = -\frac{g\,q(s)}{\vep_1 g + \vep_2 b} = 
       -\frac{q}{s} \frac{g}{\vep_0[g(\vep_r+\sigma/(\vep_0s)) +b]}\quad 
       E_2^0(s) =  \frac{b\,q(s)}{\vep_1 g + \vep_2 b} = 
       \frac{q}{s} \frac{b}{\vep_0[g(\vep_r+\sigma/(\vep_0s)) +b]}
\eeq
and performing the inverse Laplace transform gives
\beq
   E_1(t) = -\frac{g\,q}{\vep_0(\vep_r g + b)}\,e^{-t/\tau_2}  \qquad
   E_2(t) =  \frac{b\,q}{\vep_0(\vep_r g + b)}\,e^{-t/\tau_2}  \qquad 
   \tau_2 = \frac{\vep_0}{\sigma}
   \left(
  \frac{b}{g}+ \vep_r
   \right)  
\eeq
The charge is destroyed with a characteristic time constant $\tau_2$ which is equal to $\tau_1$ in case $b=g$.

\begin{figure}[ht]
 \begin{center}
 a)
  \includegraphics[width=7cm]{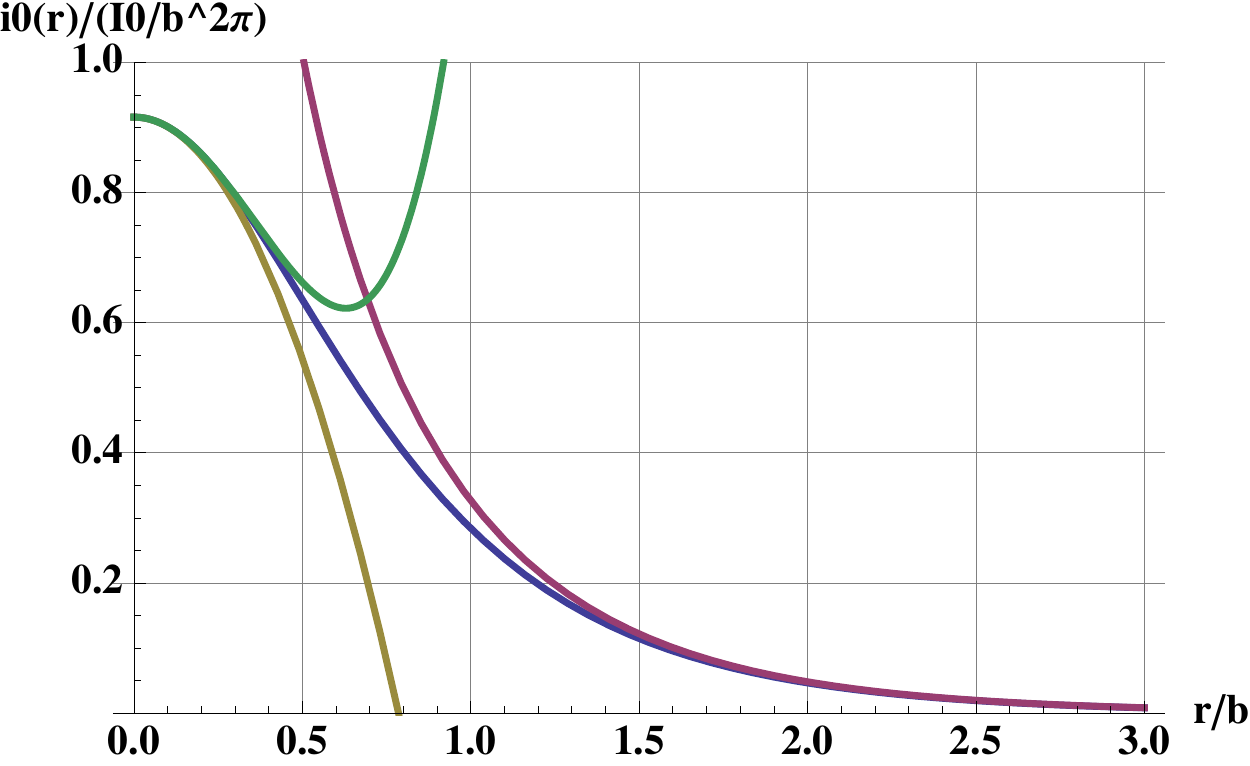}
  b)
  \includegraphics[width=7cm]{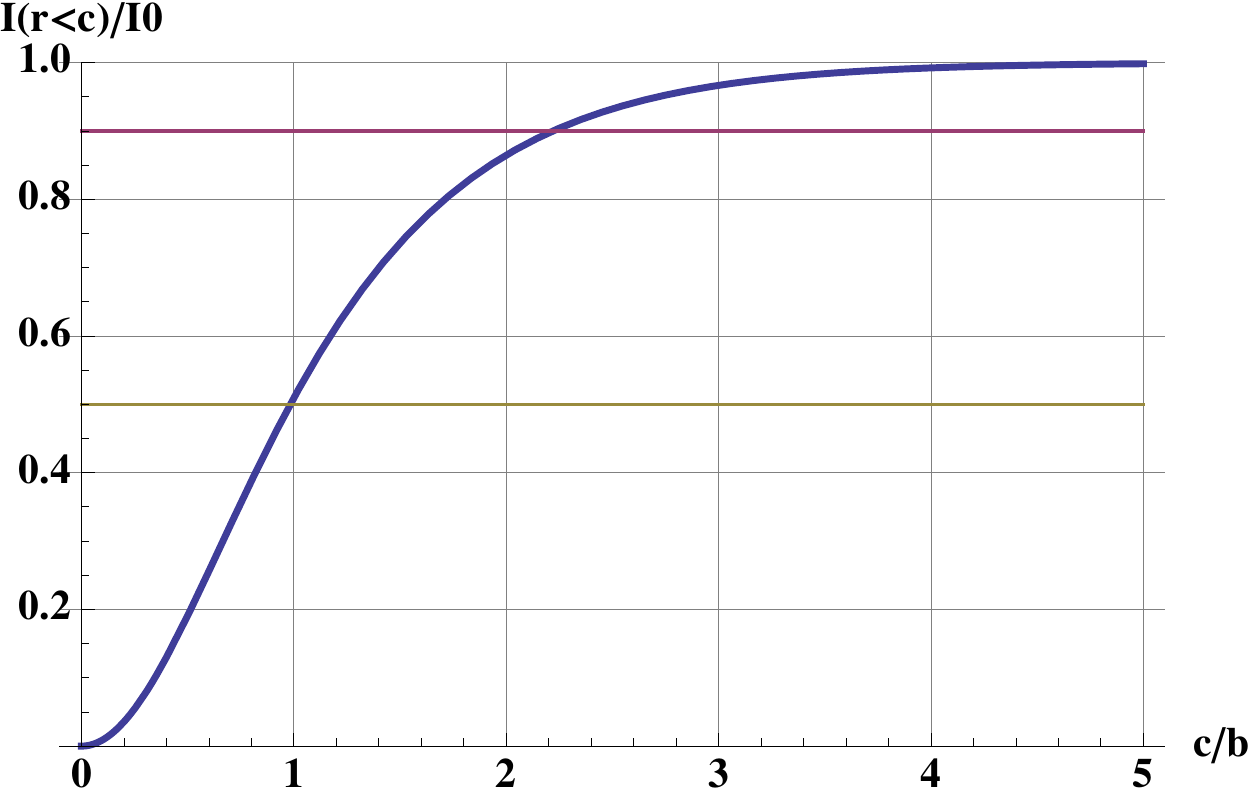}
  \caption{
  a) Current density $i_0(r)$ at $z=-b$. The exact curve together with the $2^{nd}$ order and $4^{th}$ order approximation from Eq. \ref{current_density_approx1} and the exponential approximation from Eq. \ref{current_density_approx2}.  
  b) Total current at $z=-b$ flowing inside a radius $r$ from Eq. \ref{total_current_r}.
  }
  \label{rpc_current_plots}
  \end{center}
\end{figure}

Finally we can calculate what happens when we put a point charge $Q$ on the surface of the resistive plate at $t=0$ as shown in Fig. \ref{rpc_current}a). We use Eqs. \ref{two_layer_potential1}, \ref{two_layer_potential2} with 
\beq
  \vep_1 = \vep_0\vep_r+\sigma/s \quad \vep_2 = \vep_0
   \quad Q_1=Q/s
\eeq
\bea \label{rpc_pointcharge_1}
  E_1(r,z,s) &=&
  -\frac{Q}{2\pi s} \int_0^\infty k\,J_0(kr) 
  \frac{\sinh (gk) \cosh(k(b+z))}
  {\vep_0 [\sinh (b k) \cosh (g k)+ (\vep_r+\sigma/(\vep_0s)) \cosh (b k) \sinh (g k)]}
  dk
\\ \no
  E_2(r,z,s) &=&
  \frac{Q}{2\pi s} \int_0^\infty k\,J_0(kr) 
  \frac{\sinh (bk) \cosh(k(g-z))}
  {\vep_0 [\sinh (b k) \cosh (g k)+ (\vep_r+\sigma/(\vep_0s)) \cosh (b k) \sinh (g k)]}
  dk
\eea
We find the time dependent fields by performing the inverse Laplace transforms and have 
\bea
  E_1(r,z,t) &=& -\frac{Q}{2\pi}
  \int_0^\infty k\,J_0(kr)  \frac{\sinh (gk) \cosh(k(b+z))}{\vep_0 D(k)} e^{-t/\tau(k)}dk
  \\ \no
  E_2(r,z,t) &=& \frac{Q}{2\pi}
  \int_0^\infty k\,J_0(kr)  \frac{\sinh (bk) \cosh(k(g-z))}{\vep_0 D(k)} e^{-t/\tau(k)}dk
\eea
with
\beq
  \tau(k) = \frac{\vep_0}{\sigma}\left(
  \vep_r+\frac{\tanh (bk)}{\tanh (gk)}
  \right)
\quad 
  \tau(k=\infty) = \frac{\vep_0}{\sigma}(\vep_r+1)=\tau_1 \qquad
  \tau(k=0)=\frac{\vep_0}{\sigma}\left(\vep_r+\frac{b}{g}\right) =\tau_2
\eeq
We see that the electric field is decaying to zero with a continuous distribution of time constants $\tau(k)$ in a range between two specific geometrical cases discussed before.
\\ \\
Next we are interested in the case where a DC current $I_0$ is 'impressed' on the resistive plate at $r=0$ (Fig. \ref{rpc_current}b) to find out how this current is then flowing through the resistive plate. The time dependent charge due to $I_0$ is then $Q(t)=I_0\,t$ i.e. $Q(s)=I_0/s^2$, so we have to replace $Q$ in Eq. \ref{rpc_pointcharge_1} by $I_0/s^2$. Since we want to know the stationary situation after a long time we want to know $\lim_{t\rightarrow \infty} E(r,z,t) = \lim_{s \rightarrow 0} s E(r,z,s)$ and have the expressions
\bea
  E_1(r,z) & = &  -\frac{I_0}{2\pi \sigma} \int_0^\infty k J_0(kr) \frac{\cosh(k(b+z)}{\cosh(bk)}dk  \\ \no
  E_2(r,z) & = &  \frac{I_0}{2\pi \sigma} \int_0^\infty k J_0(kr) \frac{\tanh(bk)\cosh(k(g-z)}{\sinh(gk)}dk  \\ \no  
\eea
First we note that $E_1$ does not depend on $g$ but depends only on the thickness $b$ of the resistive layer.  
This is evident from the fact that there is no $DC$ current that can flow through the gas gap, so only the geometry of the resistive layer is relevant. The current density $i_0(r)$ [A/cm$^2$] flowing into the grounded plate at $z=-b$ is related to the field on the surface of the grounded plate by
\beq
  i_0(r) = -\sigma E_1(r,z=-b)    
  =\frac{I_0}{b^2\pi}\,\int_0^\infty  \frac{1}{2} J_0\left(y\frac{r}{b}\right) \frac{y}{\cosh(y)}dy
\eeq
To evaluate this expression for small values of $r$ we can insert the series expansion for $J_0(x)$ and evaluate the integrals, which gives
\beq \label{current_density_approx1}
   \int_0^\infty \frac{1}{2}J_0\left(y\frac{r}{b}\right) \frac{y}{\cosh(y)}dy =
   \sum_{m=0}^\infty \frac{(-1)^m(2m+1)!}{(m!)^2 2^{2m+1}}
   \left[ i\mbox{Li}_{2m+2}(-i)-i\mbox{Li}_{2m+2}(i)     \right]r^{2m}   
\eeq
\beq
    \approx 0.916 \,- \,1.483\,\left(\frac{r}{b}\right)^2\,+\,1.873\left(\frac{r}{b}\right)^4-...
\eeq
where Li$_n(x)$ denotes the Polylogarithm function. For large values of $r$ the radial dependence is exponential (Section \ref{integral_1})
\beq \label{current_density_approx2}
   \int_0^\infty   \frac{1}{2}  J_0\left(y\frac{r}{b}\right) \frac{y}{\cosh(y)}dy 
   =\frac{\pi}{2}\sum_{n=0}^\infty (-1)^n (2n+1)K_0
   \left(
   \frac{(2n+1)\pi}{2}\frac{r}{b}
   \right)
   \approx 
   \frac{\pi}{2\sqrt{r/b}}\,e^{-\pi r/(2b)}
   \quad \mbox{for} \quad \frac{r}{b} \gg 1
\eeq
The current is plotted in Fig. \ref{rpc_current_plots}a) and we see that for $r/b>2$ the exponential approximation describes the situation already to very high accuracy.  The current $I(r)$ flowing within a circle of radius $r$ is given by
\beq \label{total_current_r}
   I(r) = \int_0^r 2r\pi i_0(r')dr' = 
   I_0\left[
   1-2\sum_{n=0}^\infty (-1)^n \frac{r}{b}
   K_1\left( \frac{(2n+1)\pi}{2}\frac{r}{b} \right)
   \right]
\eeq
where we have used the relation $\int_0^r r'K_0(r')dr'=1-rK_1(r)$. Fig.\ref{rpc_current_plots}b) shows this expression, and we see that the radii within which 50/90/99\% of the current a flowing are given by 
\beq
      r_{50\%} \approx b \quad 
      r_{90\%} \approx 2.3b \quad 
      r_{99\%} \approx 3.9b 
\eeq
For very large values of $r \rightarrow \infty$ we have $K_1(ar)=0$ and $I(r)=I_0$.
Using Gauss' Law $\oint \vep \vec{E}d\vec{A}=Q$ we can calculate the total charge $Q_0$ that is building up inside the RPC by integrating the electric field over the surfaces of the metal plates at $z=g, -b$ i.e. 
\beq
  Q_0=  \vep_0\int_0^\infty 2r\pi E_3(r,g)dr  -\vep_0\vep_r\int_0^\infty 2r\pi E_1(r,-b)dr  = I_0 \tau_2
\eeq
Since the electric field has a discontinuity only at $z=0$ this charge is sitting on the surface of the resistive layer and the radial distribution is given by
\bea 
  q(r) &=& \vep_0 E_3(r,z=0)-\vep_r\vep_0E_1(r,z=0)  \\ \no
  &=& \frac{I_0}{2\pi}\int_0^\infty k J_0(kr) \,
  \frac{\vep_0}{\sigma}
  \left(  \vep_r + \frac{\tanh(bk)}{\tanh(gk)}
  \right)dk 
  =\frac{I_0}{2\pi}\int_0^\infty k J_0(kr) \, \tau(k)dk
\eea
We can verify that the total charge on the surface $\int_0^\infty 2r\pi q(r) dr$ to $Q_0$. Since $\tau(k \rightarrow \infty) = \tau_1$ the integrand of the above expression diverges, so by 'adding and subtracting' $\tau_1$ and using $\int_0^\infty k J_0(kr)dk=\delta (r)/r$ we have 
\bea
   q(r) & = & \frac{I_0}{2\pi}\left[ 
   \int_0^\infty k J_0(kr) \tau_1dk+\int_0^\infty k J_0(kr) \, (\tau(k)-\tau_1)dk 
   \right]\\ \no
    & = & I_0 \tau_1\, \frac{\delta(r)}{2r\pi} + \frac{I_0}{2\pi} \, \int_0^\infty k J_0(kr) \, (\tau(k)-\tau_1)dk
\eea
so we learn that at there is a point charge of value $I_0 \tau_1$ at the place where the current is put on the surface and a distributed charge of value $I_0(\tau_2-\tau_1)$ building up on the surface of the resistive plate. In case of $g=b$ there is only a point charge, for $\tau_2>\tau_1$ i.e. for $b>g$ the distributed charge has the same sign as the point charge, while for $b<g$ it has opposite charge.

\subsection{Surface resistivity}

\begin{figure}[ht]
 \begin{center}
 a)
  \includegraphics[width=7cm]{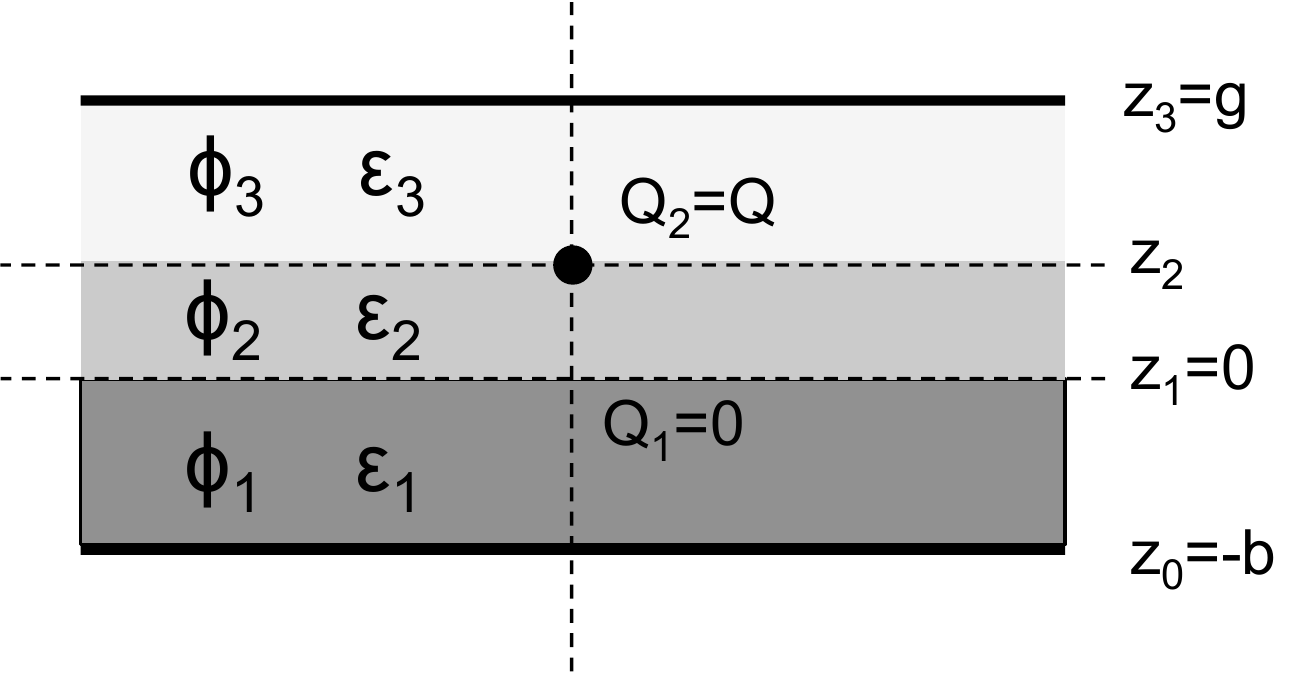}
  b)
  \includegraphics[width=7cm]{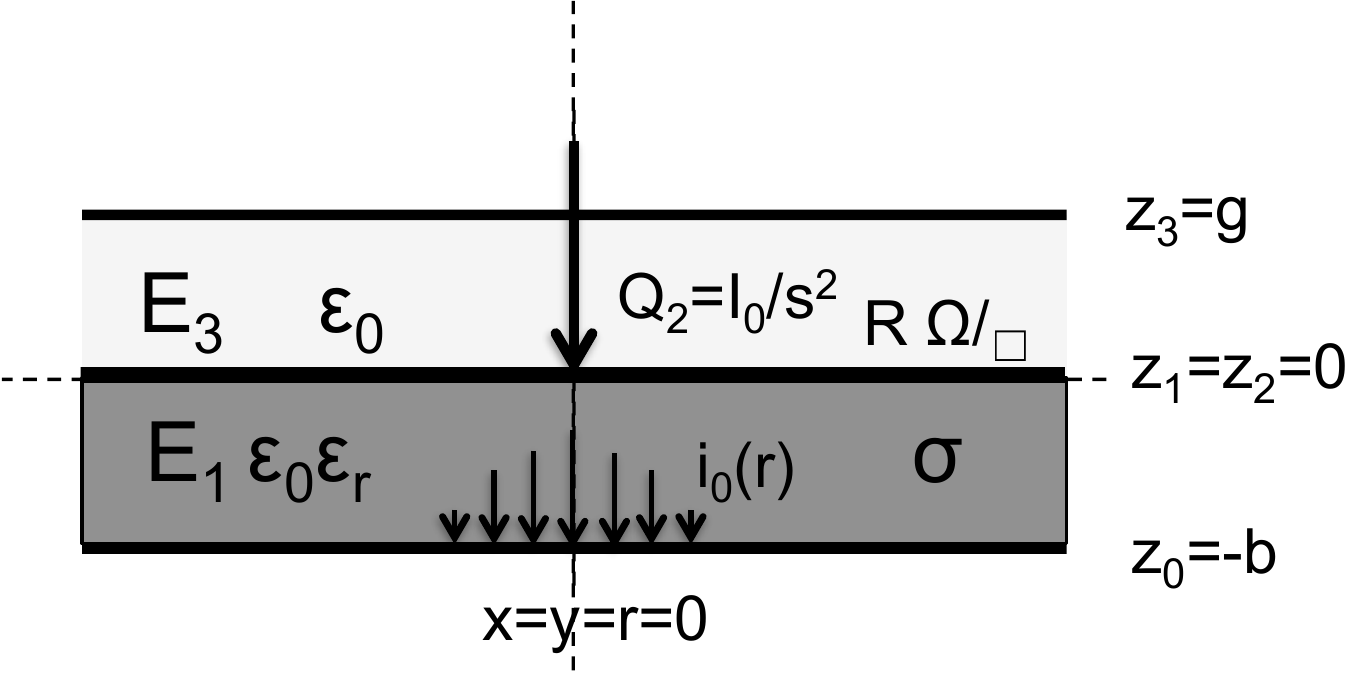}
  \caption{a) General 3 layer geometry with a point charge $Q_2$. b) A resistive plate with conductivity $\sigma$ together with an (infinitely) thin layer of surface resistivity $R \Omega/$square and an impressed current $I_0$.}
  \label{rpc_surf_resistivity}
  \end{center}
\end{figure}
It has been stated that the glass or Bakelite might develop a conductive surface once the electric field is applied. In order to predict some measurable effect of such a conductive layer we investigate the geometry shown in Fig. \ref{rpc_surf_resistivity}. We employ the formalism for a 3 layer geometry in Fig. \ref{rpc_surf_resistivity}a) and arrive at the setting shown in Fig. \ref{rpc_surf_resistivity}b) by having  
\beq
   z_0=-b \quad z_1=0 \quad z_2 \quad z_3 = g \quad \vep_1 = \vep_0\vep_r+\sigma/s \quad
   \vep_2=\vep_0+1/(sRz_2) \quad \vep_3=\vep_0 \quad Q_1=0 \quad
   Q_2=I_0/s^2
\eeq
and performing the limit $z_2 \rightarrow z_1 = 0$ to all the expressions, as well as the limit of $\lim_{s \rightarrow 0} s f_n(k, z, s)$ to find the stationary situation for long times, which yields
\beq \label{f_surf_resistivity}
   f_1(k,z) = \frac{I_0}{\sigma}\frac{\sinh(k(b+z))}{\cosh(bk)+k/(R\sigma)\sinh(bk)}
   \qquad
   f_3(k,z)=\frac{I_0}{\sigma}\frac{\sinh(bk)\sinh(k(g-z))}{\sinh(gk)[\cosh(bk)+k/(R\sigma)\sinh(bk)]}
\eeq
\bea
  E_1(r,z) & = &  -\frac{I_0}{2\pi \sigma} \int_0^\infty k J_0(kr) \frac{\cosh(k(b+z)}{\cosh(bk)+k/(R\sigma)\sinh(bk)}dk  \\ \no
  E_3(r,z) & = &  \frac{I_0}{2\pi \sigma} \int_0^\infty k J_0(kr) \frac{\sinh(bk)\cosh(k(g-z)}{\sinh(gk)[\cosh(bk)+k/(R\sigma)\sinh(bk)]}dk  \\ \no  
\eea
and the current $i_0(r)$ becomes 
\beq
   i_0(r) = -\sigma E_1(r,z=-b) = \frac{I_0}{b^2\pi}\,\int_0^\infty  \frac{1}{2} J_0\left(y\frac{r}{b}\right) \frac{y}{\cosh(y)+ \frac{y}{\beta^2} \sinh(y)}dy
   \qquad \beta^2=R\sigma b
\eeq
In the limit of very high resistivity $R \rightarrow \infty$ we recuperate the expression from the previous section without any resistive surface layer. For low resistivity $R<1/(\sigma b) \rightarrow \beta^2 \ll 1$ the integral evaluates to (Section \ref{integral_3}) 
\beq
     i_0(r) \approx  
     \frac{I_0}{b^2 \pi}
     \frac{\beta^2}{2} \left[ K_0(\beta \frac{r}{b}) + 2 \sum_{m=1}^\infty (-1)^m\, K_0(m\pi \frac{r}{b}) \right]
     \approx   \frac{I_0}{b^2 \pi}
     \frac{\beta^2}{2}  K_0(\beta \frac{r}{b})
     \approx 
     \frac{I_0}{b^2 \pi}
      \frac{\beta^2}{2}  
  \sqrt{\frac{\pi}{2}}
 \frac{e^{-\beta r/b}}{\sqrt{\beta r/b}}
\eeq
%
%
%
%
%
%
%
%
%
Comparing to Eq. \ref{current_density_approx2} we see that the radial exponential decay of the current is not any more governed by the the characteristic length $2b/\pi$ but by $b/\beta$. This becomes even more explicit by calculating the current flowing through a circle of radius $r$, which is given by
\beq
   I(r) = \int_0^r2r\pi i_0(r')dr' \approx  
 I_0
 \left[
 1-\frac{\beta r}{b}K_1\left( \beta \frac{r}{b} \right)
 \right]
\eeq
which gives
\beq
    r_{50\%} \approx 1.26\sqrt{\frac{b}{R\sigma}} \quad
    r_{90\%} \approx 3.21\sqrt{\frac{b}{R\sigma}} \quad
    r_{99\%} \approx 5.77\sqrt{\frac{b}{R\sigma}} 
\eeq
The lower the surface resistivity $R$ the more the current is spread out radially, which can easily be pictured: the low resistivity of the surface layer will result in currents flowing radially inside this thin layer, and through the contact with the thick resistive layer to the bottom ground layer this spread is visible there. 
\\
\\
By using Gauss' Law we can again calculate the total charge $Q_0$ that is building up inside the RPC and we find that it is equal to the case without the resistive layer  and is given by $Q_0=I_0 \tau_2$.

\subsection{Field of a charge disk}

\begin{figure}[ht]
 \begin{center}
 a)
  \includegraphics[width=7cm]{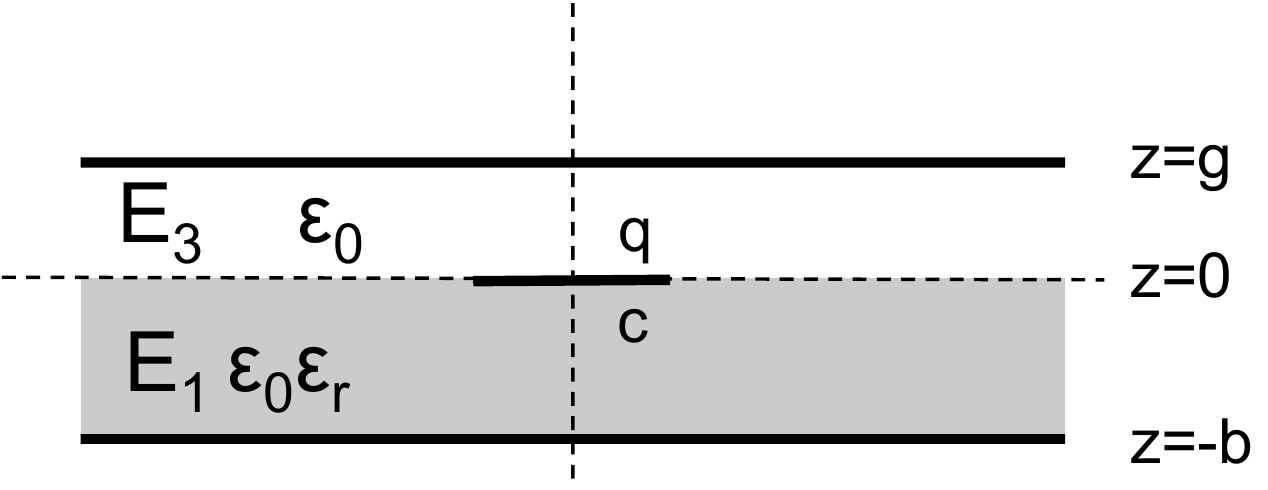}
  b)
  \includegraphics[width=7cm]{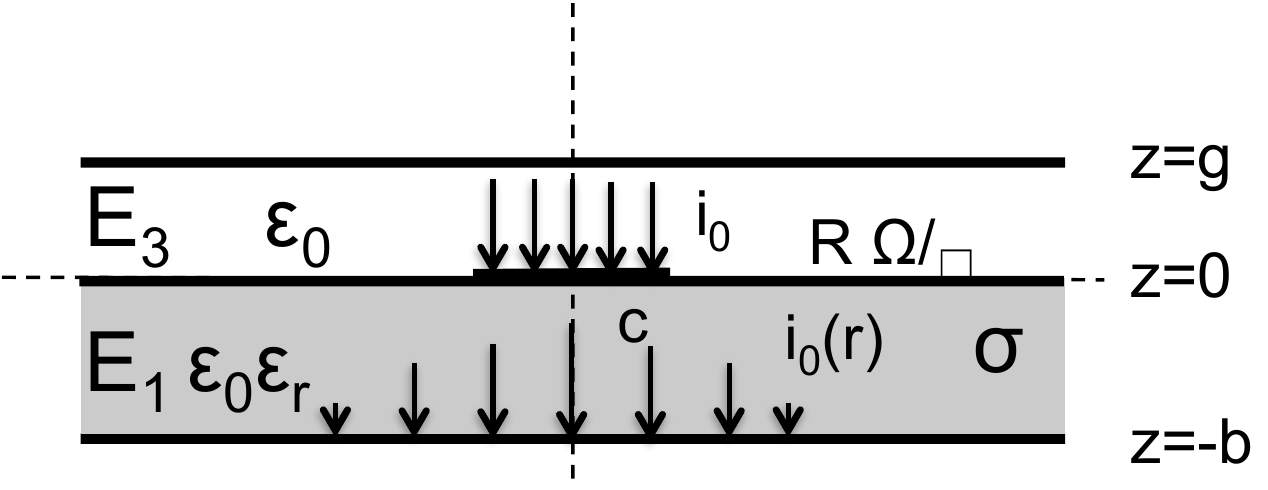}
  \caption{a) A current impressed on a circular area of the resistive plate. b) A current impressed on the surface of a resistive plate with a thin layer of surface resistivity $R$. }
  \label{rpc_circle}
  \end{center}
\end{figure}

In this section we want to investigate the situation where the current is not placed on a single point but extended over a disc of radius $c$. This represents the case where e.g. an RPC is illuminated by a particle beam in a circular subsection of it's surface. We first calculate the electric field for a disc of charge with radius $c$, centred at the origin, with charge density $q$\,[pC/cm$^2$] sitting on the resistive plate as shown in Fig. \ref{rpc_circle}a). 
We use $f_1(k,z)$ and $f_3(k,z)$ from Eq. \ref{f_rpc_geometry} with $z_2=0$ which recuperates the two layer solution. 
We replace $Q$ by the infinitesimal charge $q r_0 dr_0 d\phi_0$ sitting at point $r_0 \phi_0$, use Eq. \ref{sol1} and integrate  by $\int_0^{2\pi}d\phi_0$ and $\int_0^c dr_0$. The integration over $\phi$ leaves only the term with $m=0$ and for the $r_0$ integration we use the relation $\int_0^c r_0 J_0(kr_0) = cJ_1(kc)/k$ which gives the result
\bea
  E_1(r,z) & = &
  -c q \int_0^\infty J_0(kr) J_1(kc) \frac{\sinh (gk) \cosh(k(b+z))}{\vep_0 D(k)} dk \\
  E_3(r,z) & = &
  c q \int_0^\infty J_0(kr) J_1(kc) \frac{\sinh (bk) \cosh(k(g-z))}{\vep_0 D(k)} dk
\eea
In order to arrive at the solution for the situation shown in \ref{rpc_circle}b) we proceed as outlined in the previous section and using $f_1(k, z)$ and $f_3(k,z)$ from Eq. \ref{f_surf_resistivity} we find 
\bea 
  E_1(r,z) & = &  -\frac{i_0}{\sigma} \int_0^\infty  J_0(kr) c J_1(kc)\frac{\cosh(k(b+z)}{\cosh(bk)+k/(R\sigma)\sinh(bk)}dk  \\ 
  E_3(r,z) & = &  \frac{i_0}{ \sigma} \int_0^\infty  J_0(kr) cJ_1(kc)\frac{\sinh(bk)\cosh(k(g-z)}{\sinh(gk)[\cosh(bk)+k/(R\sigma)\sinh(bk)]}dk  
\eea
%
%
%
%
%
%
%
We first verify that by making the illuminated area infinitely large we find the expected electric fields for uniform illumination. We use the relation $\int_0^\infty r J_0(kr)dr=\delta (k)/k$ which means that $\lim_{c \rightarrow \infty}cJ_1(kc)=\delta(k)$ and find 
\beq \label{rpc_voltage_drop}
      E_1^0 = -\frac{i_0}{\sigma} \qquad  E_3^0 = \frac{i_0 b}{\sigma g} =  \frac{i_0 b}{g} \, \rho
\eeq
Using $q(t)=i_0 t$ and therefore $q(s)=i_0/s^2$ in Eq. \ref{uniform_surface_charge} for the uniform charge $q$ on the surface and performing  $E(t\rightarrow \infty)=\lim_{s \rightarrow 0} s E(s)$ we find back exactly the same field from above. $E^0_3$ is the 'standard' expression for the decrease of the electric field in an RPC due to the resistivity of the material.
Using Gauss' Law as before we find the total charge density that is building up on the resistive surface to be
\beq
   q_{surf} = -\vep_0\vep_r E_1^0 + \vep_0 E_2^0 = i_0\frac{\vep_0}{\sigma}
   \left(
   \vep_r+\frac{b}{g}
   \right)  = i_0 \,\tau_2
\eeq
We finally evaluate the field $E_3(r,z)$ in the 'gas gap' of an RPC for irradiation of a circular area of radius $c$ and relate it to the field $E_3^0$ due to uniform illumination 
\beq
   \frac{E_3(r,z)}{E^0_3} = \frac{g}{b} \int_0^\infty \frac{c}{b}  
   J_0\left( y\frac{r}{b} \right) J_1\left(y\frac{c}{b}\right)\frac{\sinh(y)\cosh(y\frac{g-z}{b})}{\sinh(y\frac{g}{b})[\cosh(y)+\frac{y}{\beta^2}\sinh(y)]} dk
\eeq 
Figure \ref{rpc_circle}a) shows the electric field inside the gas gap at $z=g/2$ for an illuminated disc of radius $c=10b$. For values of $\beta =R\sigma b<1$, i.e. for small values of $R$ the electric field in the gas gap due to the impressed current is reduced since the current is distributed over a larger area of the resistive plate. It shows the principle possibility to locally increase the rate capability of an RPC by a thin resistive layer that spreads out the current through the resistive plate. 

\begin{figure}[ht]
 \begin{center}
 a)
  \includegraphics[width=7cm]{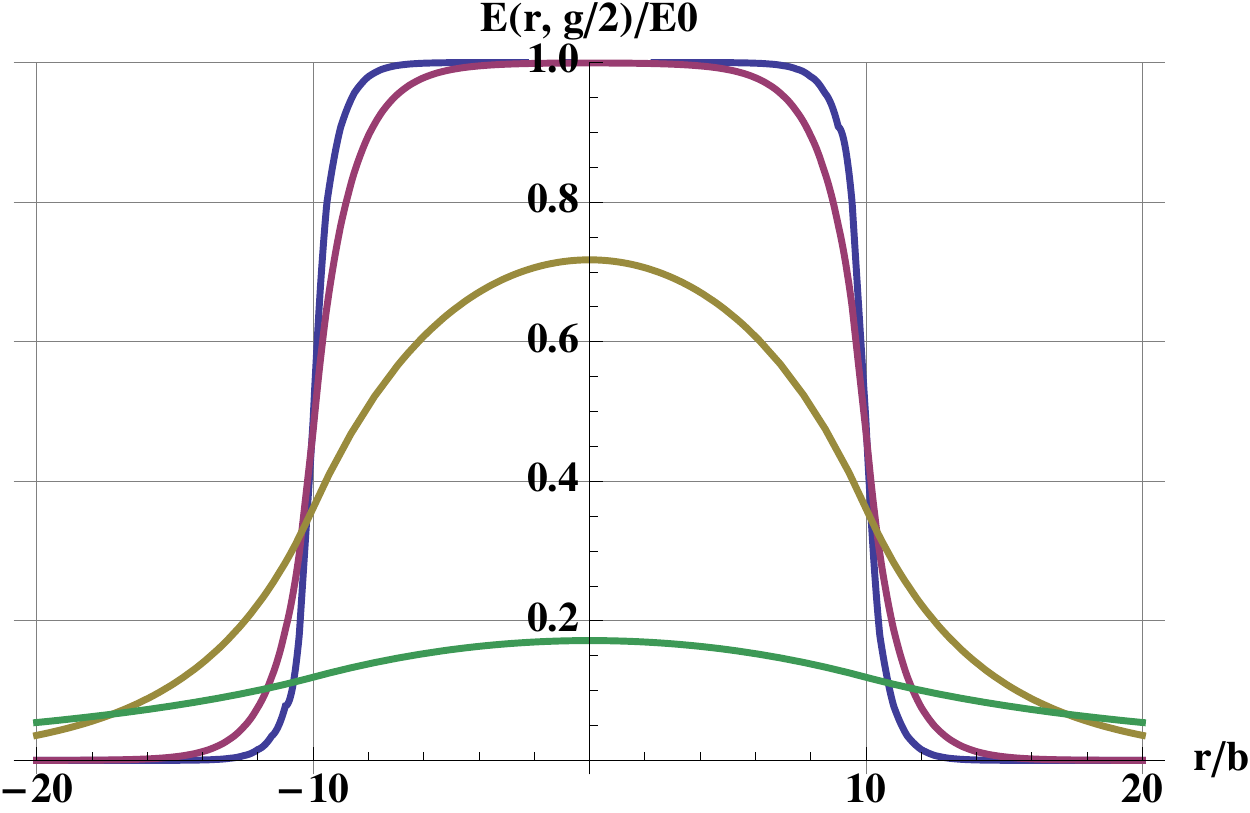}
  b)
  \includegraphics[width=7cm]{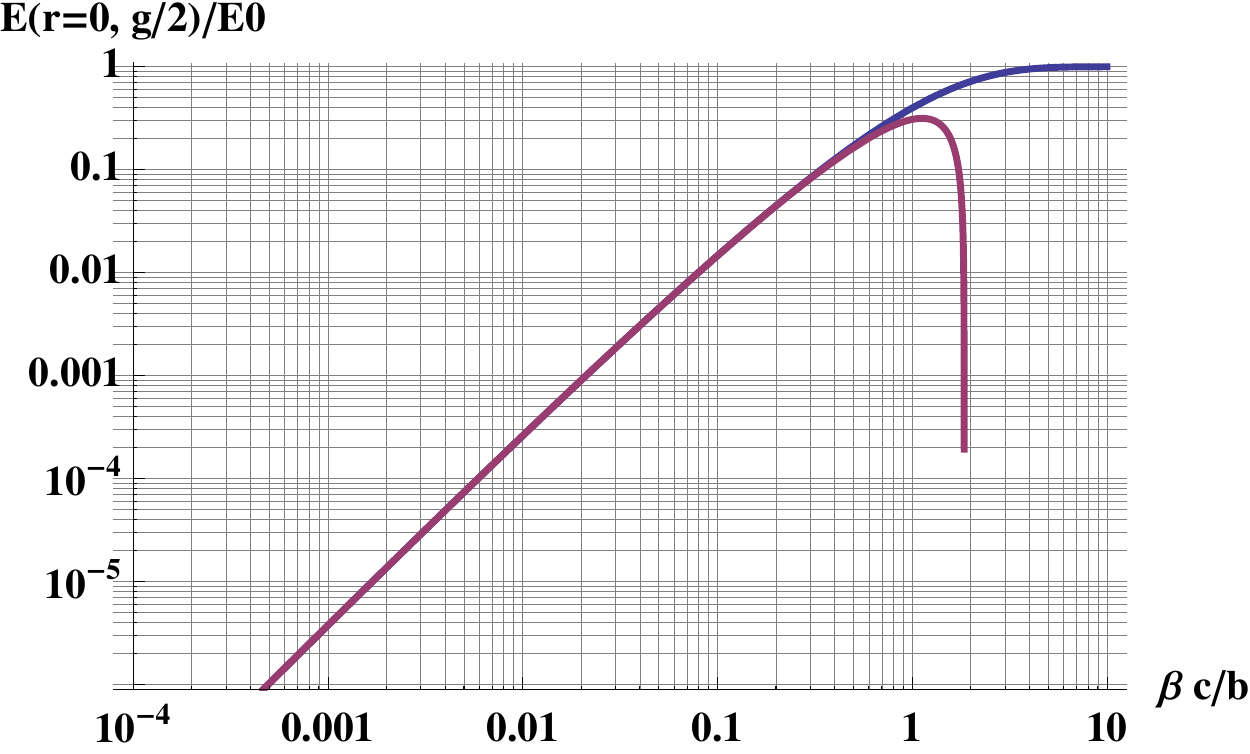}
  \caption{a) Electric field in the gas gap for values of $\beta = \infty, 1, 0.2, 0.05$. b) Values of the electric field in the gas gap at $r=0$ for different values of $\beta \frac{c}{b}$. The the line approaching unity represents Eq. \ref{disc_drop1}, the other one represents the approximation of Eq. \ref{disc_drop2}.}
  \label{rpc_circle2}
  \end{center}
\end{figure}

To approximate the electric field in the center of the disc i.e. at $r=0$ we proceed in he following way: in case the radius of the disc is larger than the thickness of the resistive plate we have $\frac{c}{b} \gg 1$ and $\frac{c}{b}J_1(y\frac{c}{b})$ is close to $\delta(y)$ i.e only small values of $y$ contribute to the integral. We can therefore expand the expression for small values of $y$ and have 
\beq \label{disc_drop1}
    \frac{E_3(r=0,z)}{E^0_3}  \approx \int_0^\infty \frac{c}{b}  J_1\left(y\frac{c}{b}\right)\frac{1}{1+\frac{y^2}{\beta^2}} dy = 1 - \beta \frac{c}{b}
    K_1\left( \beta \frac{c}{b} \right)
\eeq
For values of $\beta \frac{c}{b} \ll 1 $ we have 
\beq  \label{disc_drop2}
      \frac{E_3(r=0,z)}{E^0_3}  \approx 1 - \beta \frac{c}{b}
    K_1\left( \beta \frac{c}{b} \right) \approx 
    \left(\beta \frac{c}{b}\right)^2 \left[ 0.308-\frac{1}{2} \ln  \left(\beta \frac{c}{b}\right) \right]
\eeq
The expressions are shown in Fig. \ref{rpc_circle2}b). 
Finally for completeness, the expression for the total current $I_1(r)$ flowing into the ground plate ar $z=-b$ within a radius $r$ is given by 
\beq
  I_1(r)=-\int_0^r 2 r' \pi \sigma  E_1(r',z=-b)dr' =   2 \pi b^2 I_0 \int_0^\infty
   \frac{1}{y} \frac{r}{b} J_1\left(y\frac{r}{b}\right)\frac{c}{b}J_1
   \left( y\frac{c}{b}\right)\frac{1}{\cosh(y)+\frac{y}{\beta^2}\sinh(y)}
   dy
\eeq

%
%
%
%
%
%
%
%
%
%
%
%

\clearpage

\section{Single thin resistive layer\label{thin_resistive_layers}}

\begin{figure}[ht]
 \begin{center}
 a)
   \includegraphics[width=7cm]{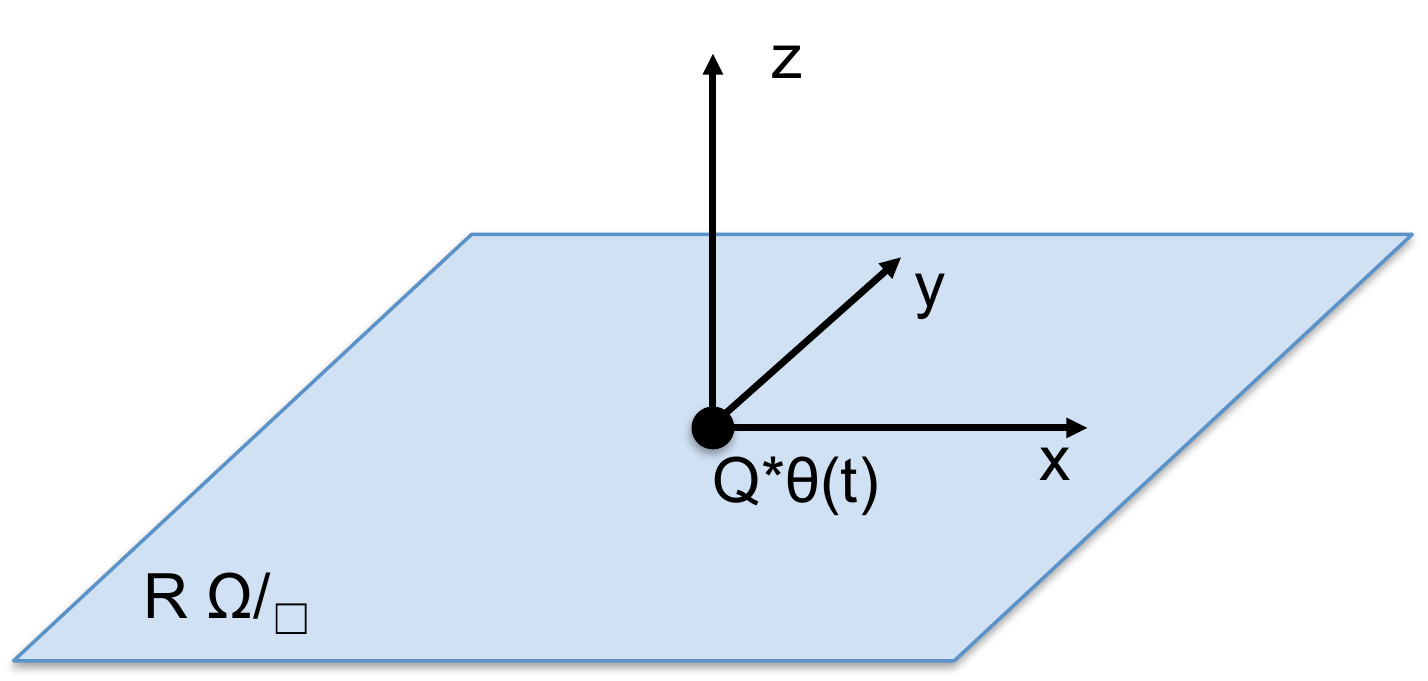}
 b)
  \includegraphics[width=7cm]{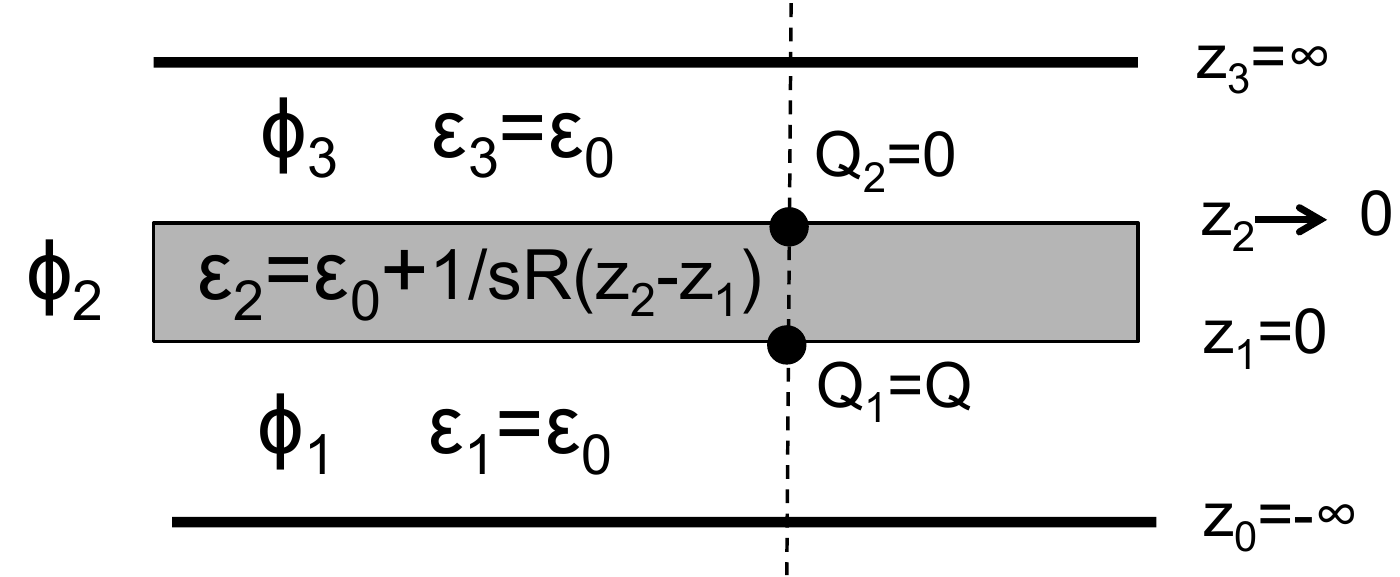}
  \caption{a) A resistive layer with surface resistance $R\,[\Omega$/square]. b) The fields for this single layer can be calculated from the indicated 3-layer geometry by performing the indicated limits of the expressions for $z_0, z_2, z_3$.}
  \label{1layer_sketch}
  \end{center}
\end{figure}
In this section we want to study the fields of a single layer of surface resistivity $R\,[\Omega$/square] at $z=0$ where we place a charge $Q$ at $r=0$ at $t=0$ as shown in Fig. \ref{1layer_sketch}a).  We write $Q(t)=Q\Theta(t)$ where $\Theta(t)$ is the Heaviside step function. In the Laplace domain this reads as $Q(s)=Q/s$. The fields can be derived from the 3-layer geometry shown in Fig. \ref{1layer_sketch}b)  with 
\beq \label{single_thin_layer_onditions}
    \vep_1 = \vep_0 \qquad \vep_2 = \vep_0+\frac{1}{sz_2R}  \qquad \vep_3 = \vep_0 \qquad Q_1=\frac{Q}{s} \qquad Q_2 = 0
\eeq
and taking the appropriate limits 
\bdi    
    z_0 \rightarrow -\infty \qquad  z_1=0 \qquad z_2 \rightarrow 0 \qquad z_3 \rightarrow \infty 
\edi
Since we have shrunk layer 2 to zero thickness we only have the coefficients $A_1, B_1$ for the layer $z<0$ and  $A_3, B_3$ for the layer $z>0$ and get 
\beq
   f_1(k,z) = \frac{QR}{k+2\vep_0Rs}\,e^{kz} 
   \qquad 
   f_2(k,z) = \frac{QR}{k+2\vep_0Rs}\,e^{-kz}
\eeq
In the time domain they read as
\beq
  f_1(k,z) = \frac{Q}{2\vep_0}\,e^{-k(vt-z)} 
   \qquad 
   f_2(k,z) = \frac{Q}{2\vep_0}\,e^{-k(vt+z)}
\eeq
\subsection{Infinitely extended resistive layer}

\begin{figure}[ht]
 \begin{center}
  a)
  \includegraphics[width=7cm]{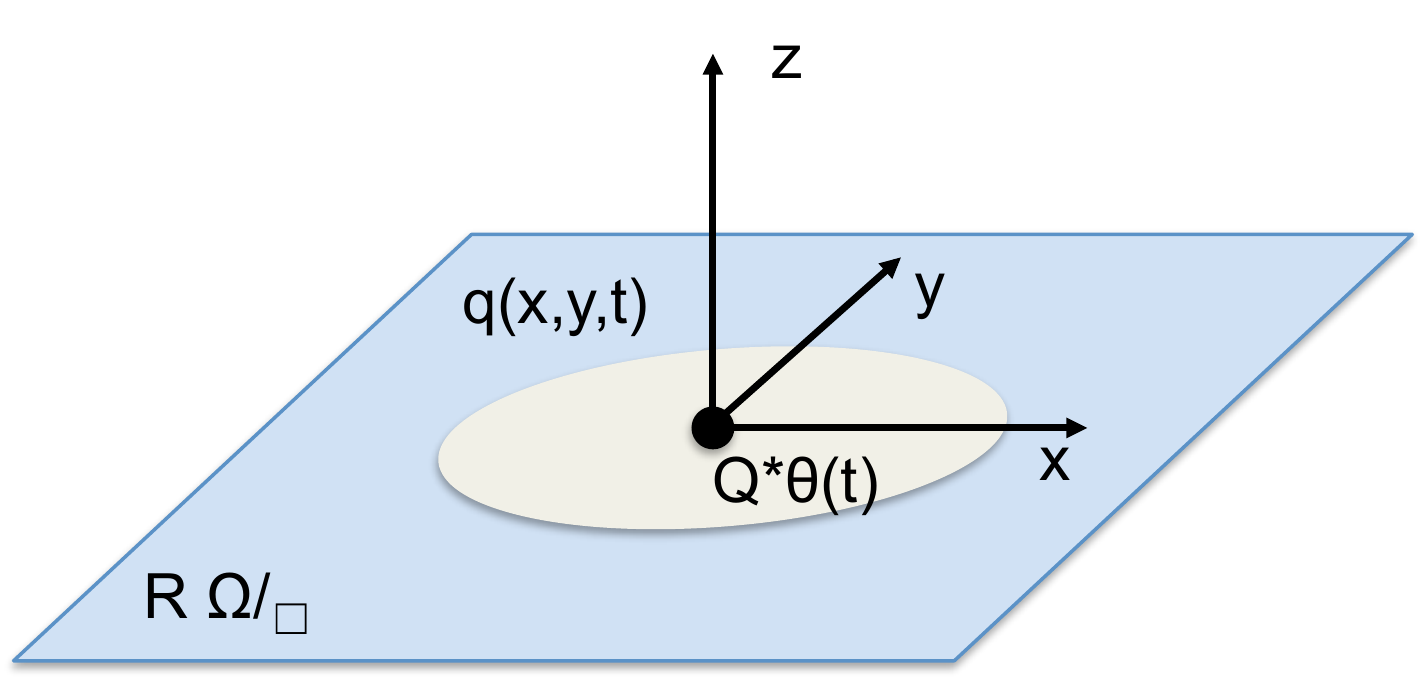}
  b)
   \includegraphics[width=4.5cm]{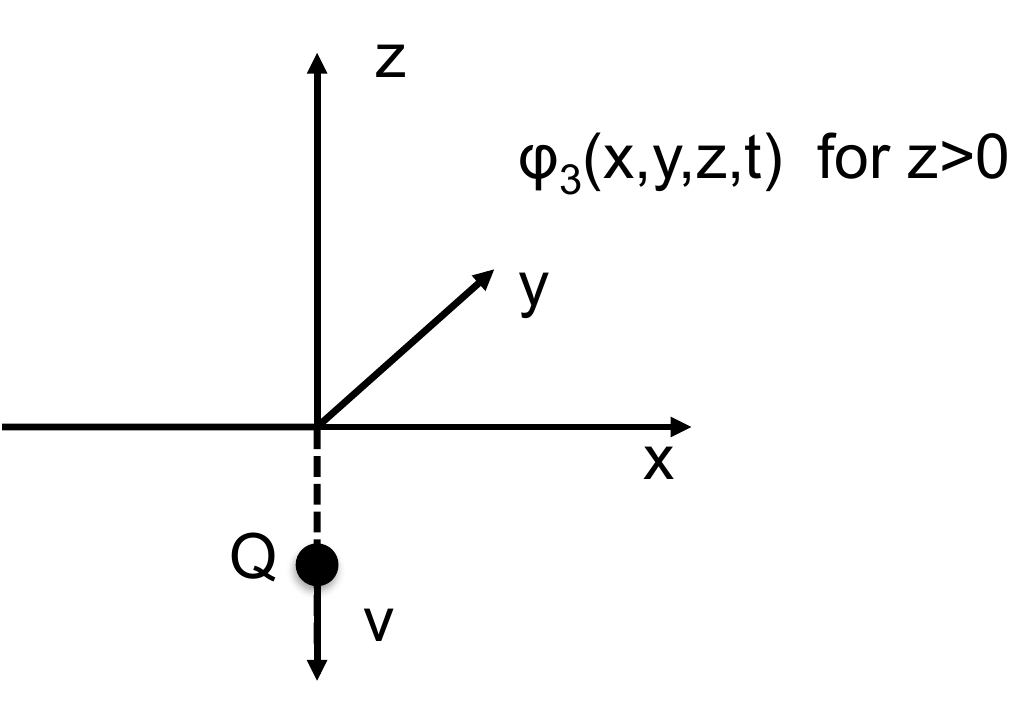}
  \caption{a) A point charge placed at an infinitely extended resistive layer at $t=0$. b) The solution for the time dependent potential is equal to a point charge moving with velocity $v$ olong the $z$-axis.}
  \label{charge_diff1}
  \end{center}
\end{figure}

First we investigate an infinitely extended layer as shown in Fig. \ref{charge_diff1}a.
The charge $Q$ will cause currents to flow inside the resistive layer that are 'destroying' it. The solution for the potential is given by
\beq
     \phi_1(r,z,t) = \frac{Q}{4\pi \vep_0} \int_0^\infty J_0(kr) e^{-k(vt-z)} \quad
     \phi_3(r,z,t) = \frac{Q}{4\pi \vep_0} \int_0^\infty J_0(kr) e^{-k(vt+z)}
\eeq
and using Eq. \ref{inverse_square_identity} this becomes
\beq
  \phi_1(r,z,t) = \frac{Q}{4\pi\vep_0} \, \frac{1}{\sqrt{r^2+(-z+vt)^2}} \quad 
   \phi_3(r,z,t) = \frac{Q}{4\pi\vep_0} \, \frac{1}{\sqrt{r^2+(z+vt)^2}} 
\eeq
We see that the potential due to the point charge placed on the infinitely extended resistive layer at $t=0$ is equal to the potential of a charge $Q$ that is moving with a velocity $v=1/2\vep_0 R$ away from the layer along the $z-$axis as indicated in Fig. \ref{charge_diff1}b).  As an example for a surface resistivity of $R=1\,$M$\Omega$/square the velocity is 5.6\,cm/$\mu$s. The time dependent surface charge density $q(r, t)$ on the resistive layer is calculated through Gauss law as 
\beq
   q(r,t) = \vep_0   \frac{\partial \phi_1}{\partial z} \vert_{z=0} - \vep_0  \frac{\partial \phi_3}{\partial z} \vert_{z=0}
\eeq
which evaluates to
\beq \label{dist_no_ground}
  q(r,t) = \frac{Q}{2 \pi} \, \frac{vt}{\sqrt{(r^2+v^2t^2)^3}}
\eeq
The total charge on the resistive surface $Q_{tot} = \int_0^\infty 2r\pi q(r,t)dr$ is equal to $Q$ at any time, as expected. The peak and the FWHM of the charge density are given by
\beq
    q_{max} = \frac{Q}{2\pi} \, \frac{1}{v^2 t^2} \qquad FWHM = 2 (4^{1/3}-1)^{1/2} \approx 1.53 vt
\eeq
The charge is therefore 'diffusing' with a velocity $v$, but does not assume a Gaussian shape as expected from a diffusion effect but behaves as $1/r^3$ for large values of $r$.
The total current $I(r)$ flowing radially through a circle of radius $r$ is given by 
\beq
       I(r) = \frac{2r\pi}{R} E(r) = -\frac{2r\pi}{R}\,\frac{\partial \phi_1}{\partial r}\vert_{z=0} = \frac{Q v r^2}{ (r^2+v^2t^2)^{3/2}}
\eeq
It is easily verified that the rate of change of the total charge inside a radius $r$ i.e. $dQ_r(t)/dt = d/dt\,\int_0^r 2r' \pi q(r',t),dr'$ is equal the the current $I(r)$.

\clearpage

\subsection{Resistive layer grounded on a circle}

\begin{figure}[ht]
 \begin{center}
 a)
  \includegraphics[width=7cm]{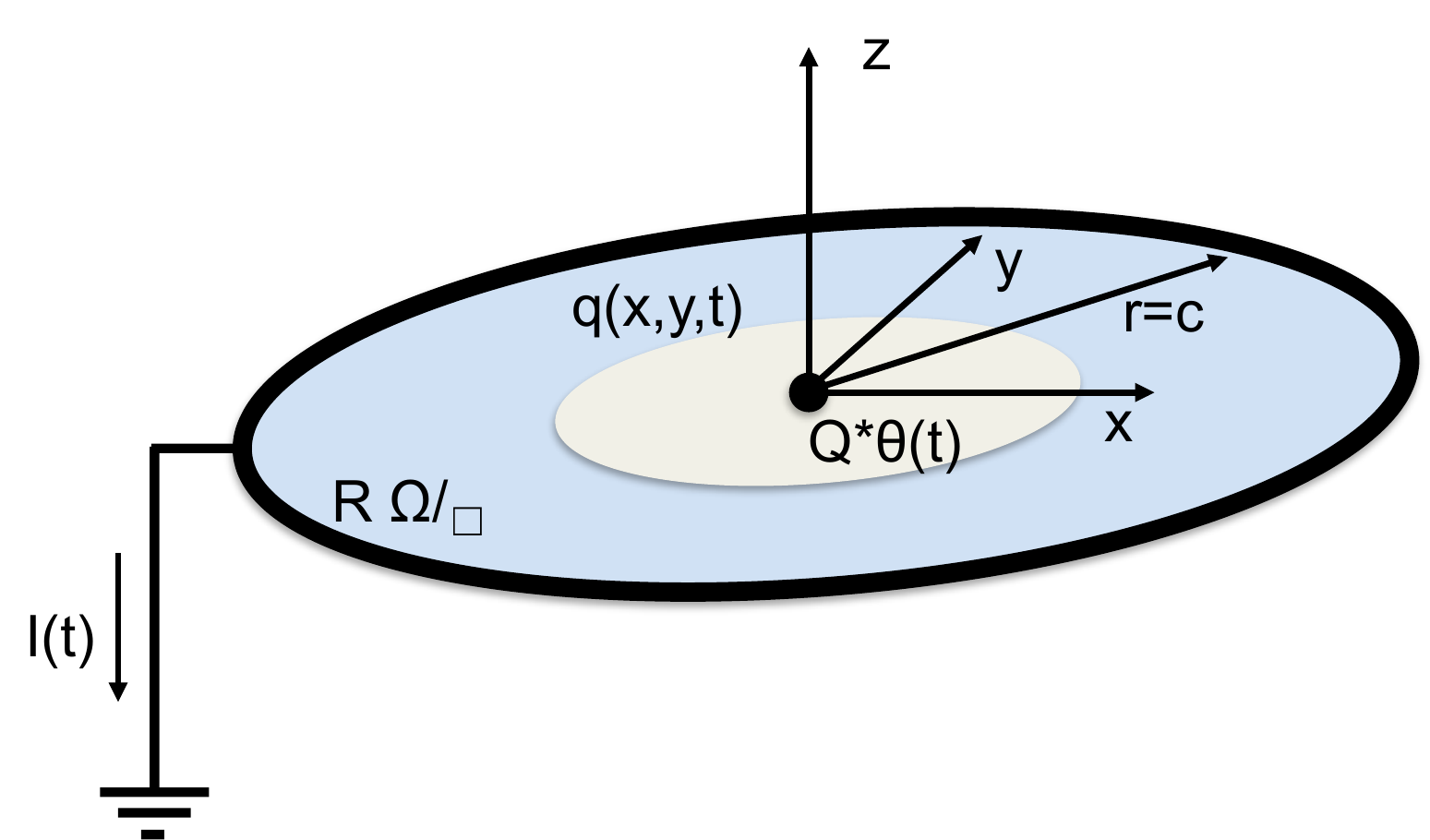}
b)
   \includegraphics[width=7cm]{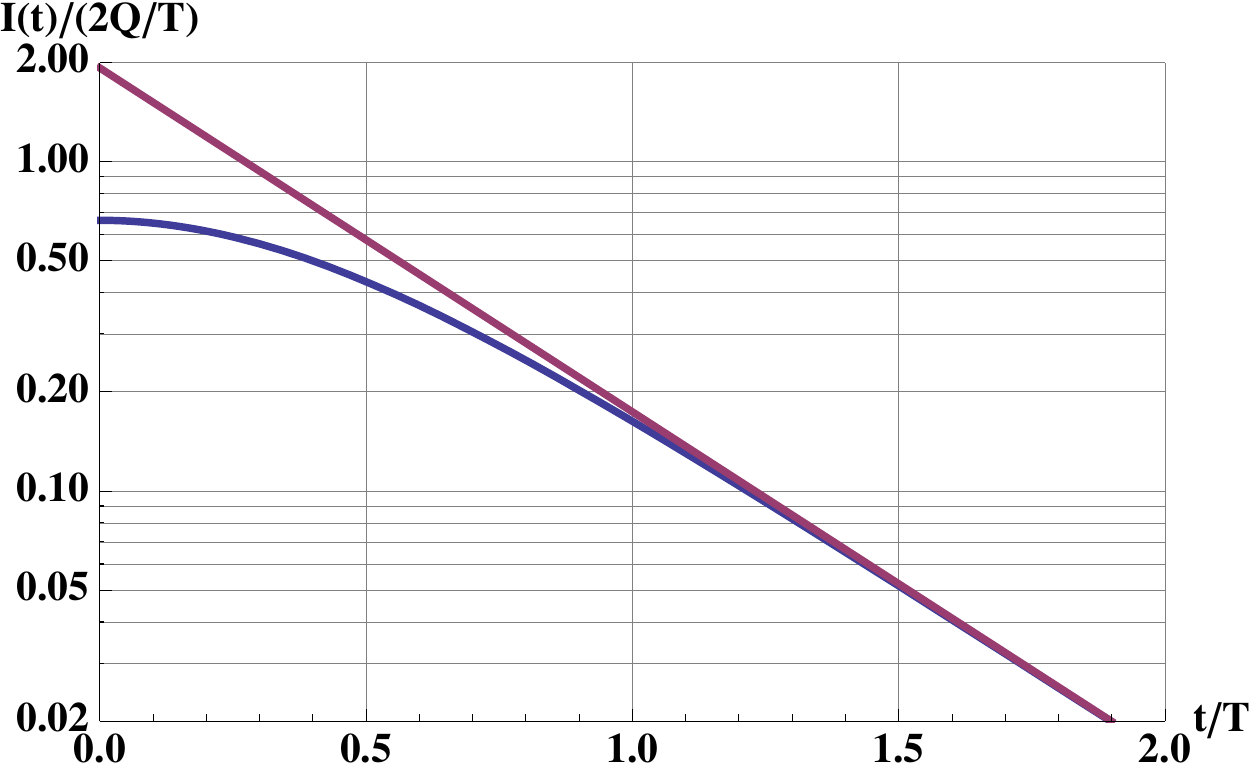}
   \caption{a) A point charge placed in the center of a resistive layer that is  grounded at $r=c$. b) Current $I(t)$ flowing to ground, where the straight line corresponds to the approximation from Eq. \ref{ring_approx}.}
  \label{charge_diff25}
  \end{center}
\end{figure}

We now assume the geometry to be grounded at a radius $r=c$ as shown in Fig. \ref{charge_diff25}a. Using Eq. \ref{sol2} with $r_0=0$ we have the solution
\beq
   \phi_1(r, z, t) = \frac{Q}{2\pi\vep_0 c} \sum_{l=1}^{\infty} \frac{J_0(j_{0l}\frac{r}{c})}{j_{0l}J_1^2(j_{0l})} \, e^{-j_{0l} (t/T-z/c)}
   \qquad
   T=c/v
\eeq
and $\phi_3(r,z,t) = \phi_1(r,-z,t)$. The charge inside the radius $c$ is not a constant but it will disappear with a characteristic time constant $T=c/v$ by currents flowing into the 'grounded' ring at $r=c$. As before we can calculate the surface charge density and charge inside the radius $r$, which evaluate to
\beq
   q(r,t) = \frac{Q}{c^2 \pi} \sum_{l=1}^\infty \frac{J_0(j_{0l} r/c)}{J_1^2(j_{0l})} e^{-j_{0l}t/T} \quad
   Q_{tot}(t) = 2Q\sum_{l=1}^\infty \frac{1}{j_{0l}J_1(j_{0l})} e^{-j_{0l}t/T}
\eeq
The current flowing into the 'grounded' ring is then 
\beq
  I(t)= -\frac{dQ_{tot}}{dt} = \frac{2r\pi}{R} E_r(r,t) = \frac{2Q}{T} \sum_{l=1}^\infty \frac{1}{J_1(j_{0l})} e^{-j_{0l}t/T}  
\eeq
One can verify that the total amount of charge flowing to ground $\int_0^{\infty} I(t)dt$ is equal to $Q$ as required.  The current can be pictured to decay with an infinite number of time constants $\tau_l=T/j_{0l}$, so for long times the largest one i.e. $T/j_{01}\approx 0.42\,T$ will dominate and the current decays as 
\beq \label{ring_approx}
               I(t) \approx  \frac{2Q}{TJ_1(j_1)}\,e^{-j_{01} t/T} \qquad t \gg T
\eeq
The exact and approximate expressions for $I(t)$ are shown in Fig. \ref{charge_diff25}b.

\subsection{Resistive layer grounded on a rectangle}

\begin{figure}[ht]
 \begin{center}
 a)
  \includegraphics[width=7cm]{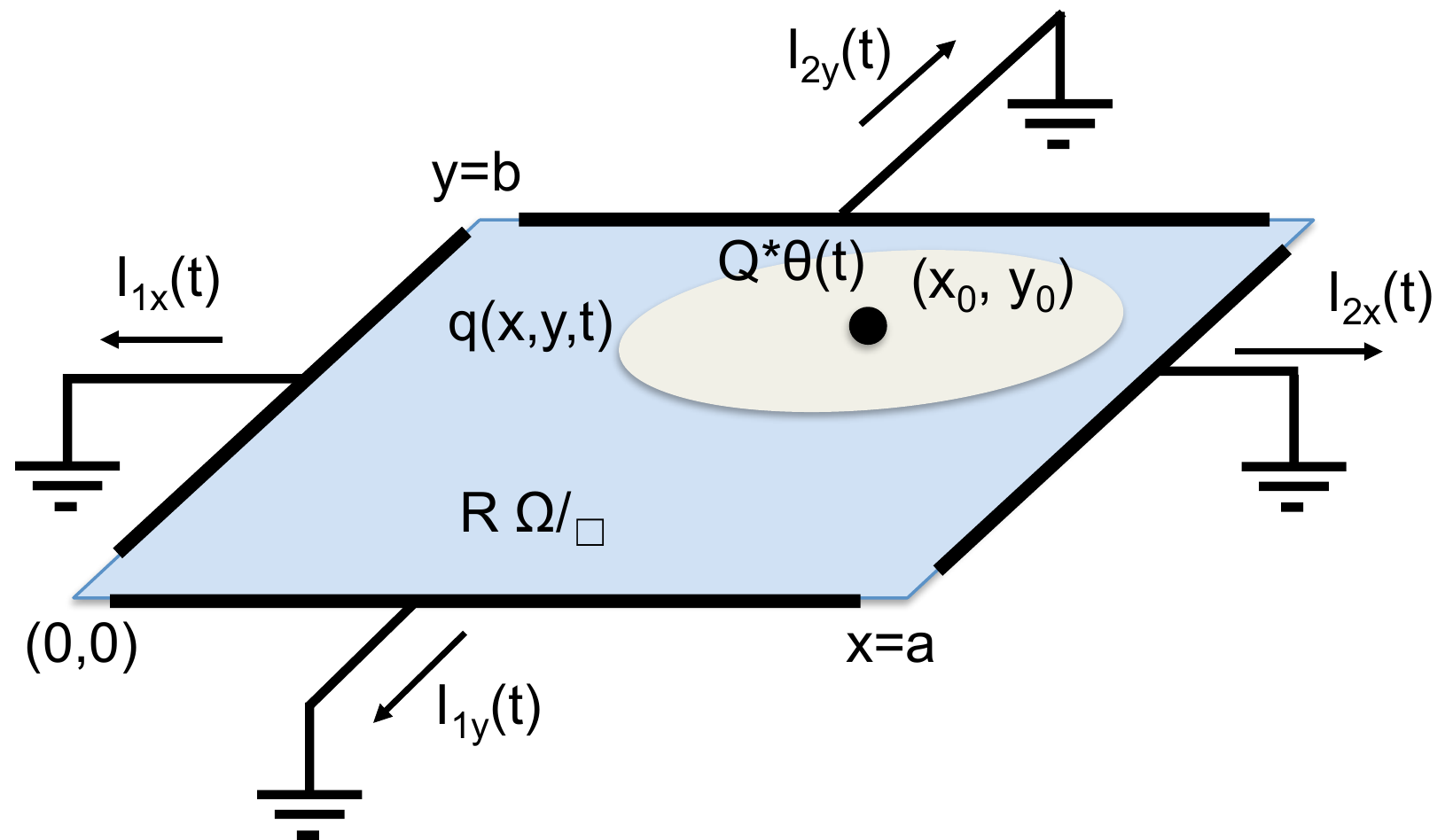}
b)
   \includegraphics[width=7cm]{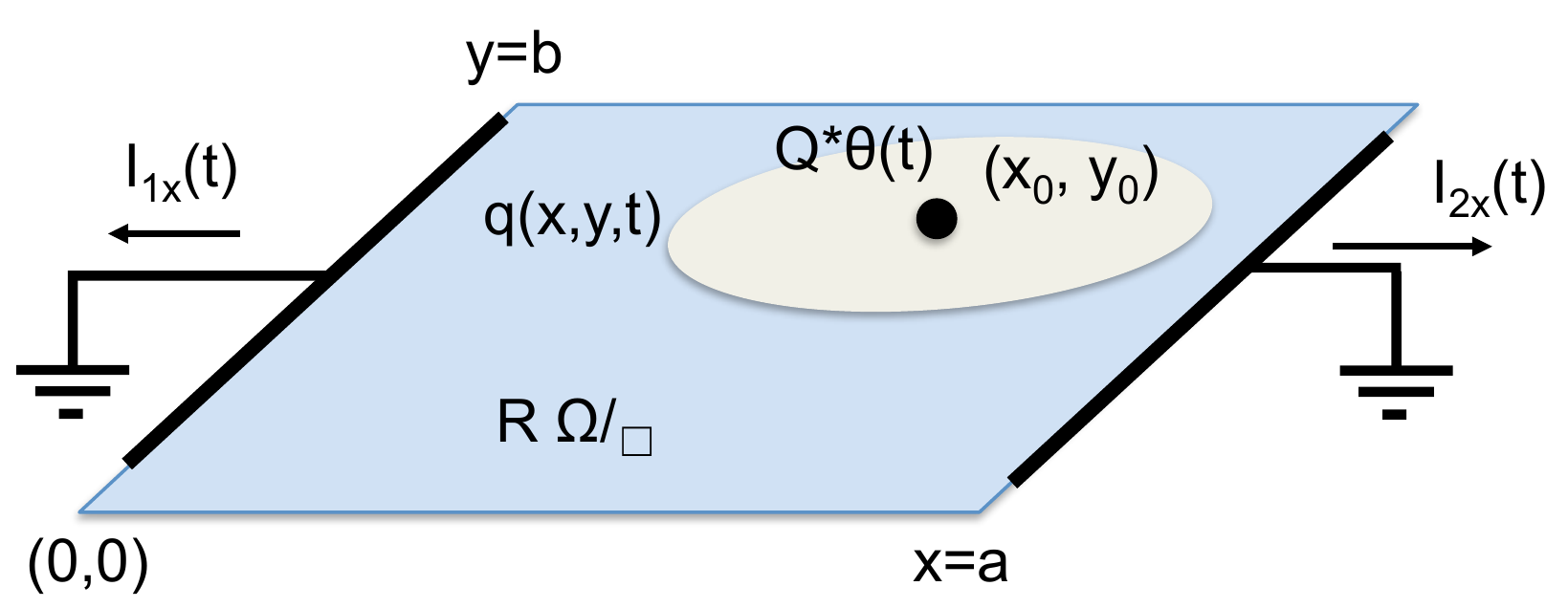}
   \caption{a) A point charge placed on a resistive layer that is grounded on a rectangle. b) A point charge placed on a resistive layer that is grounded on at $x=0$ and $x=a$ but insulated on the other borders.}
  \label{charge_diff4}
  \end{center}
\end{figure}

\begin{figure}[ht]
 \begin{center}
  \includegraphics[width=7cm]{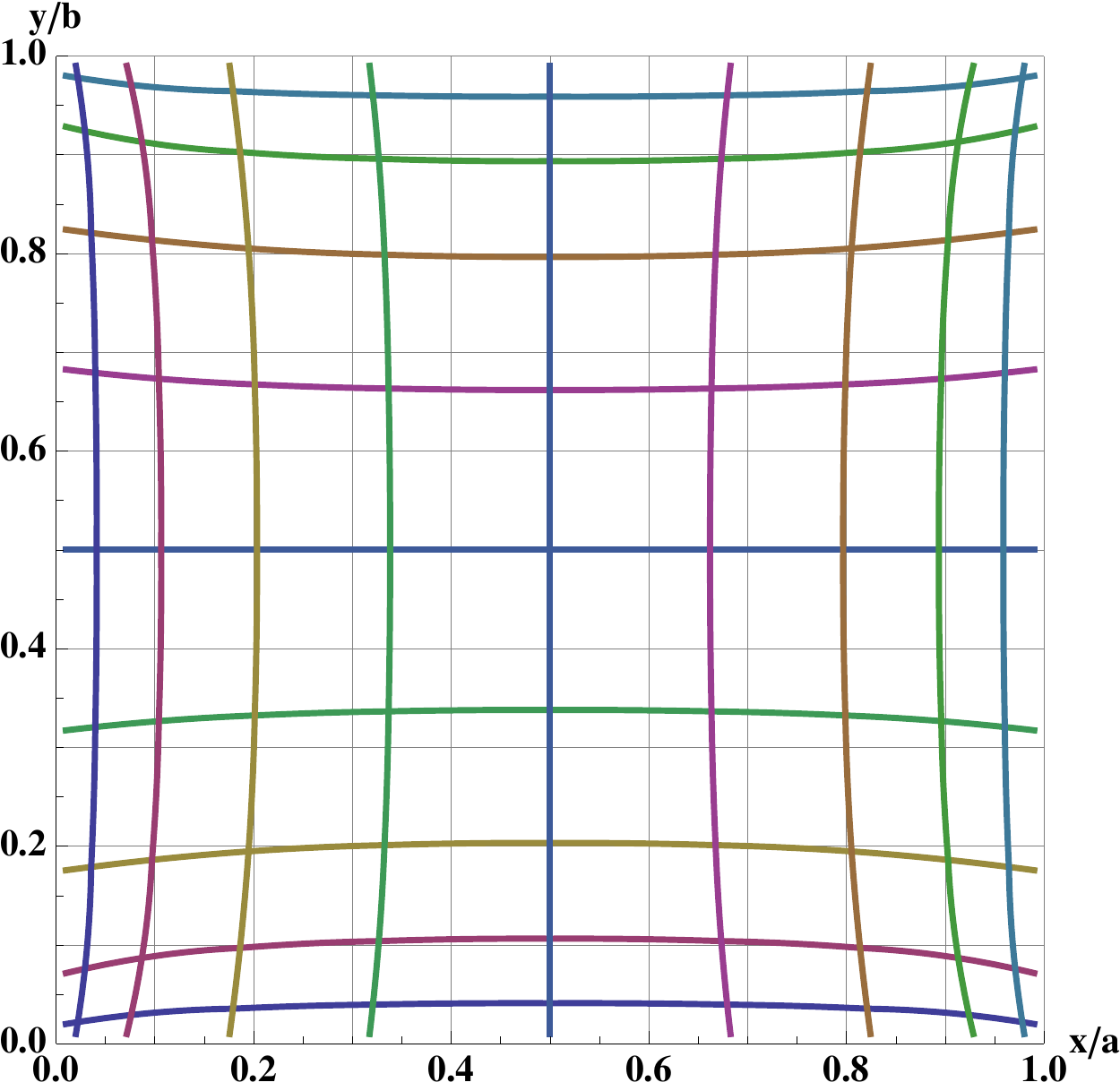}
  \caption{Correction map for the case where the position of the charge is determined by linear interpolation between the measured charges on the 4 boundaries of the geometry in Fig. \ref{charge_diff4}a. }
  \label{kissenverzerrung}
  \end{center}
\end{figure}

Next we assume a rectangular grounded boundary at $x=0, x=a$ and $y=0, y=b$ and place a charge $Q$ at position $x_0, y_0$ at $t=0$ as indicated in Fig. \ref{charge_diff4}a).
The potentials $\phi_1$ and $\phi_3$ are given by Eq. \ref{sol3}. Assuming the currents pointing to the outside of the boundary, the currents flowing through the 4 boundaries are
\beq
    I_{1x} = -\frac{1}{R}\,\int_0^b -\frac{\partial \phi_1}{\partial x}\vert_{x=0} dy \qquad 
    I_{2x} =  \frac{1}{R}\,\int_0^b -\frac{\partial \phi_1}{\partial x}\vert_{x=a} dy 
\eeq
\beq
    I_{1y} = -\frac{1}{R}\,\int_0^a -\frac{\partial \phi_1}{\partial x}\vert_{y=0} dx \qquad 
    I_{2y} =  \frac{1}{R}\,\int_0^a -\frac{\partial \phi_1}{\partial x}\vert_{y=b} dx 
\eeq
which evaluates to
\beq
  I_{1x}(t) = \frac{4Qv}{a^2}\sum_{l=1}^\infty\sum_{m=1}^\infty
  \frac{l}{m}\frac{1}{k_{lm}}
  \left[1-(-1)^m \right]
  \sin \frac{l\pi x_0}{a}\sin \frac{l\pi y_0}{b}
  e^{-k_{lm}vt}
\eeq
\beq
  I_{2x}(t) = \frac{4Qv}{a^2}\sum_{l=1}^\infty\sum_{m=1}^\infty
  \frac{l}{m}\frac{1}{k_{lm}}
  (-1)^l\left[(-1)^m -1\right]
  \sin \frac{l\pi x_0}{a}\sin \frac{l\pi y_0}{b}
  e^{-k_{lm}vt}
\eeq
\beq
  I_{1y}(t) = \frac{4Qv}{b^2}\sum_{l=1}^\infty\sum_{m=1}^\infty
  \frac{m}{l}\frac{1}{k_{lm}}
  \left[1-(-1)^l \right]
  \sin \frac{l\pi x_0}{a}\sin \frac{l\pi y_0}{b}
  e^{-k_{lm}vt}
\eeq
\beq
  I_{2y}(t) = \frac{4Qv}{b^2}\sum_{l=1}^\infty\sum_{m=1}^\infty
  \frac{m}{l}\frac{1}{k_{lm}}
  (-1)^m\left[(-1)^l -1\right]
  \sin \frac{l\pi x_0}{a}\sin \frac{l\pi y_0}{b}
  e^{-k_{lm}vt}
\eeq
In case we want to know the total charge flowing through the grounded sides we have to integrate the above expressions from $t=0$ to $\infty$ which results in the same expressions and just $e^{-k_{lm}vt}$ replaced by $1/(k_{lm}v)$. These measured currents or charges can be used to find the position $x_0, y_0$ where the charge $q$ was deposited. This principle is e.g. applied in the MicroCAT detector \cite{microcat}. As an example, Fig. \ref{kissenverzerrung} shows an evaluation of the above expressions for the total charges measured on the four sides for the positions $[x_0, y_0] = [(a/10, 2a/10, ..., 9a/10), (b/10, 2b/10, ..., 9b/10)]$ and using linear charge interpolation to determine the position. This means that the figure represents the 'correction map' to arrive at the correct position from the linear interpolation of the measured charges.

\subsection{Resistive layer grounded at $\pm a$ and insulated at $\pm b$.}

\begin{figure}[ht]
 \begin{center}
  \includegraphics[width=7cm]{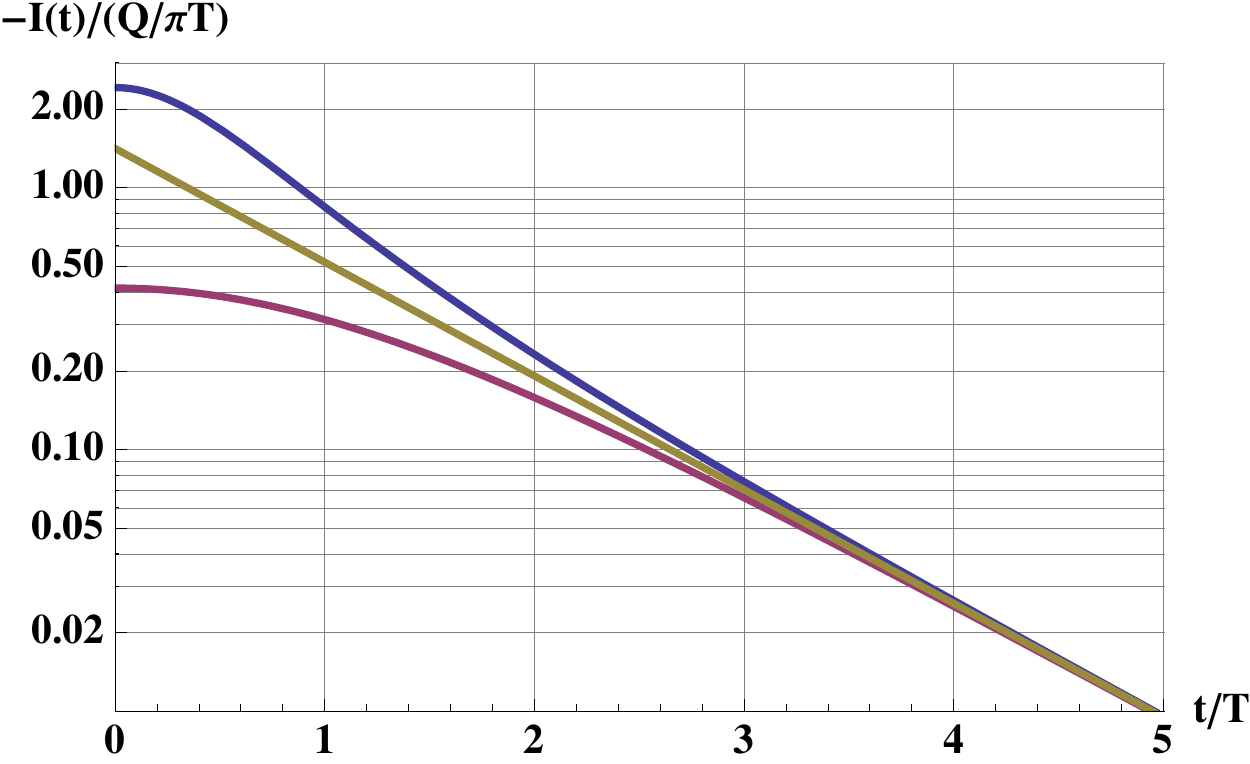}
  \caption{Currents $I_{1x}(t)$ (top) and $I_{2x}(t)$ (bottom) for the geometry of Fig. \ref{charge_diff4}b for $x_0=a/4$. The straight line in the middle refers to the approximation from Eq. \ref{asymptotic_charge_sharing}.}
  \label{square_currents}
  \end{center}
\end{figure}

In case the resistive layer is grounded at $x=0, x=a$ and insulated at $y=0, y=b$, as shown in Fig. \ref{charge_diff4}b), the currents are only flowing into the grounded elements at $x=0$ and $x=a$. We use Eq. \ref{sol4} and with some effort the summation can be achieved and evaluates to
 \beq
     I_{1x}(t) = -\frac{1}{R}\,\int_0^b -\frac{\partial \phi_1}{\partial x}\vert_{x=0} dy =
      -\,\frac{Q}{\pi T} \frac{\sin(\pi\frac{x_0}{a})}{\cosh(\frac{t}{T})-\cos( \pi\frac{x_0}{a})} 
\eeq
\beq
      I_{2x}(t) = \frac{1}{R}\,\int_0^b -\frac{\partial \phi_1}{\partial x}\vert_{x=a} dy =
       -\,\frac{Q}{\pi T} \frac{\sin(\pi\frac{x_0}{a})}{\cosh(\frac{t}{T})+\cos(\pi\frac{x_0}{a})}  
\eeq
with $T = \frac{2a\vep_0R}{\pi}  = \frac{a}{\pi v}$. For large times both expressions tend to
\beq \label{asymptotic_charge_sharing}
   I_{1x}(t) = I_{2x}(t) \approx -\frac{2Q}{\pi T} \sin\left( \pi \frac{x_0}{a} \right) \,e^{-t/T}
\eeq
Fig. \ref{square_currents} shows the two currents for a charge deposit at position $x_0=a/4$ together with the asymptotic expression from Eq. \ref{asymptotic_charge_sharing}.
The total charges $q_1$ and $q_2$ that are flowing through the grounded ends are given by
\beq
  q_1=\int_0^\infty I_{1x}(t) dt = Q \frac{a-x_0}{a}  \qquad  q_2=\int_0^\infty I_{2x}(t) dt = Q \frac{x_0}{a}
\eeq
so we learn that the charges are just shared in proportion to the distance from the grounded boundary, equal to the resistive charge division.

\section{Resistive layer parallel to a grounded plane}

\begin{figure}[ht]
 \begin{center}
 a)
   \includegraphics[width=7cm]{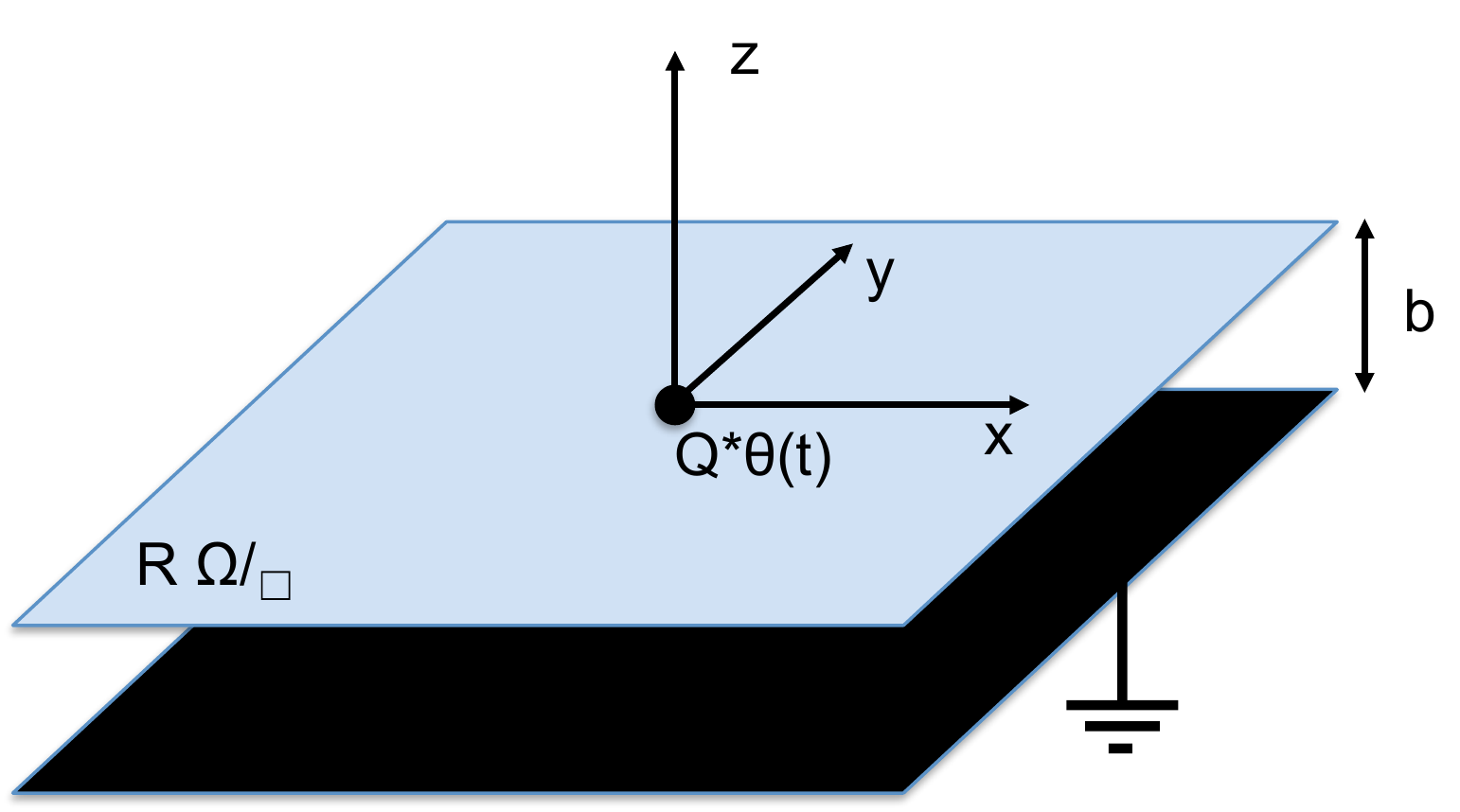}
  b)
  \includegraphics[width=7cm]{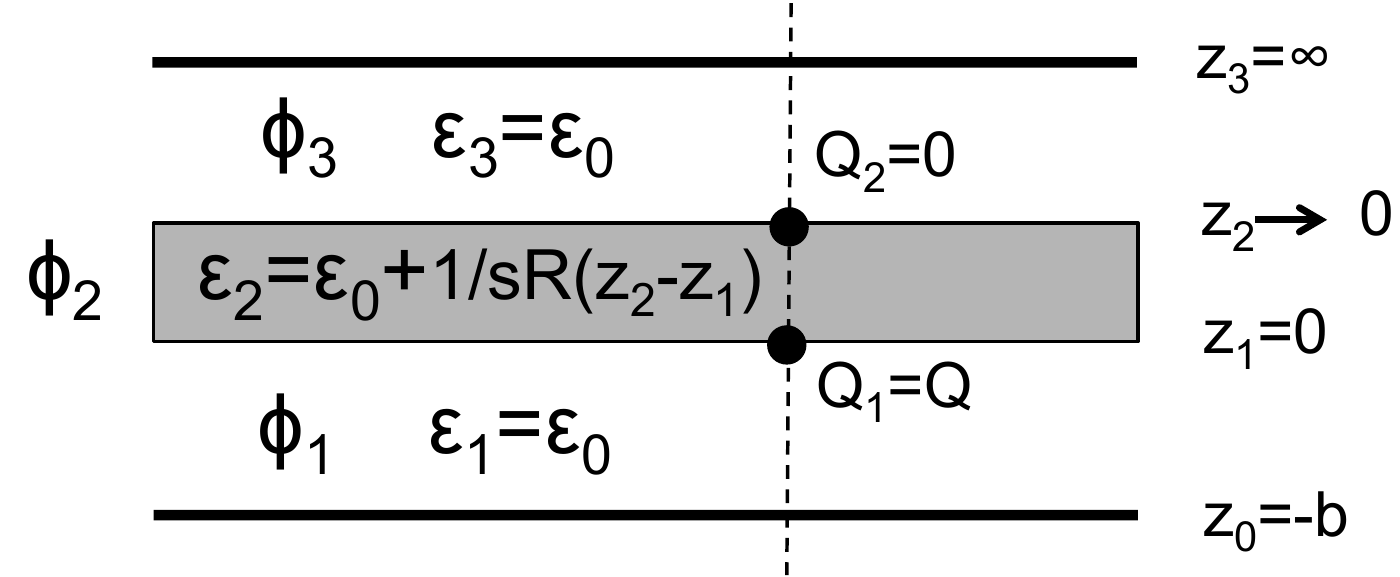}
  \caption{a) A resistive layer with surface resistance $R\,[\Omega$/square] in presence of a ground layer at distance $b$. b) The fields for this geometry can be calculated from the 3-layer geometry by performing the indicated limits of the expressions for $z_2, z_3$}
  \label{1layer_sketch_ground}
  \end{center}
\end{figure}

In this section we want to study the fields and charges in a layer of surface resistivity $R\,[\Omega$/square] at $z=0$ where we place a charge $Q$ at $r=0$ at $t=0$ in presence of a grounded plane at $z=-b$ as shown in Fig. \ref{1layer_sketch_ground}.
The solution can again be derived from the 3-layer geometry with 
\bdi
    \vep_1 = \vep_0  \qquad 
     \vep_2 = \vep_0+\frac{1}{sz_2R} \qquad 
     \vep_3 = \vep_0 \qquad
    Q_1=\frac{Q}{s} \qquad 
    Q_2 = 0
\edi
and taking the limits 
\beq
       z_0=-b \qquad z_1=0 \qquad z_2 \rightarrow 0 \qquad z_3 \rightarrow \infty
\eeq
Since we have shrunk layer 2 to zero we only have the coefficients $A_1, B_1$ for the layer $-b<z<0$ and  $A_3, B_3$ for the layer $z>0$.
\beq
        A_1 = \frac{QR\,e^{kb}}{2D(k)} \quad   B_1 = -\frac{QR\,e^{-kb}}{2D(k)}  \quad
       A_3 = 0 \quad B_3 = -\frac{QR \sinh(kb)}{D(k)}
\eeq
\beq
       D(k)=k\sinh(kb)+e^{kb}\vep_0 Rs
\eeq
In the time domain they read as
\beq
              A_1 =  \frac{Q}{2\vep_0} \exp[{-(1-e^{-2kb})kvt}] \quad
              B_1 =  \frac{Q}{2\vep_0} e^{-2kb}\,\exp[{-(1-e^{-2kb})kvt}] \quad
\eeq
\beq
              A_3 = 0      \quad  B_3 =  \frac{Q}{2\vep_0} (1-e^{-2kb})\,\exp[{-(1-e^{-2kb})kvt}] \quad
\eeq

\subsection{Infinitely extended geometry}

\begin{figure}[ht]
 \begin{center}
 a)
  \includegraphics[width=7cm]{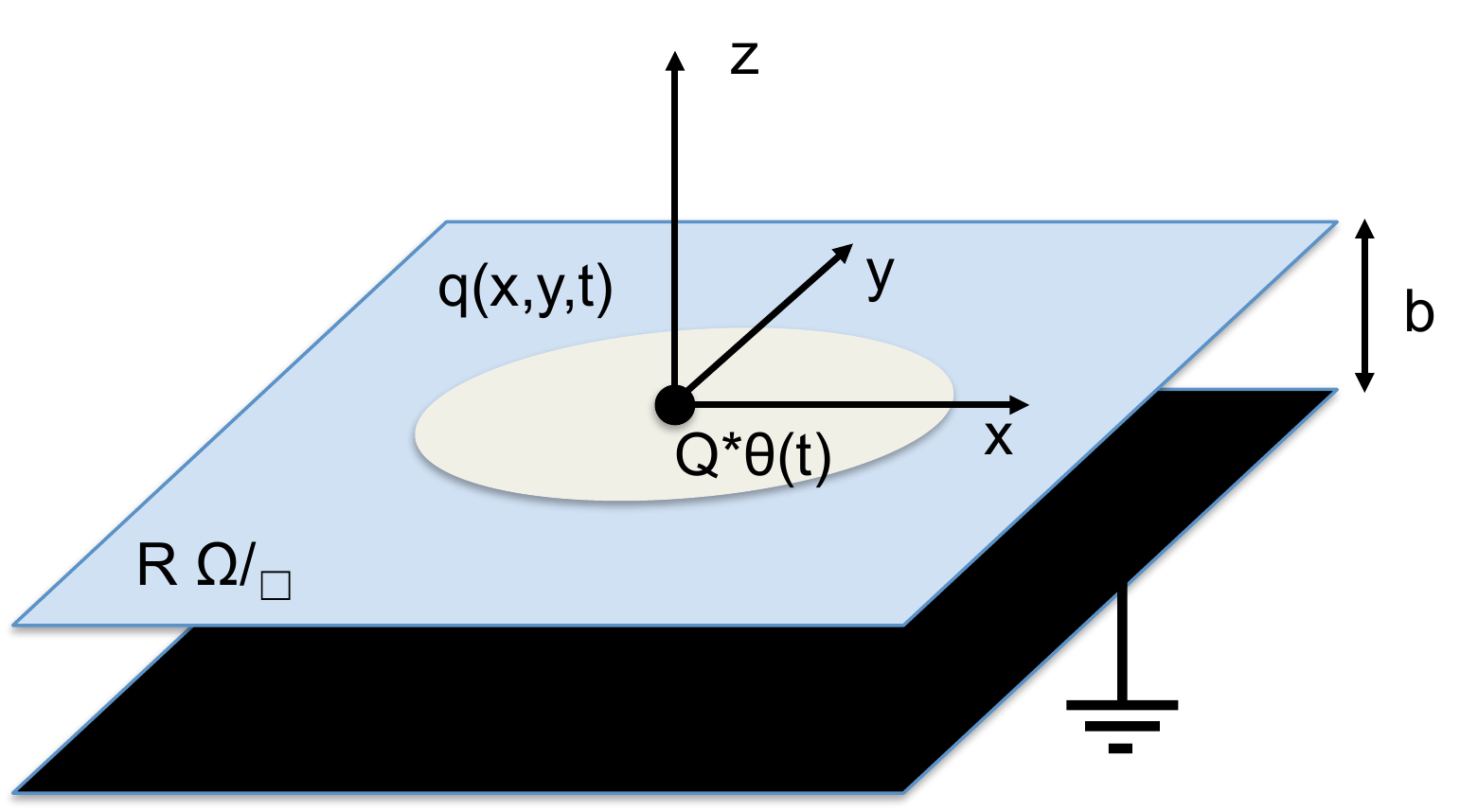}
b)
   \includegraphics[width=7cm]{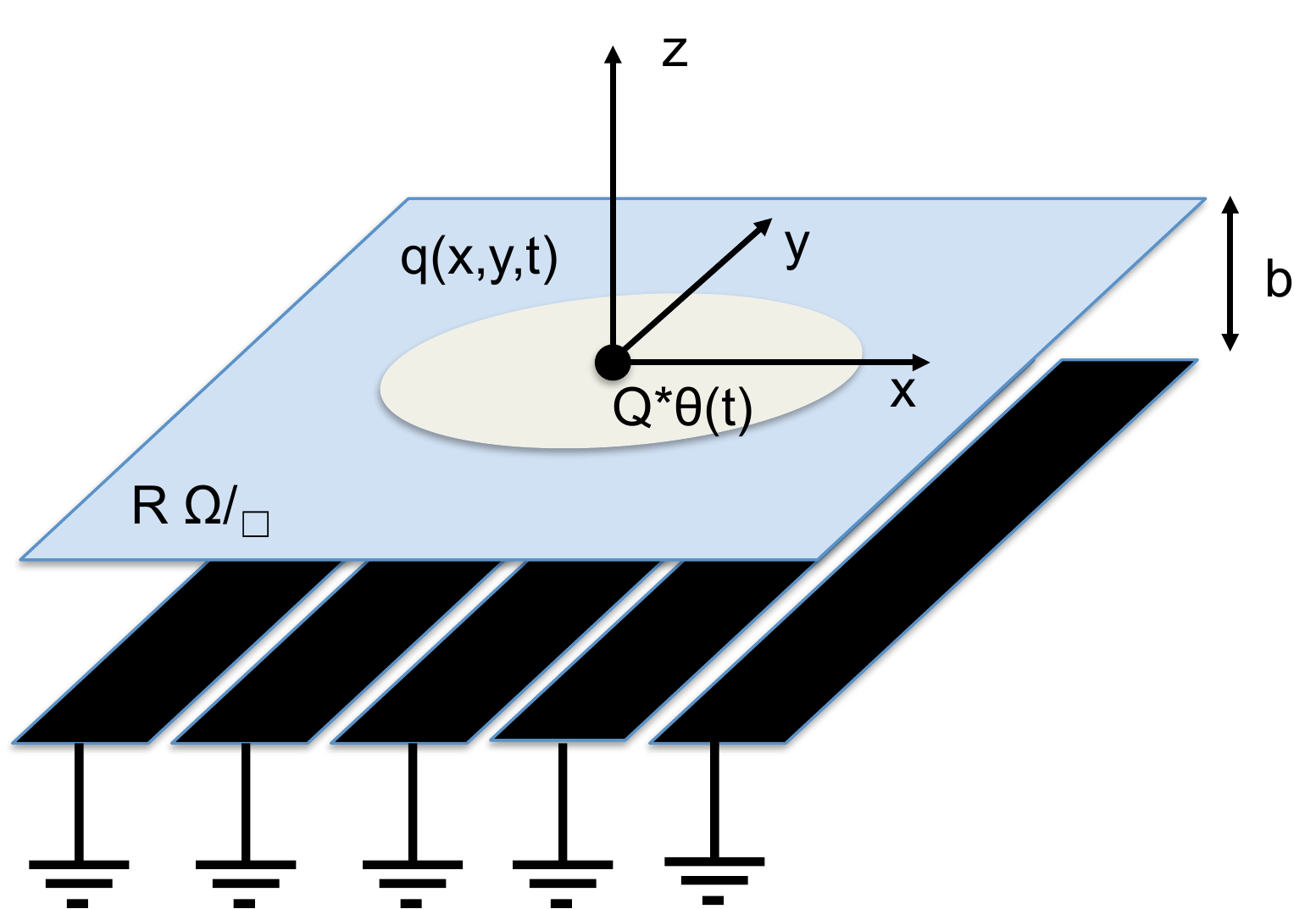}
   \caption{a) A point charge $Q$ placed on an infinitely extended resistive layer in presence of a grounded layer. b) In case the ground plane is segmented, the time dependent charge distribution $q(r,t)$ does induce charges on the strips.}
  \label{charge_diff6}
  \end{center}
\end{figure}
\begin{figure}[ht]
 \begin{center}
 a)
  \includegraphics[width=6.5cm]{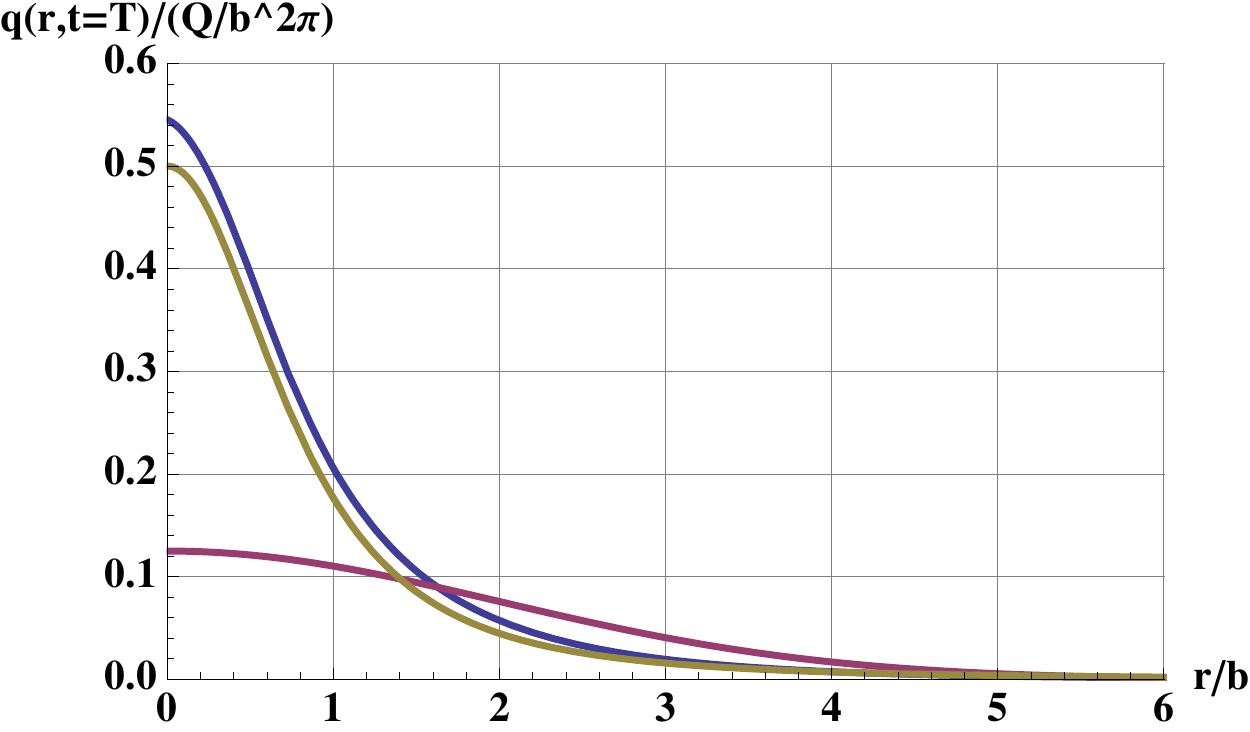}
 b)
  \includegraphics[width=6.5cm]{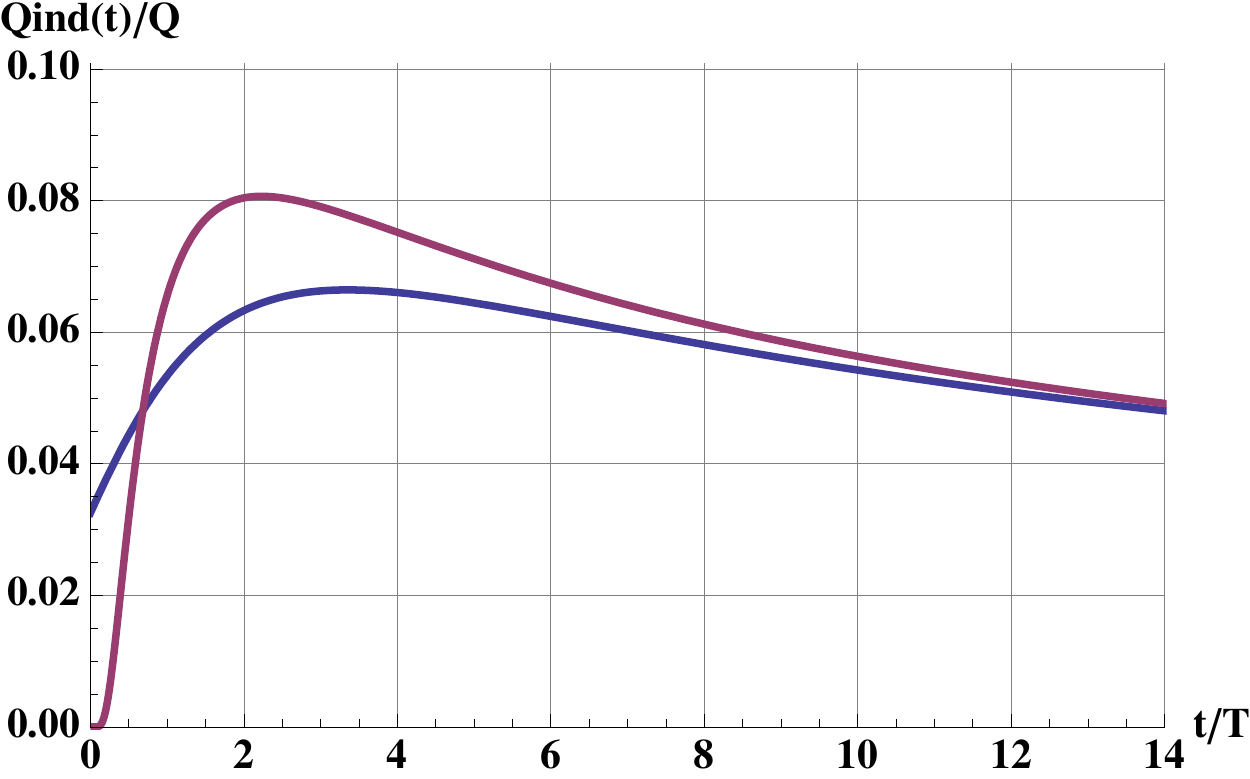}
  \caption{a) Charge distribution from  Eq. \ref{dist_correct}  (top) compared to the Gaussian approximation for a time of $t=T$ (bottom). The middle curve that closely tracks the correct one is from Eq. \ref{dist_no_ground} and refers to the geometry without grounded plate. b) Induced charge on a strip of width $b$ at position of $x=2b$. The curve staring from zero refers to the Gaussian approximation.}
  \label{gauss_and_correct}
  \end{center}
\end{figure}

Assuming an infinitely extended geometry, the time dependent charge density evaluates to
\beq \label{dist_correct}
   q(r,t) = \frac{Q}{b^2 \pi}  \,\frac{1}{2} \int_0^\infty \kappa J_0(\kappa \frac{r}{b})\,
   \exp \left[ {-\kappa(1-e^{-2\kappa})\frac{t}{T}} \right]  d\kappa
   \qquad T=\frac{b}{v} = 2b\vep_0R
\eeq
It can be verified that $\int_0^\infty 2r\pi q(r,t)dr = Q$ at any time. For long times i.e. large values of $t/T$, the integrand contributes only for small values of $\kappa$ and we can approximate the exponent by
\beq
    -\kappa(1-e^{-2\kappa})\frac{t}{T} \, \rightarrow  \, -2 \kappa^2 \frac{t}{T}
\eeq
and the integral evaluates to
\beq \label{gaussian_ground}
   q(r,t) = \frac{Q}{b^2 \pi} \, \frac{1}{8t/T}\,e^{-\frac{r^2}{8b^2t/T}}
\eeq
We see that the charge distribution does assume a Gaussian shape for long times, in contrast to the situation discussed in the previous section where the ground plane is absent. This fact can be understood by investigating the equation defining this specific geometry: the current  $\vec{j}(x,y,t)$ flowing inside the resistive layer is related to the electric field $\vec{E}(x,y,t)$ in the resistive layer by $\vec{j}=\vec{E}/R$. The relation between the current and the charge density $q(x,y,t)$ is $\grad \vec{j} = -\partial q/\partial t$. With $\vec{E}=-\grad \phi$ we then get
\beq
      \frac{\partial q}{\partial t} = 
   \frac{1}{R} \left(
           \frac{\partial^2 \phi}{\partial x^2} +  \frac{\partial^2 \phi}{\partial y^2}    \right)
\eeq
If we set $q=C\phi$ we get the diffusion equation 
\beq \label{diffusion_equation}
   \frac{\partial q}{\partial t} = 
   h \left(
           \frac{\partial^2 q}{\partial x^2} +  \frac{\partial^2 q}{\partial y^2}    \right)
           \quad h=1/RC \quad C=\frac{\vep_0}{b}
\eeq
where $C$ is the capacitance per unit area between two metal plates at distance $b$. The solution of this equation for a point charge $Q$ put at $r=0, t=0$ evaluates exactly to the above Gaussian expression from Eq. \ref{gaussian_ground}. This relation between voltage and charge($Q=CU$) is however only a good approximation if the charge distribution does not have a significant gradient over distances of the order of $b$. For small times when the charge distribution is very peaked around zero this is certainly not a good approximation. In Fig. \ref{gauss_and_correct} a) the charge distribution from Eq. \ref{dist_correct} at time $t=T$ is compared to the above Gaussian as well as Eq. \ref{dist_no_ground} for the geometry without a ground plane.  We see that for small times the solution of the diffusion equation does not work very well. 
\\ \\
Equation \ref{diffusion_equation} is often written in analogy of the one dimensional transmission line equation, which for negligible transconductance $G$ reads as 
\beq
   \frac{\partial^2 V(x,t)}{\partial x^2} = LC  \frac{\partial^2 V(x,t)}{\partial t^2} + RC  \frac{\partial V(x,t)}{\partial t} 
\eeq
In case '$LC\partial / \partial t \ll RC$', meaning that the $RC$ time constant is much larger than the signal propagation time along the transmission line, the equation is approximated as  
\beq
   \frac{\partial^2 V(x,t)}{\partial x^2} = RC  \frac{\partial V(x,t)}{\partial t} 
\eeq
which corresponds to the one dimensional diffusion equation. The transmission line equation is however derived from the simplified 'lumped' transmission line model, where a sequence of $R, L, C$ elements is assumed and the continuous limit is taken. It means that at any point $x$ the relation between the voltage $V$ and the charge density $q$ at this point is given by $q=CV$, which again is only a good approximation when the gradient of the charge density is small over the transverse dimension of the transmission line. The above equation for the one dimensional problem and Eq. \ref{diffusion_equation} for the two dimensional problem are therefore both bad approximations for times $t<RC$ in case a point charge is placed somewhere in the geometry.
%
%
%
%
%
%
%
%
%
\\ \\
The presence of the charge on the resistive layer induces a charge on the grounded metal plane. If we assume that the metal plane is segmented into strips, as shown in Fig. \ref{charge_diff6}b, we can calculate the induced charge through the electric field on the surface of the plane. Assuming a strip centred at $x=x_p$ with a width of $w$ and infinite extension in $y$ direction, we find the induced charge to
\beq
     Q_{ind}(t) = \int_{x_p-w/2}^{x_p+w/2}\int_{-\infty}^{\infty}-\vep_0\frac{\partial \phi_1}{\partial z}\vert_{z=-b}\,dydx 
     \qquad 
\eeq
which evaluates to
\beq \label{qind_correct}
   Q_{ind}(t) =    \frac{2Q}{ \pi} \,  \int_0^\infty \frac{1}{\kappa}\cos(\kappa \frac{x_p}{b}) \sin (\kappa \frac{w}{2b})
   \exp \left[ -\kappa-\kappa(1-e^{-2\kappa})\frac{t}{T} \right]  d\kappa
\eeq
Approximating the integrand for large values of $t/T$ as above, the expression evaluates to 
\beq \label{qind_approx}
   Q_{ind}(t)=  
   \frac{Q}{2}\left[
     \mbox{erf} \left( \frac{2x_p+w}{4b\sqrt{2t/T}}\right) - \mbox{erf}  \left( \frac{2x_p-w}{4b\sqrt{2t/T}}\right)
   \right]
\eeq
The same solution is found by using the relation of a capacitor where the ground plate should just carry the charge density $-q(x,y,t)$, with $q(x,y,t)$ from Eq. \ref{gaussian_ground}, and integrating it over the strip area 
\beq
     Q_{ind}(t)=\int_{x_p-w/2}^{x_p+w/2}\int_{-\infty}^{\infty} q(x,y,t)dx dy 
\eeq
Both expressions (Eq. \ref{qind_correct} and Eq. \ref{qind_approx}) are shown in Fig. \ref{gauss_and_correct}b. Although there are significant differences at small times the curves approach each other for longer times when the charge distribution becomes broad. The solutions still do not represent a detector signal due to the unphysical assumption that the charge is created 'out of nowhere' at $t=0$. The correct signal on a strip due to a pair of charges $\pm Q$ moving in a detector will be discussed in Section \ref{signal_spread}. 

\subsection{Geometry grounded on circle}

\begin{figure}[ht]
 \begin{center}
 a)
  \includegraphics[width=7cm]{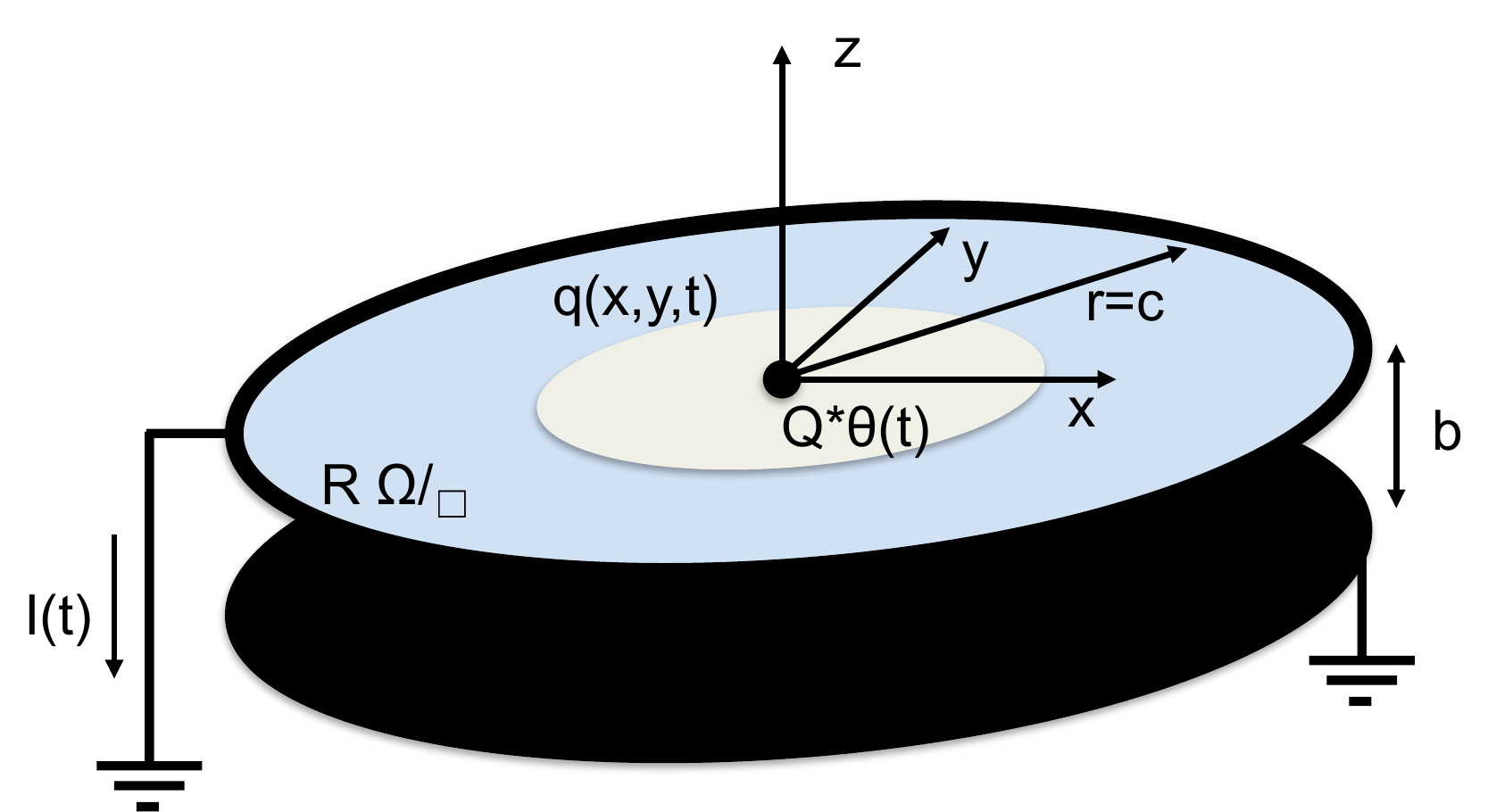}
   \caption{ b) The same geometry grounded at a radius $r=c$.}
  \label{charge_diff6}
  \end{center}
\end{figure}
To conclude we assume the geometry to be grounded at $r=0$ as shown in Fig. \ref{charge_diff6}b. We proceed as above and the charge $Q_{tot}$ inside the radius $c$ is given by
\beq
   Q_{tot}(t) = 2Q \sum_{l=1}^\infty \frac{1}{j_{0l}J_1(j_{0l})}\,
   \exp \left[ -j_{0l}(1-e^{-2j_{0l} b/c})\frac{t}{T} \right]
\eeq
The charge disappears with and infinite number of time constants
\beq
   \tau_l = \frac{T}{j_{0l}(1-e^{-2j_{0l}b/c})}
\eeq
If the radius of the circle $c$ is much larger than the distance $b$ the longest time constant approximates to $\tau_1 \approx T/j_{01}\approx 0.42T$ which is equal to the case where no ground plane is present. In case $c \gg b$ we have $\tau_1 \approx 0.1T\,c/b$, which tells us that the closer the resistive layer is to the grounded plane the slower the charge will disappear.

\section{Uniform currents on thin resistive layers}

\begin{figure}[ht]
 \begin{center}
  \includegraphics[width=7cm]{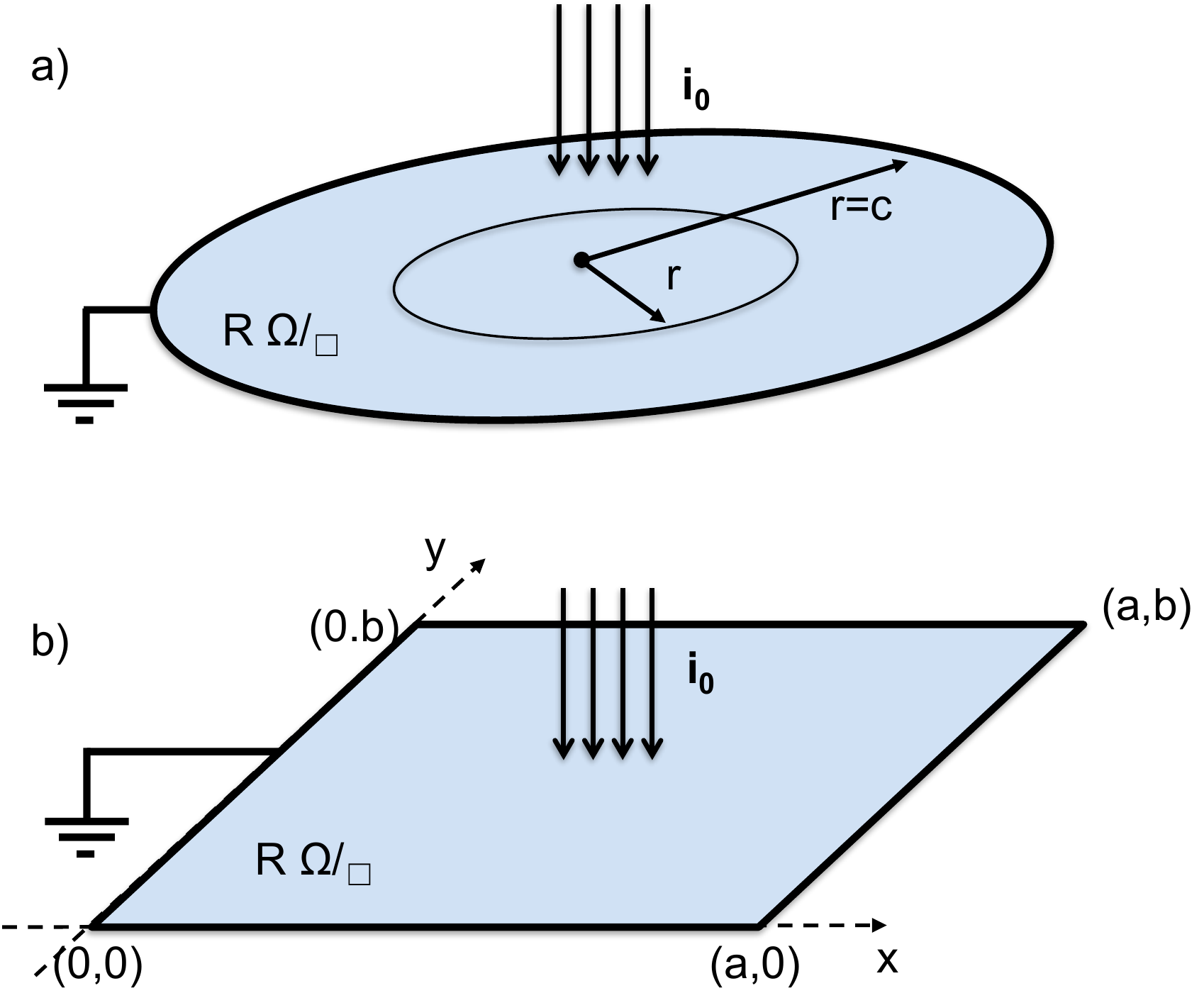}
   \includegraphics[width=7cm]{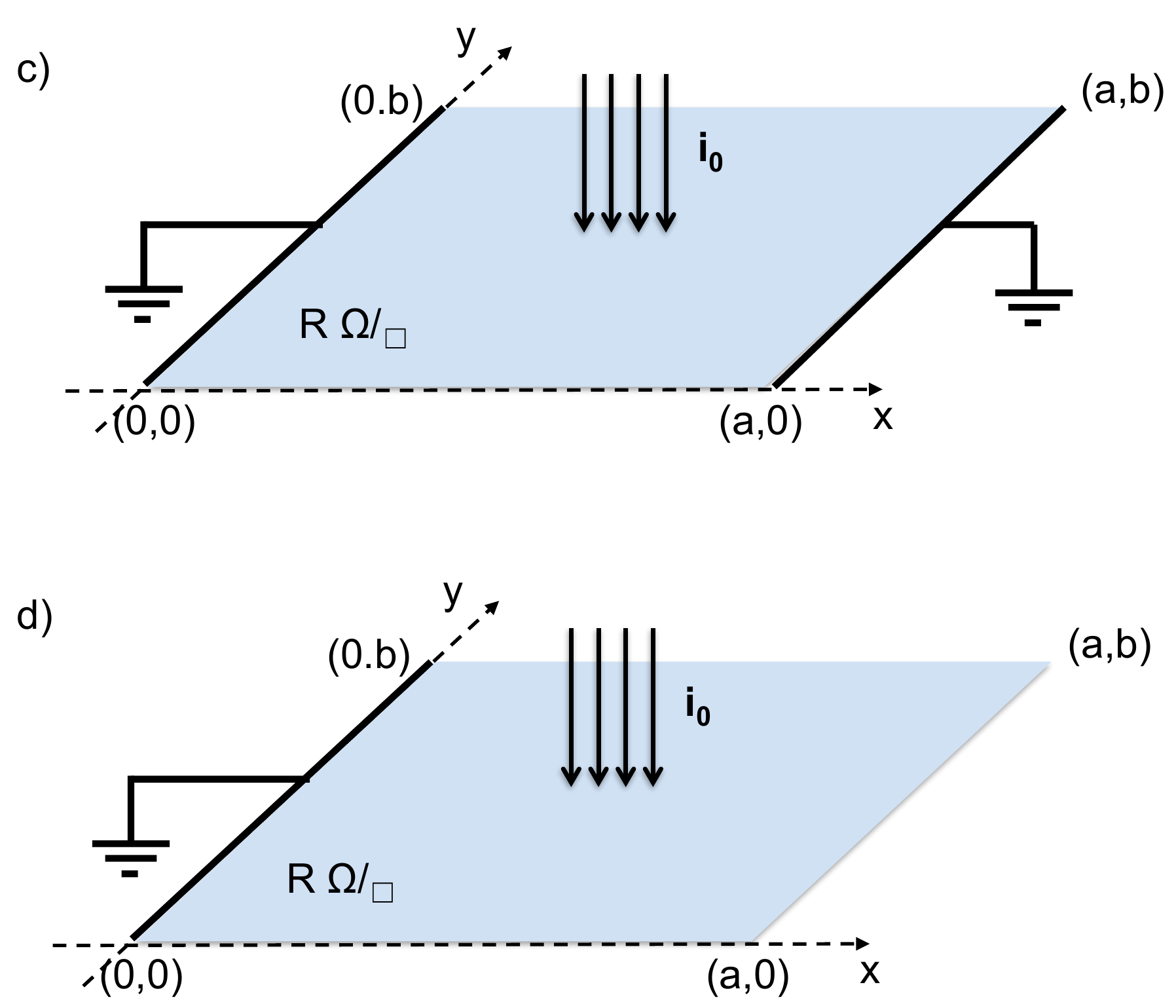}
  \caption{A uniform current 'impressed' on the resistive layer will result in a potential distribution that depends strongly on the boundary conditions. The 4 geometries shown in this figure are discussed.}
  \label{uniform_current}
  \end{center}
\end{figure}
\begin{figure}[ht]
 \begin{center}
  \includegraphics[width=7cm]{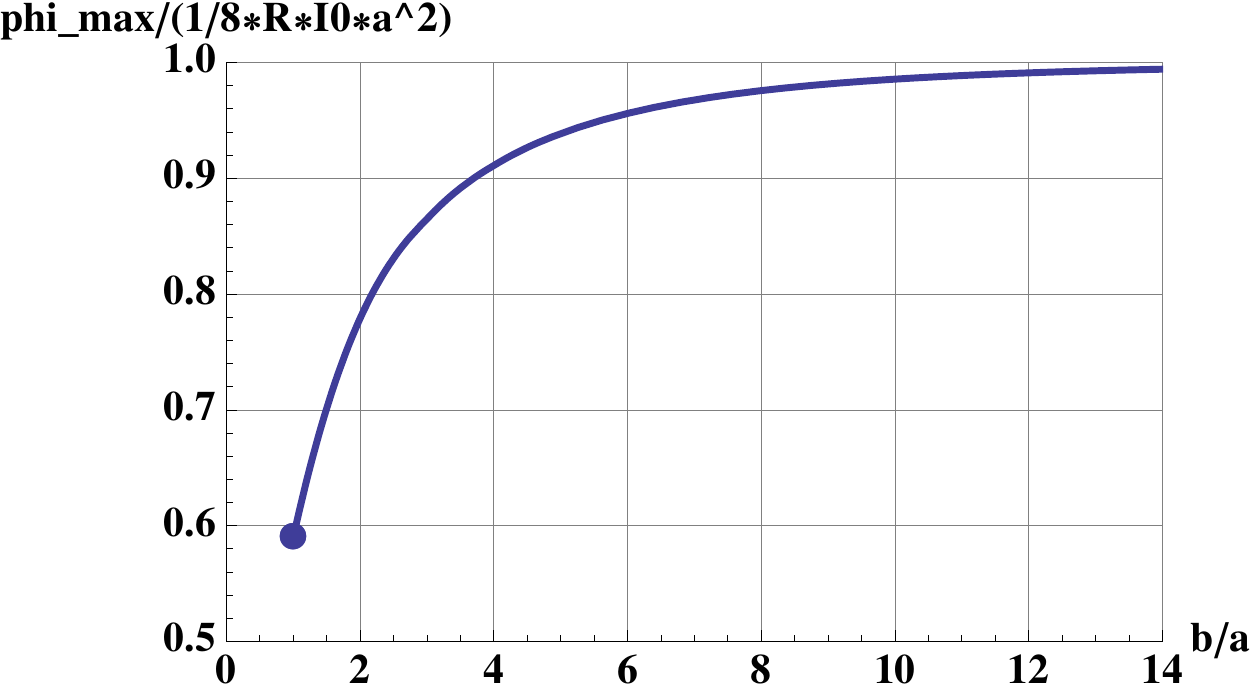}
  \caption{Ratio between the maximum potential of the geometry with 4 grounded edges and the geometry with 2 grounded edges. For large values of $b/a$ the ratio has to approach unity, and we see that already at $b/a=4$ the expressions differ only by 10\%.}
  \label{phimaxratio}
  \end{center}
\end{figure}

In this section we want to discuss the potentials that are created on thin resistive layers for uniform charge deposition. In detectors like RPCs and Resistive Micromegas such resistive layers are used for application of the high voltage and for spark protection. The resistivity must be chosen small enough to ensure that potentials that are established on these layers due to charge-up are not influencing the applied electric fields responsible for the proper detector operation. If such detectors are in an environment of uniform particle irradiation the situation can be formulated by placing a uniform 'externally impressed' current per unit area $i_0$\,[A/cm$^2$] on the resistive layer. For illustration we use the example of a resistive layer an absence of any grounded planes from Section \ref{thin_resistive_layers}.
First we want to investigate the geometry shown in Fig. \ref{uniform_current}a) where the layer is grounded on a circle at $r=c$.  The charge  $dq$ placed on an infinitesimal area at position $r_0, \phi_0$ after time $t$ is given by $dq(t) = i_0r_0dr_0d\phi_0 t$, or in the Laplace domain $dq(s) = i_0r_0dr_0d\phi_0/s^2$. We therefore have to replace $Q/s$ in Eq. \ref{single_thin_layer_onditions} by $q(s)$, which results in 
\beq 
   f_1(k,z,s) = \frac{i_0}{s}\frac{Rr_0dr_0d\phi_0}{k+2\vep_0Rs}\,e^{kz} 
   \qquad 
   f_2(k,z,s) = \frac{i_0}{s}\frac{Rr_0dr_0d\phi_0}{k+2\vep_0Rs}\,e^{-kz}
\eeq
Since we want to know the steady situation for long times i.e. for $t\rightarrow \infty$ we $f(k,z,t \rightarrow \infty) = \lim_{s \rightarrow 0} sf(k,z,s)$ and have
\beq \label{f_uniform_current}
   f_1(k,z) = \frac{Ri_0 r_0dr_0d\phi_0}{k}\,e^{kz} 
   \qquad 
   f_2(k,z) = \frac{Ri_0 r_0dr_0d\phi_0}{k}\,e^{-kz}
\eeq
Inserting this into Eq. \ref{sol2} and integration over the area of the disk $\int_0^cdr_0\int_0^{2\pi}d\phi_0$ we find that only the coefficients for $m=0$ are different from zero and get
\beq
  \phi_1(r,z) = \phi_3(r,-z)=2c^2Ri_0 \sum_{l=1}^\infty \frac{J_0(j_{0l}r/c)}{j_{0l}^3 J_1(j_{0l})}\,e^{j_{0l}z/c}
\eeq
For $z=0$ i.e. on the surface of the resistive layer, the expression can be summed and we have
\beq
    \phi_1(r,z=0)=\phi_3(r,z=0) = \frac{1}{4}R i_0 (c^2-r^2)
\eeq
This expression can also be derived in an elementary way: the total current on a disc of radius $r$ i.e. $r^2 \pi i_0$, is equal to the total radial current flowing at radius $r$ i.e. $2r\pi E_r/R$. This defines the radial field inside the layer to $E_r=Ri_0r/2$. With the boundary condition $\phi(c)=\int_0^cE_r(r)dr=0$ we find back the above expression. The maximum potential is therefore in the centre of the disc and is equal to
\beq \label{circle_max}
     \phi(r=0) = \frac{c^2 \pi R i_0 }{4\pi} = \frac{1}{4\pi} R  I_{tot}\approx 0.08 \, R  I_{tot}
\eeq
To find the potentials in the rectangular geometry of Fig. \ref{uniform_current}b we again have $f_1, f_2$ from Eq. \ref{f_uniform_current} we just have to replace $r_0 dr_0 d\phi_0$ by $dx_0dy_0$ and perform the integration $\int_0^a dx_0 \int_0^b dy_0$ of Eq. \ref{sol3}, which results in
\beq \label{phi_uniform_square}
  \phi_1(x,y,z) =  \phi_3(x,y,-z)= abR i_0 \frac{4}{\pi^4}\sum_{l=1}^\infty \sum_{m=1}^\infty
  \frac{[1-(-1)^l][1-(-1)^m] \sin(l\pi x/a) \sin(m\pi y/b)}
  {l^3m b/a + m^3 l a/b}\,e^{k_{lm}z}
\eeq
The expression cannot be written in closed form but converges quickly, so numerical evaluation is straight forward. The peak of the potential can be found by setting $d \phi_1/dx=0, d\phi_1/dy=0$ and is found at $x=a/2, y=b/2$, which is also evident by the symmetry of the geometry. The  maximum potential on the resistive layer is then 
\beq
   \phi_{max} = \phi(a/2,b/2,z=0) = \frac{1}{8}R i_0 a^2 b^2\sum_{l=1}^\infty \sum_{m=1}^\infty\frac{128}{\pi^4}\
  \frac{(-1)^{l+m}}{b^2(2l-1)^3(2m-1)+a^2(2m-1)^3(2l-1)}
\eeq
For a square geometry ($b=a$) the sum evaluates to $\approx 0.59$ so the peak voltage in the center is
\beq \label{square_max}
   \phi_{max} \approx 0.074 R i_0 a^2 = 0.074 \, R I_{tot}  
\eeq
We see that the value is only less than 10\,\% different from the peak voltage for the circular boundary in Eq. \ref{circle_max}. 
\\ \\
For uniform illumination of the geometry Fig. \ref{uniform_current}c that is grounded at $x=0,a$ and insulated at $y=0,b$ we use expression Eq. \ref{sol4} and proceed as before and find 
\beq
   \phi_1(x, z) =  \phi_3(x, -z)=2 R i_0  a^2 \sum_{l=1}^\infty \frac{(1-(-1)^l) \sin(l\pi x/a)}{l^3\pi^3}\,e^{l \pi z/a}
\eeq
The potential is is independent of $y$ and for $z=0$ the sum can be written inclosed form 
\beq
    \phi_1(x, z=0) = \frac{1}{2}Ri_0(ax-x^2) \qquad \phi_{max}=\frac{1}{8}a^2R i_0 
\eeq
Again this expression can be found in an elementary way by the fact that due to symmetry the currents can only flow in $x$-direction and the current at $x=a/2$ must be zero. The total current arriving on the area of $x=a/2 \pm \Delta x$ i.e. $2\Delta x b i_0$ is equal to the total current flowing at distance $s$ i.e. $2E(s)/Rb$. With $x=a/2+\Delta x$ we find back the above expression. The potential is therefore independent of $b$. For large values of $b/a$ the expression Eq. \ref{phi_uniform_square} must therefore approach the same value. Indeed for $a/b=0$ the sum evaluates to unity and the expression agree. From Fig. \ref{phimaxratio} we see that for an aspect ratio $b/a = 4$ the expressions agree already within 10\,\% of the  
\\
\\
Finally, in case the layer is only grounded at $x=0$ and all other boundaries are insulated, the maximum potential is at $x=a$ and the results are
\beq
   \phi_1(x) = \frac{1}{2}Ri_0 (2ax-x^2) \qquad \phi_{max} = \frac{1}{2} R i_0 a^2
\eeq

\clearpage
\section{Signals and charge spread in detectors with resistive elements \label{signal_spread}}

\begin{figure}[ht]
 \begin{center}
a)
\includegraphics[width=7cm]{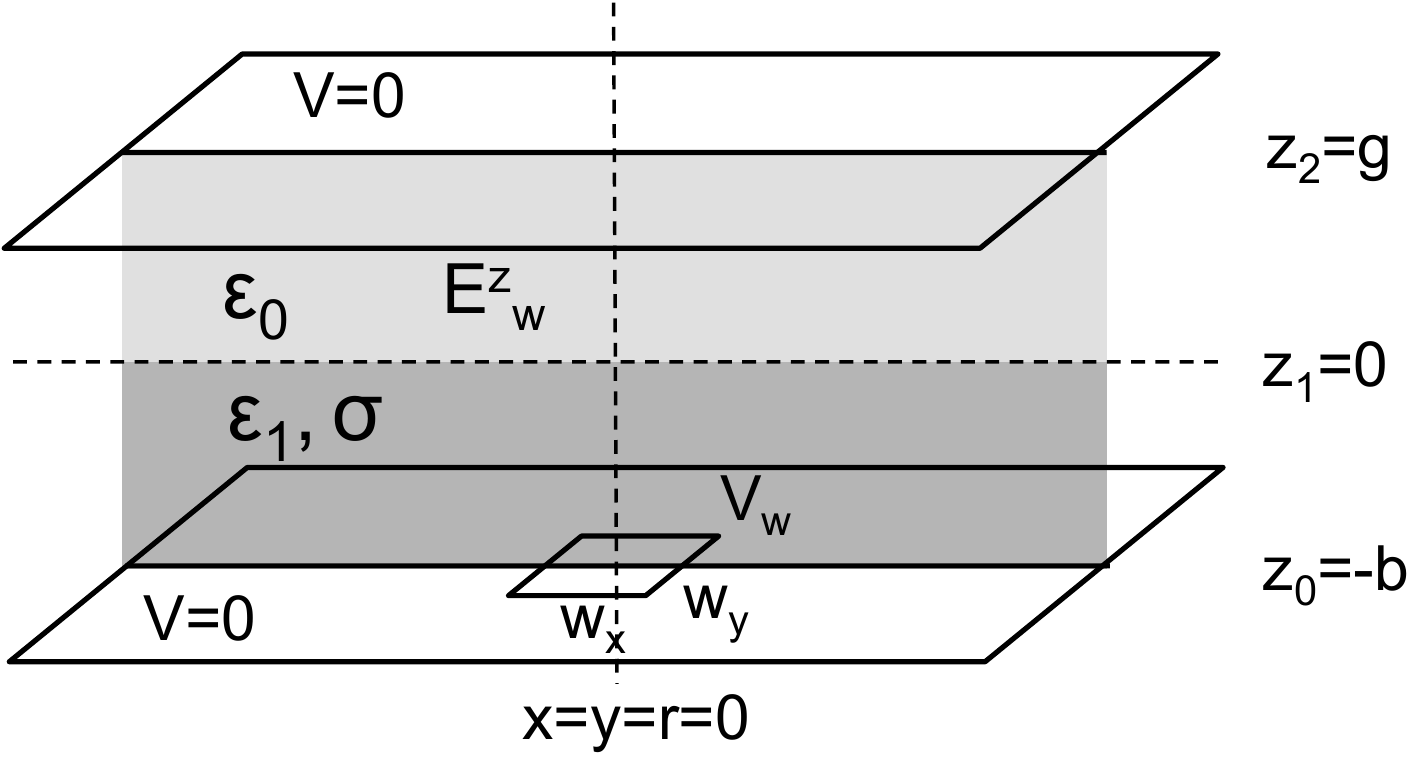}
$\qquad$
b)
\includegraphics[width=7cm]{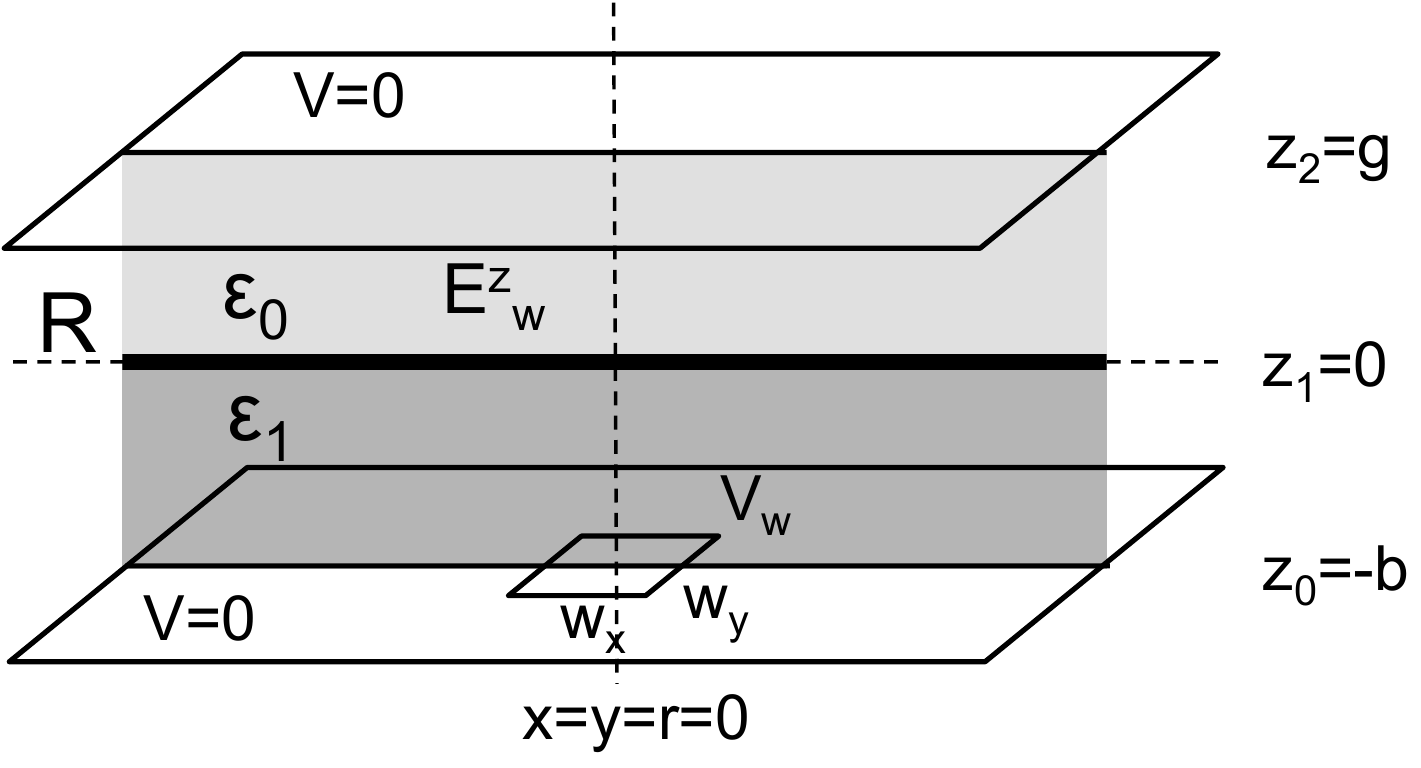}
c)
\includegraphics[width=7cm]{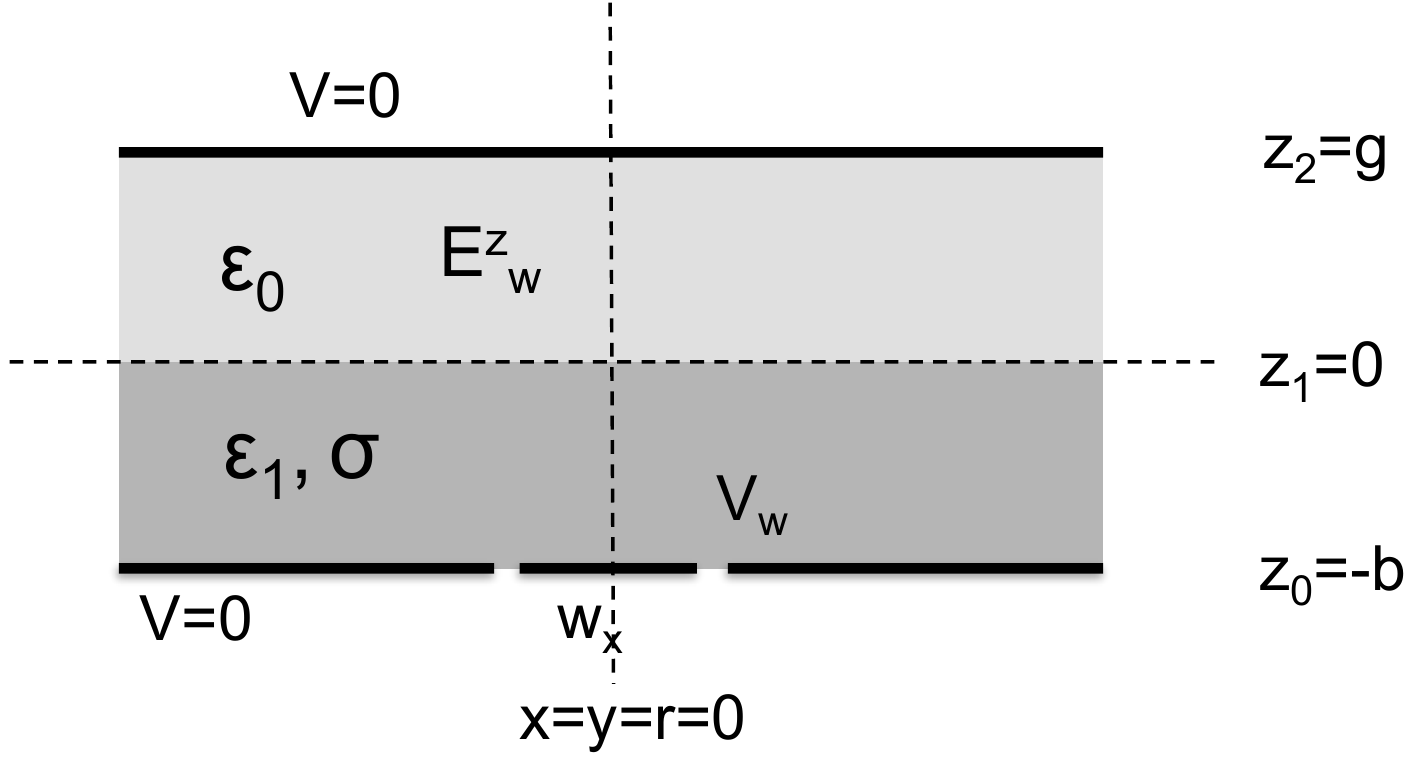}
$\qquad$
d)
\includegraphics[width=7cm]{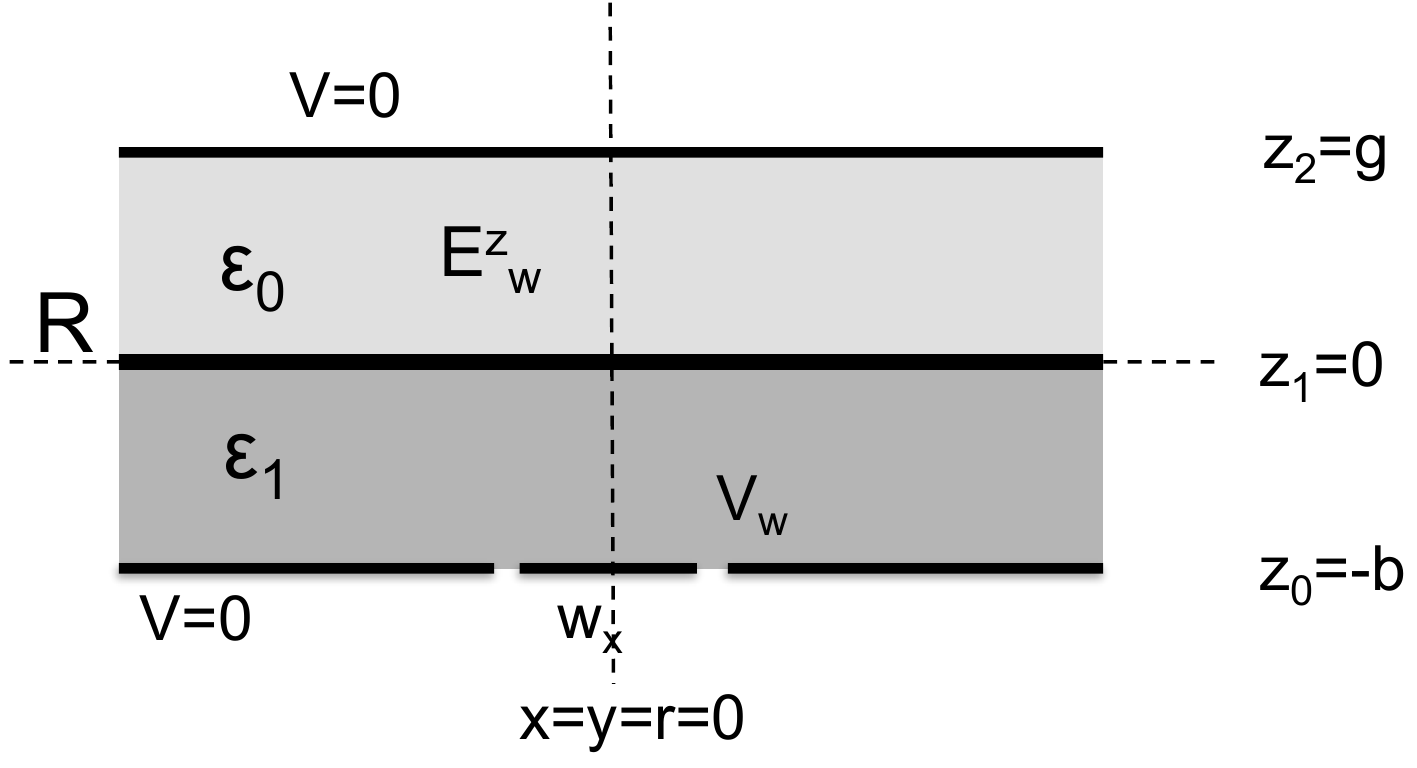}
e)
\includegraphics[width=7cm]{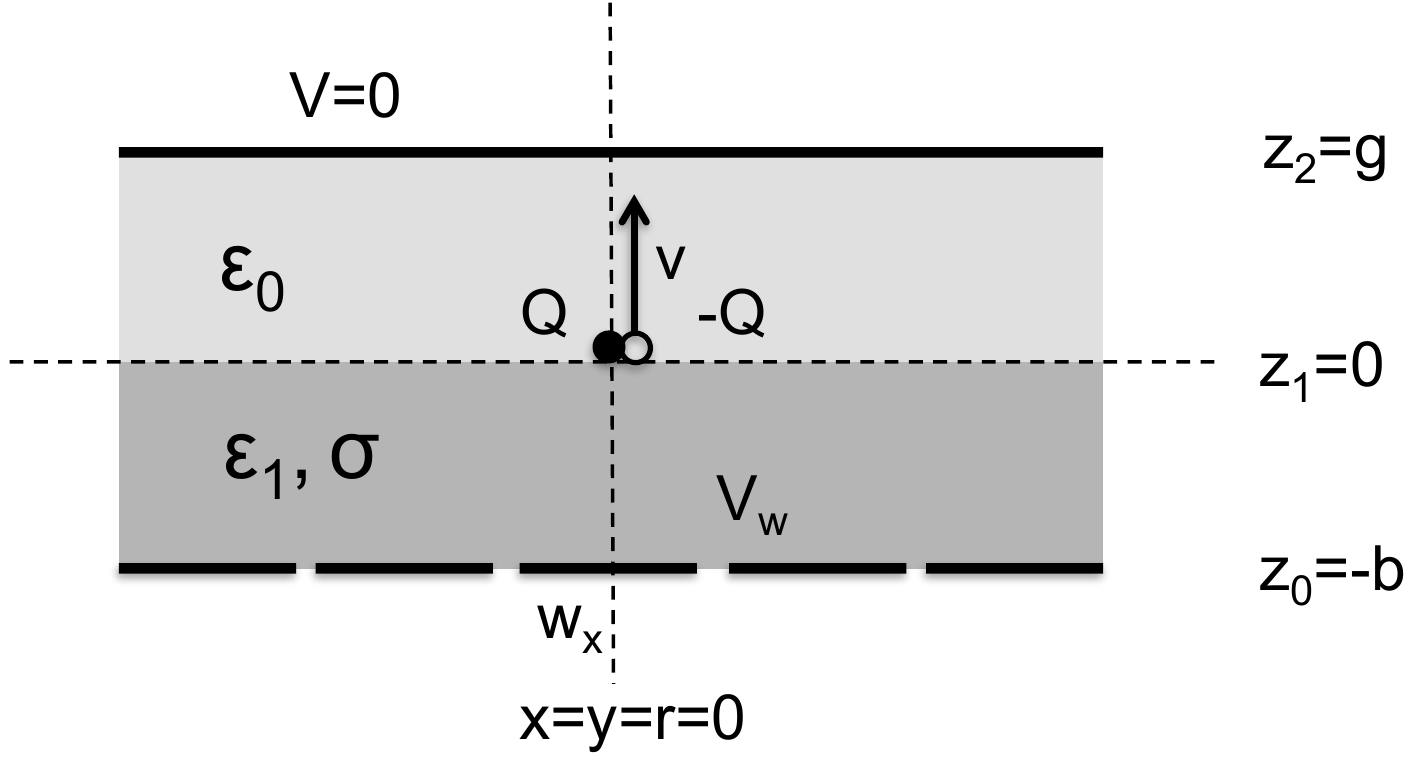}
$\qquad$
f)
\includegraphics[width=7cm]{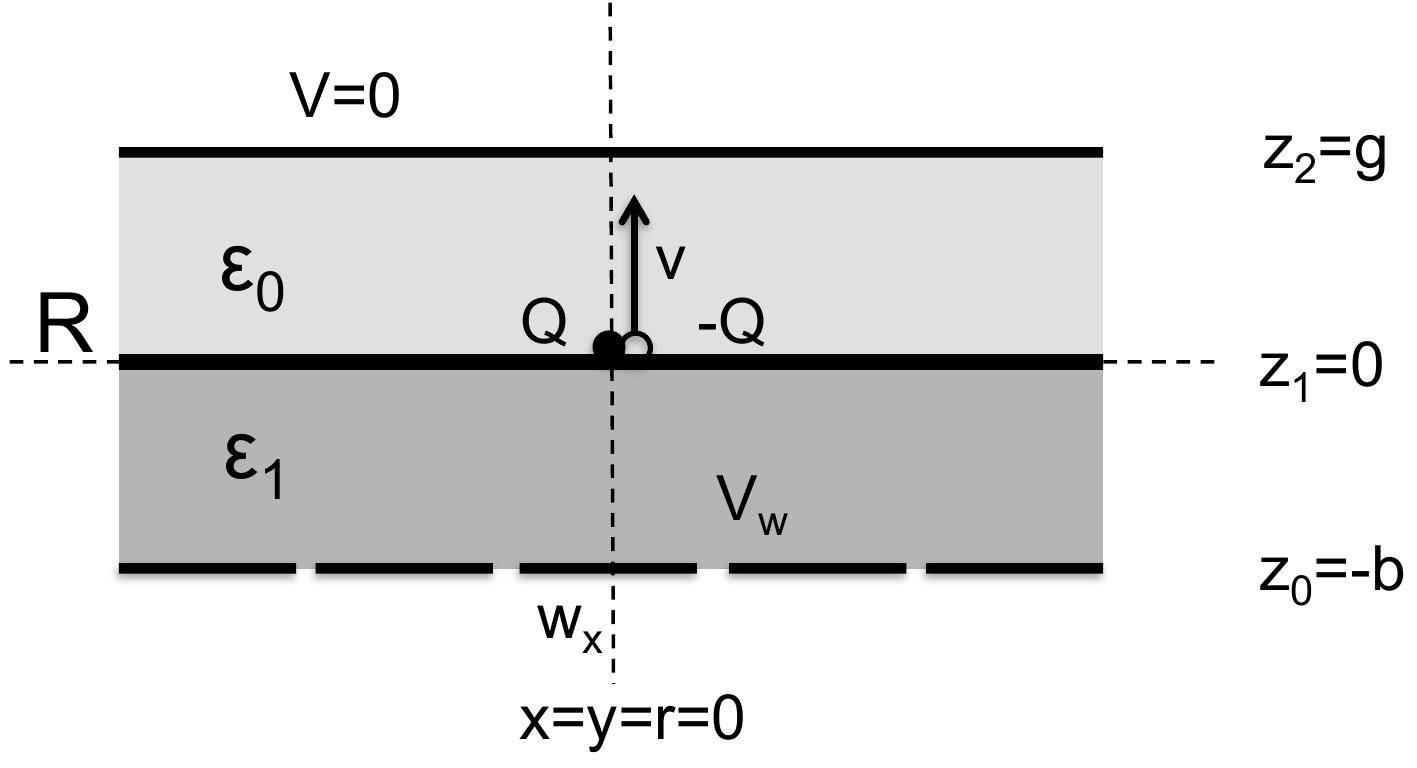}
  \caption{Weighting field for a geometry with a resistive layer having a bulk resistivity of $\rho=1/\sigma$[$\Omega$cm] (left) and a geometry with a thin resistive layer of value $R\,[\Omega$/square] (right).}
  \label{weighting_field_resistive}
  \end{center}
\end{figure}
%
%
%
In this section we finally want to calculate the signals induced on a readout pad or readout strip in presence of 
a resistive layer, either as a bulk resistive layer touching the readout structure (Fig. \ref{weighting_field_resistive}a) or as a thin resistive layer that is insulated from the readout pads (Fig. \ref{weighting_field_resistive}b). The situation could e.g. represent the geometry of a Micromegas detector with a resistive layer either for spark protection or for charge spreading. We can picture the layer $0<z<g$ as the amplification gap of such a device.
\\ \\
Following Section \ref{weighting_field_section}, the time dependent weighting fields for a pad of dimension $w_x$ and $w_y$ centred at zero and an infinitely long strip of width $w_x$ centred at zero, can be written as 
\bea
    E^z_w(x,y,z,t) =  \frac{V_w}{g} \frac{4}{\pi^2}\int_0^\infty \int_0^\infty
    \cos (k_x \frac{x}{g}) \sin (k_x \frac{w_x}{2g}) \cos (k_y \frac{y}{g}) \sin (k_y \frac{w_y}{2g})
    \frac{h(k,z,t)}{k_x k_y} dk_x dk_y
\eea
\bea
    E^z_x(x,z) = \frac{V_w}{g}\frac{2}{\pi}\int_0^\infty
    \cos (k \frac{x}{g}) \sin (k \frac{w_x}{2g}) \frac{h(k,z,t)}{k} dk
\eea
for both geometries. They are discussed in the next sections.

\subsection{Layer with bulk resistivity}

\begin{figure}[ht]
 \begin{center}
 a)
  \includegraphics[width=6cm]{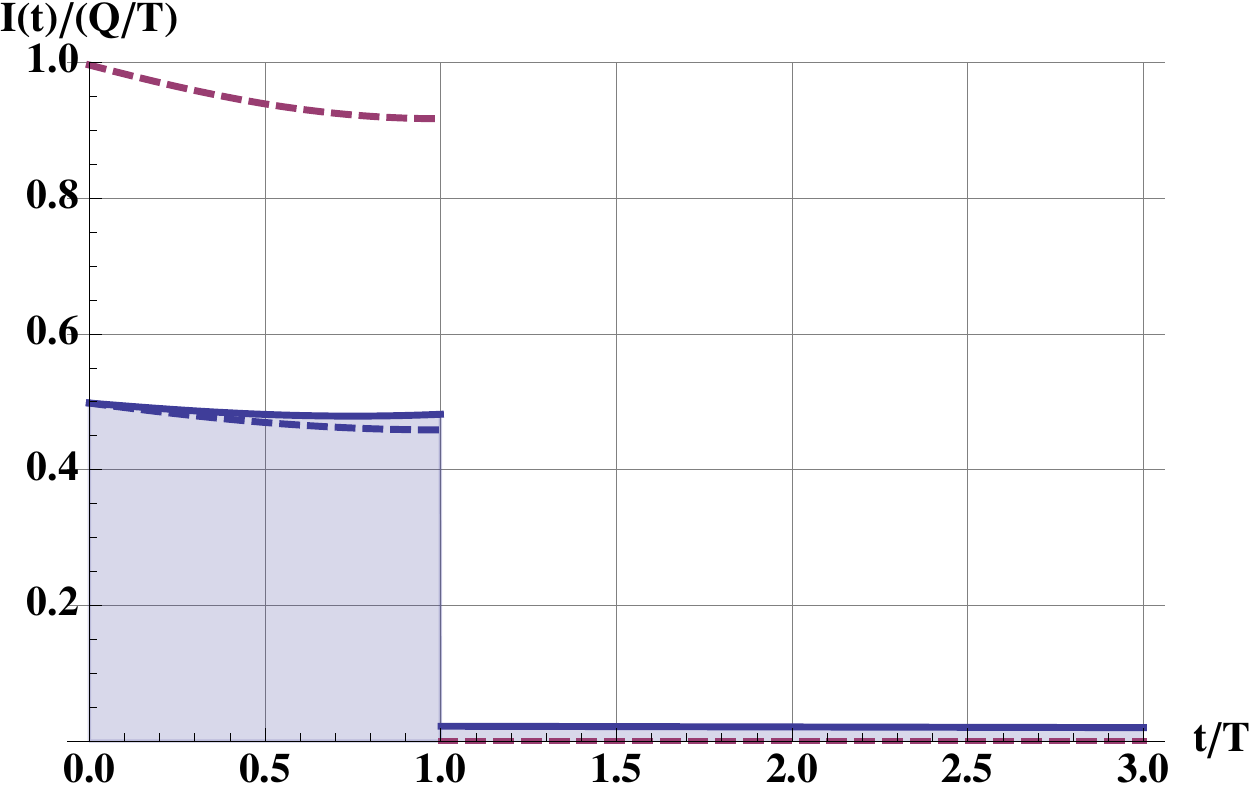}
  b)
  \includegraphics[width=6cm]{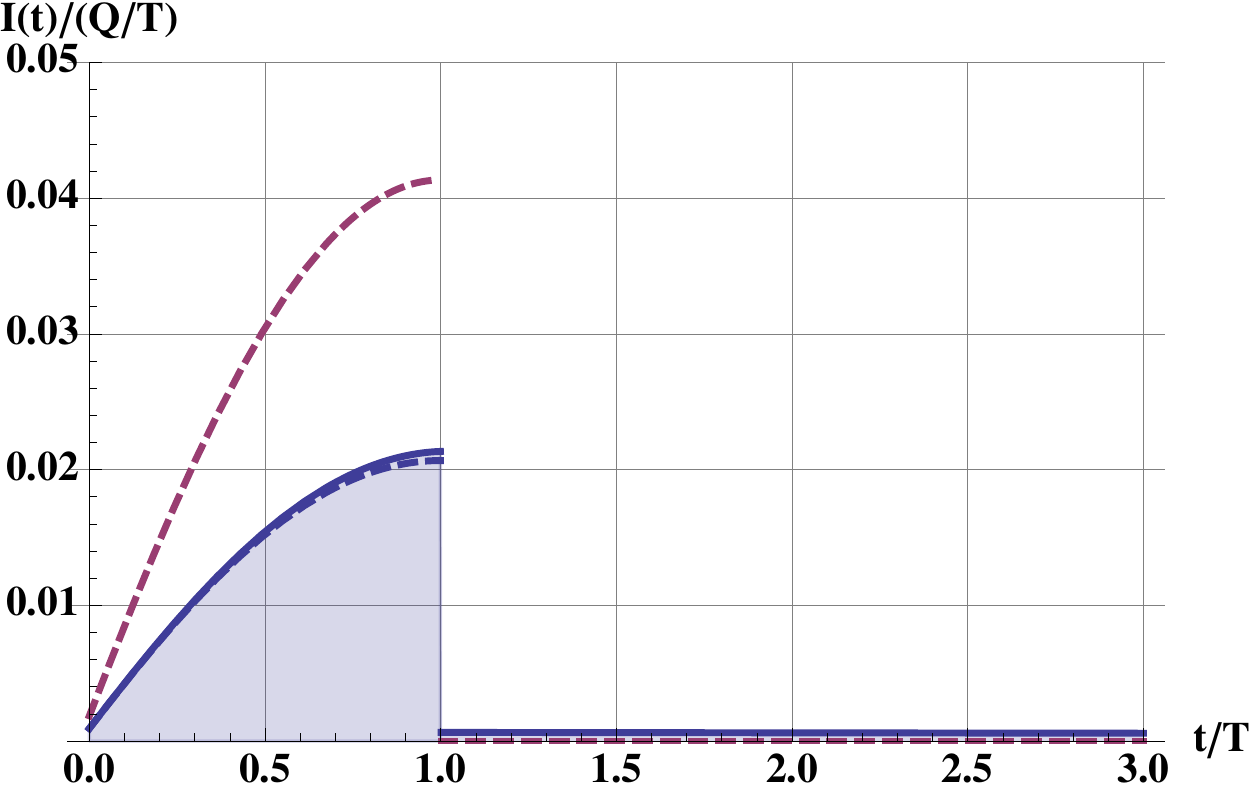}
  \caption{Uniform charge movement from $z=0$ to $z=g$, with $\vep_r=1, w_x=4g, b=g, \tau_0=10T$ for a)$x=0$ and b) $x=4g$.}
  \label{bulk_signals1}
\vspace{0.5cm}
 a)
  \includegraphics[width=6cm]{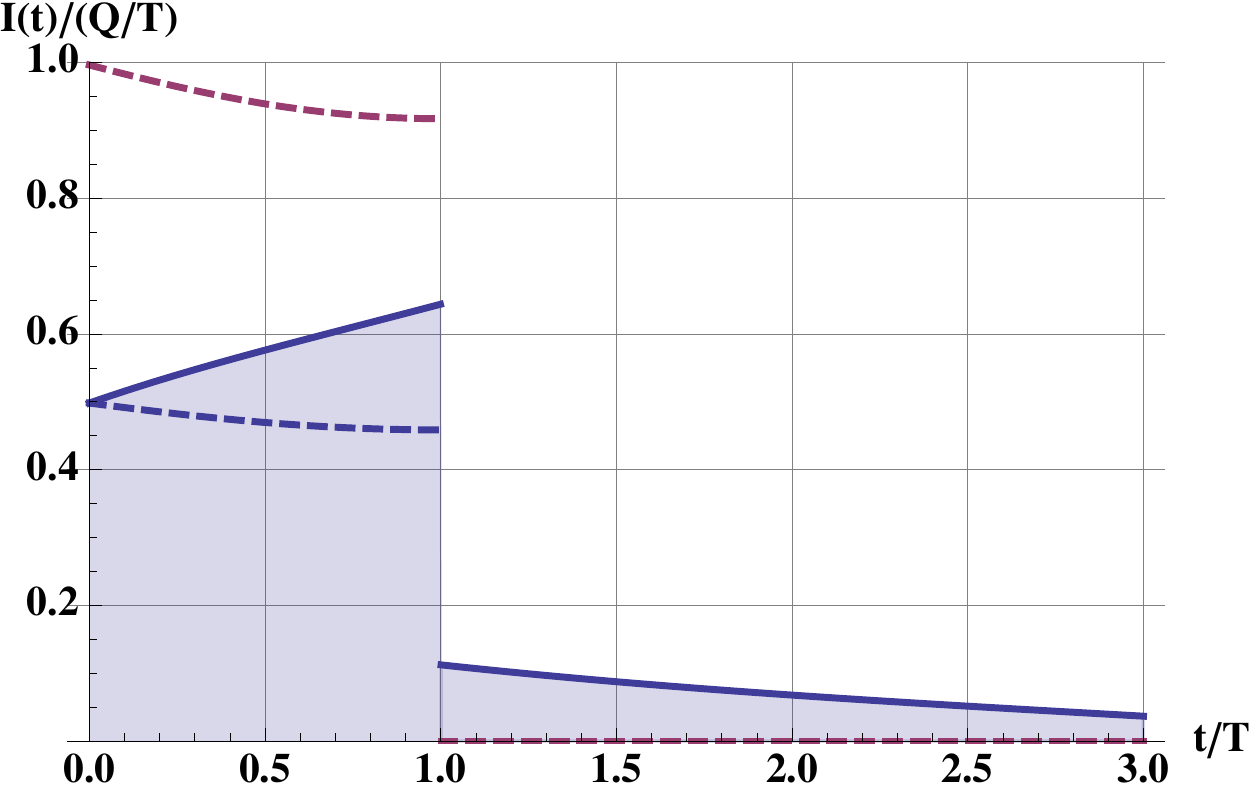}
  b)
  \includegraphics[width=6cm]{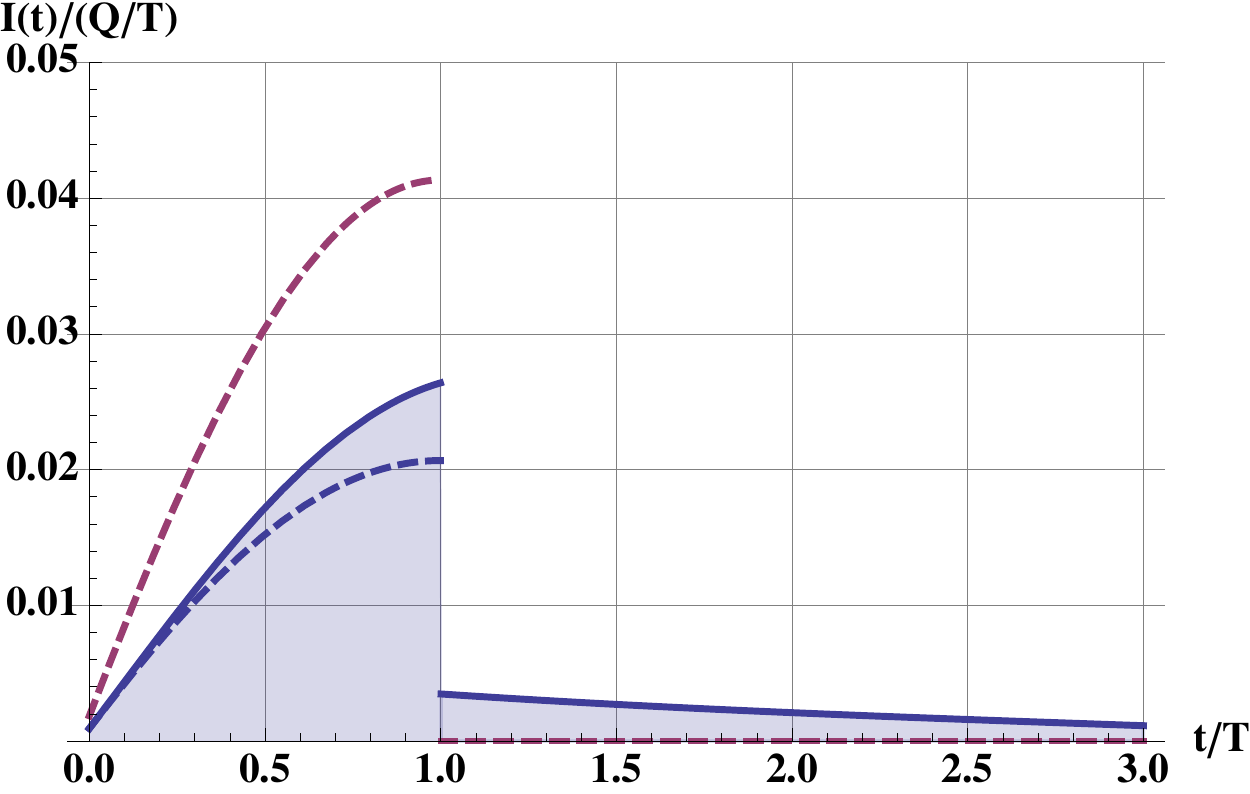}
  \caption{Uniform charge movement from $z=0$ to $z=g$, with $\vep_r=1, w_x=4g, b=g, \tau_0=T$ for a)$x=0$ and b) $x=4g$.}
  \label{bulk_signals2}
\vspace{0.5cm}
 a)
  \includegraphics[width=6cm]{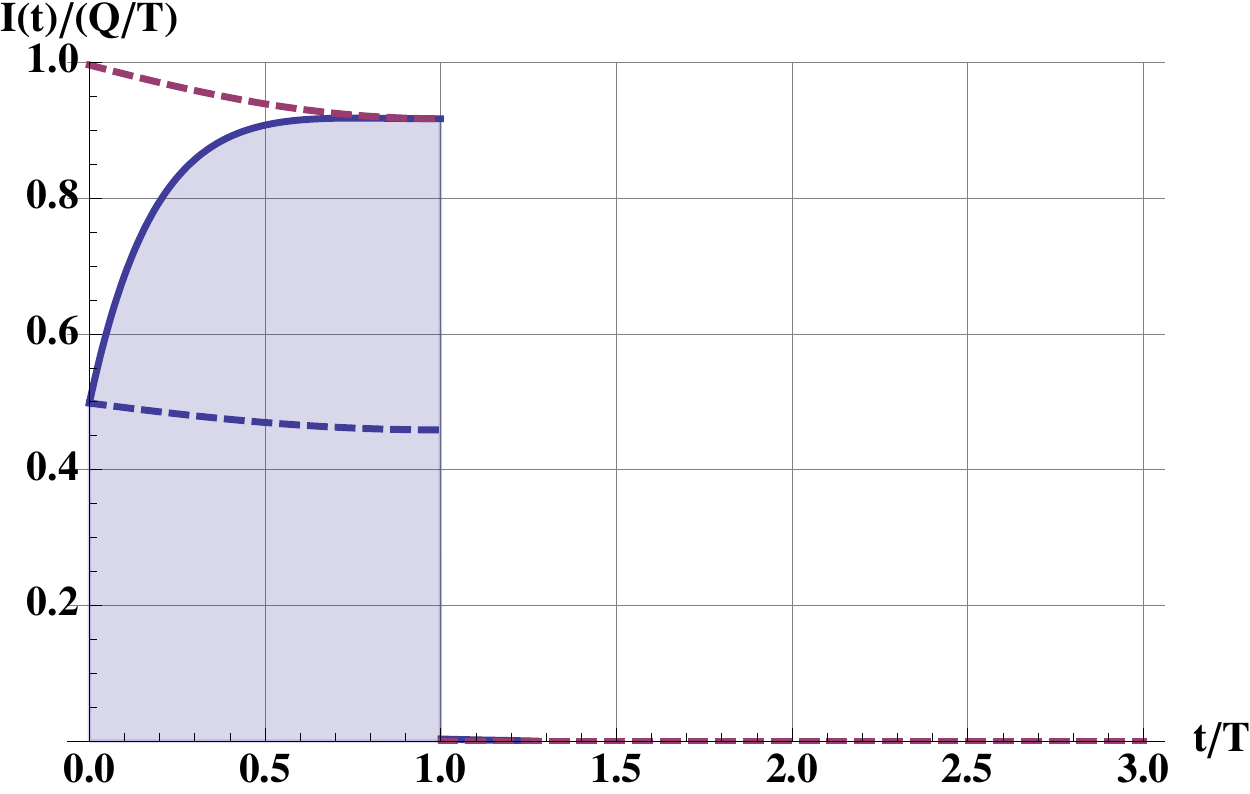}
  b)
  \includegraphics[width=6cm]{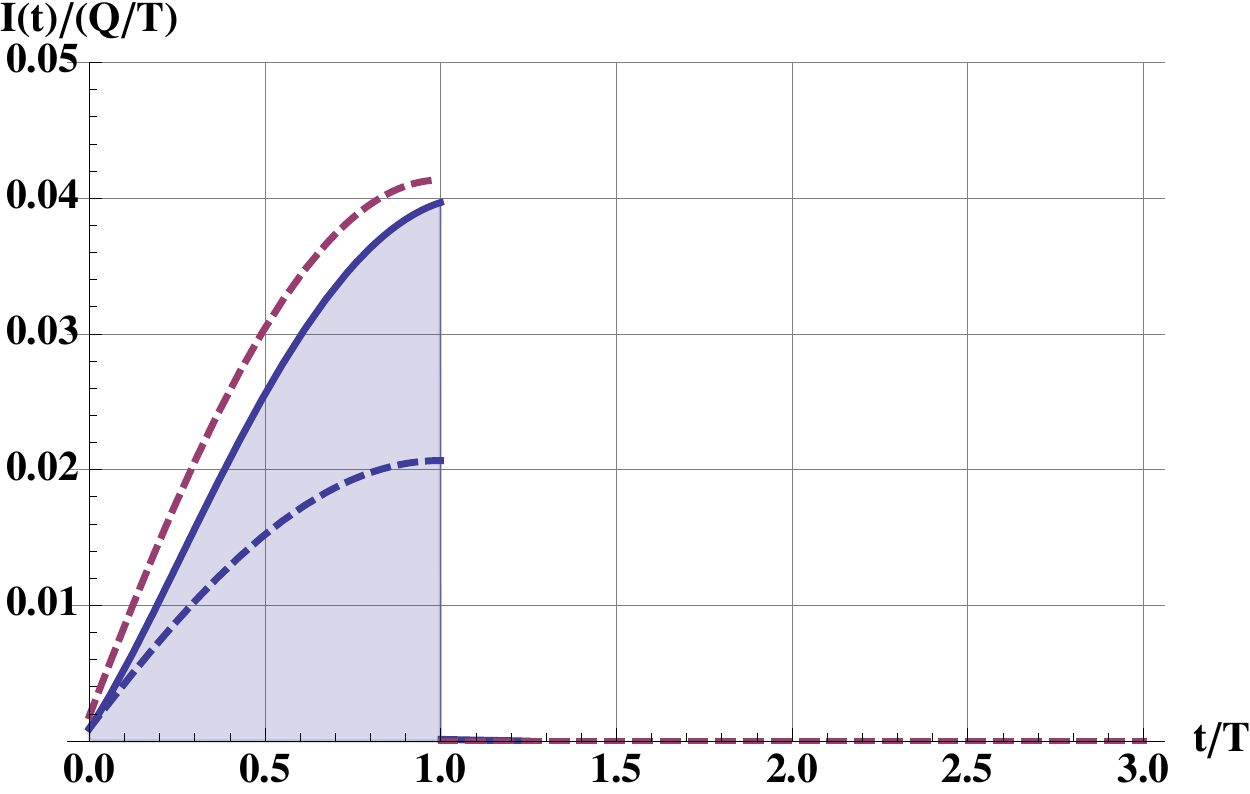}
  \caption{Uniform charge movement from $z=0$ to $z=g$, with $\vep_r=1, w_x=4g, b=g, \tau_0=0.1T$ for a)$x=0$ and b) $x=4g$.}
  \label{bulk_signals3}
  \end{center}
\end{figure}

If the layer has a bulk resistivity of $\rho=1/\sigma$ (Fig. \ref{weighting_field_resistive}a) the expression for $h(k,z)$ is $(0<z<g)$ is
\beq
  h(k,z,t) = k \cosh(k(1-\frac{z}{g}))
  \left[
  \frac{\vep_r \delta(t)}{D(k)}+\frac{1}{\tau_0}b_1(k)e^{-\frac{t}{\tau_0}f_1(k)}
  \right]
\eeq
\beq
  D(k) = \sinh(k\frac{b}{g}) \cosh(k)+\vep_r\cosh(k\frac{b}{g}) \sinh(k)
\eeq
\beq
  b_1(k) = \frac{\sinh(k\frac{b}{g}) \cosh(k)}{D(k)^2} \qquad 
  f_1(k) = \frac{\sinh(k) \cosh(k\frac{b}{g})}{D(k)}
\eeq
with $\tau_0 = \vep_0/\sigma = \vep_0 \rho$. We investigate the geometry where the ground plane at $z=-b$ is segmented into infinitely long strips of width $w_x$ (Fig. \ref{weighting_field_resistive}c). We also assume a pair of charges $Q, -Q$ produced at $t=0$ at $z=0$, the charge $Q$ does not move and the charge $-Q$ moves from $z=0$ to $z=g$ with uniform velocity $v$ i.e. $z(t)=vt=g\,t/T$, $0<t<T$, $T=g/v$ (Fig. \ref{weighting_field_resistive}e). The current is then calculated to 
\beq
   I(t)=-\frac{-Q}{V_w} \int_0^t E_w(x, z(t'),t-t')\dot z(t') dt' 
   = \frac{Q}{V_w} \int_0^t E_w(x, g\,t'/T,t-t')g/T dt'
   \qquad t<T
\eeq
\beq
   I(t)=-\frac{-Q}{V_w} \int_0^T E_w(x, z(t'),t-t')\dot z(t') dt' 
   = \frac{Q}{V_w} \int_0^T E_w(x, g\,t'/T,t-t')g/T dt'
   \qquad t>T
\eeq
This results in 
\bea
     I(t<T) & = & \frac{Q}{T} \int_0^\infty \frac{2}{\pi}      \cos (k \frac{x}{g}) \sin (k \frac{w_x}{2g}) 
      \times 
      [ 
     \frac{\vep_r \cosh(k-k\frac{t}{T})}{D(k)}  
     \\ && \no
       + b_1
     \frac{e^{-\frac{t}{\tau_0} f_1}(f_1\cosh(k)+\frac{\tau_0}{T}k\sinh (k))
     -f_1\cosh(k-k\frac{t}{T}) -k \frac{\tau_0}{T}\sinh(k-k\frac{t}{T})   }
     {k^2\frac{\tau_0^2}{T^2}-f_1^2}
     ] 
     dk
\eea
\bea
     I(t>T) =\frac{Q}{T} \int_0^\infty \frac{2}{\pi}      \cos (k \frac{x}{g}) \sin (k \frac{w_x}{2g}) 
         b_1\,e^{-\frac{t-T}{\tau_0}f_1}
     \frac{e^{-\frac{T}{\tau_0} f_1}(f_1\cosh(k)+k\frac{\tau_0}{T}\sinh(k))-f_1 }{k^2\frac{\tau_0^2}{T^2}-f_1^2}
\eea
In the limiting case of very high resistivity i.e. $\tau_0 \rightarrow \infty$ the layer represents and insulator and we find 
\beq
  \lim_{\tau_0\rightarrow \infty} I(t<T) = 
       \frac{Q}{T} \int_0^\infty \frac{2}{\pi}      \cos (k \frac{x}{g}) \sin (k \frac{w_x}{2g})  
       \frac{\vep_r\cosh(k-k\frac{t}{T})}{D(k)} dk
       \qquad
         \lim_{\tau_0\rightarrow \infty} I(t>T) = 0
\eeq
For the case where the layer represents a perfect conductor the expression becomes
\beq
  \lim_{\tau_0\rightarrow 0} I(t<T) = 
       \frac{Q}{T} \int_0^\infty \frac{2}{\pi}      \cos (k \frac{x}{g}) \sin (k \frac{w_x}{2g})  
       \frac{\cosh(k-k\frac{t}{T})}{\sinh(k)\cosh(k\frac{b}{g})} dk
       \qquad
        \lim_{\tau_0\rightarrow 0} I(t>T) = 0
\eeq
This last expression is correct if the strips are truly grounded. For any realistic setup where the strips are connected to readout electronics and therefore have a finite resistance to ground, the signal will spread to all the strips since the strips together with the bulk behave as one single node. The result is therefore correct only to levels of conductivity $\sigma$ where the impedance between the strips is significantly larger than the input resistance of the amplifier. If this is not the case, the impedance matrix of the strips has to be calculated and current signal $I(t)$ has to be placed on the full network to evaluate the signals \cite{riegler2}. Figures \ref{bulk_signals1}, \ref{bulk_signals2}, \ref{bulk_signals3} show the induced current signals given above on a central strip of width $w_x=4g$ and the first neighbouring strip centred at $x=4g$ for different values of conductivity, i.e. for different time constants $\tau_0$. The figures show in dashed lines also the limiting cases for very large and very small values of $\tau_0$.
\\ \\
First we observe that all signals are unipolar, which is due to the fact that the charge that is flowing in the resistive bulk layer in order to compensate the charge $-Q$ sitting on the surface of the resistive plate, is truly coming out of the readout strips. In case the time $T$ of charge movement is equal to the time constant $\tau_0$ (Fig. \ref{bulk_signals2}), the signal is significantly affected and develops a long tail for $t>T$ due to the flow of charge compensating the point charge on the surface. The smaller the conductivity, the longer (but smaller) is the tail of the signal as shown in Fig. \ref{bulk_signals1}  for $\tau_0=10\,T$. For short time constants of the resistive layer the signal on the central strip is large and has a short tail, and the crosstalk to the neighbour strips increases as shown in Fig. \ref{bulk_signals3} for $\tau_0=0.1T$.
For completeness we give the results for the case where the pair of charges $Q, -Q$ is created at $z=g$ and the charge $Q$ is moving from $z=g$ to $z=0$ with uniform velocity during a time $T$. 
\bea
     I(t<T)  = \frac{Q}{T} \int_0^\infty \frac{2}{\pi}      \cos (k \frac{x}{g}) \sin (k \frac{w_x}{2g}) 
      \left[
     \frac{\vep_r \cosh(k\frac{t}{T})}{D(k)} + b_1
     \frac{f_1e^{-\frac{t}{\tau_0} f_1}-f_1\cosh(k\frac{t}{T}) +k \frac{\tau_0}{T}
     \sinh(k\frac{t}{T})   }{k^2\frac{\tau_0^2}{T^2}-f_1^2}
     \right] dk
\eea
\bea
     I(t>T)  =  \frac{Q}{T} \int_0^\infty \frac{2}{\pi}    \cos (k \frac{x}{g}) \sin (k \frac{w_x}{2g}) 
        b_1\,e^{-\frac{t-T}{\tau_0}f_1}
     \frac{f_1e^{-\frac{T}{\tau_0} f_1}-f_1\cosh(k) +k \frac{\tau_0}{T}
     \sinh(k)   }{k^2\frac{\tau_0^2}{T^2}-f_1^2}
    dk
\eea
In the limiting case of very high resistivity i.e. $\tau_0 \rightarrow \infty$ the layer represents an insulator and we find 
\beq
  \lim_{\tau_0\rightarrow \infty} I(t<T) = 
       \frac{Q}{T} \int_0^\infty \frac{2}{\pi}      \cos (k \frac{x}{g}) \sin (k \frac{w_x}{2g})  
       \frac{\vep_r\cosh(k\frac{t}{T})}{D(k)} dk
       \qquad
        \lim_{\tau_0\rightarrow \infty} I(t>T) = 0
\eeq
In the limiting case where the resistivity is zero i.e. $\tau_0 \rightarrow 0$ the layer represents a perfect conductor we have
\beq
  \lim_{\tau_0\rightarrow 0} I(t<T) = 
       \frac{Q}{T} \int_0^\infty \frac{2}{\pi}      \cos (k \frac{x}{g}) \sin (k \frac{w_x}{2g})  
       \frac{\cosh(k\frac{t}{T})}{\sinh(k)\cosh(k\frac{b}{g})} dk
       \qquad
       \lim_{\tau_0\rightarrow 0} I(t>T) = 0
\eeq
%
%
%

\subsection{Layer with surface resistivity}

\begin{figure}[ht]
 \begin{center}
  \includegraphics[width=5cm]{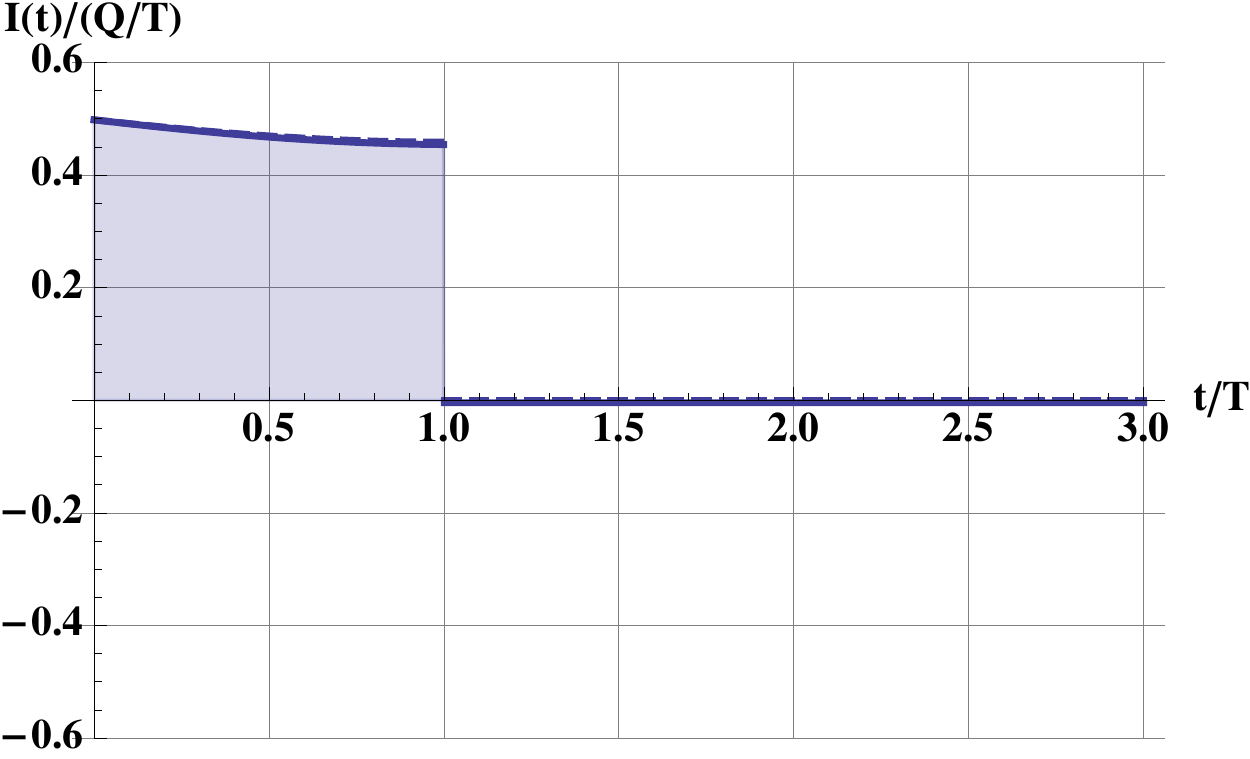}
  \includegraphics[width=5cm]{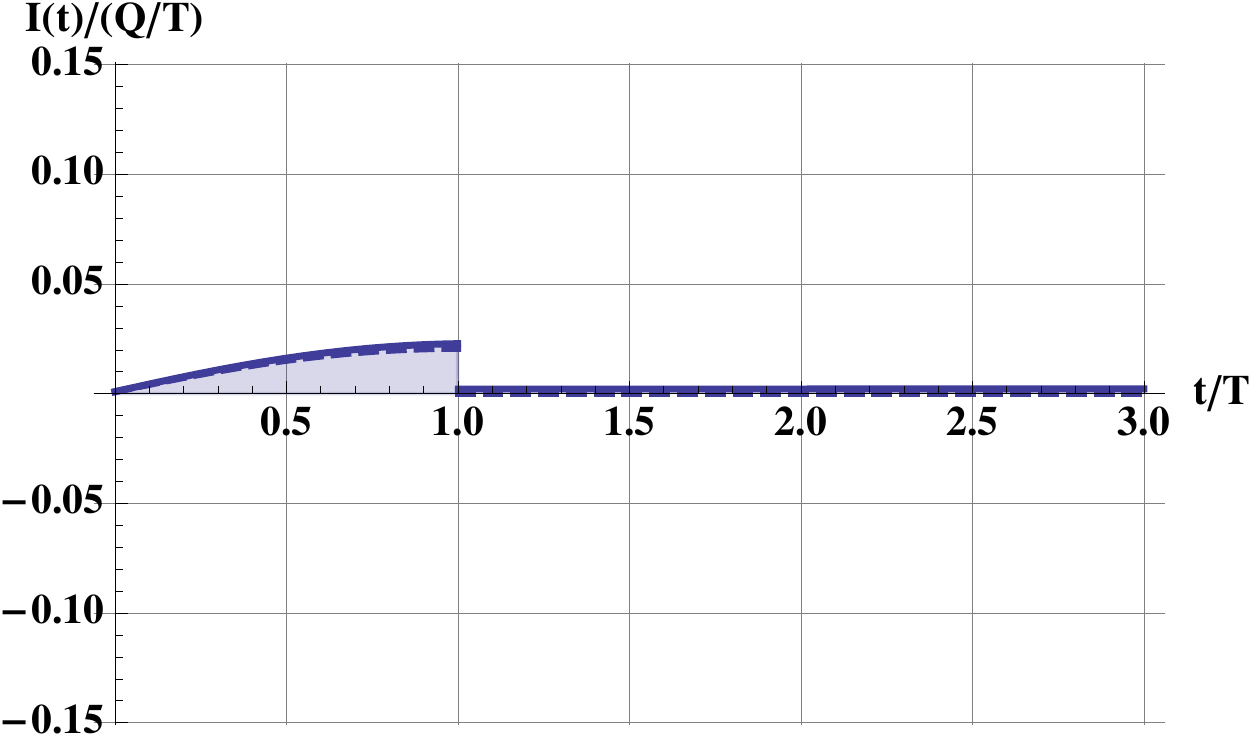}
  \includegraphics[width=5cm]{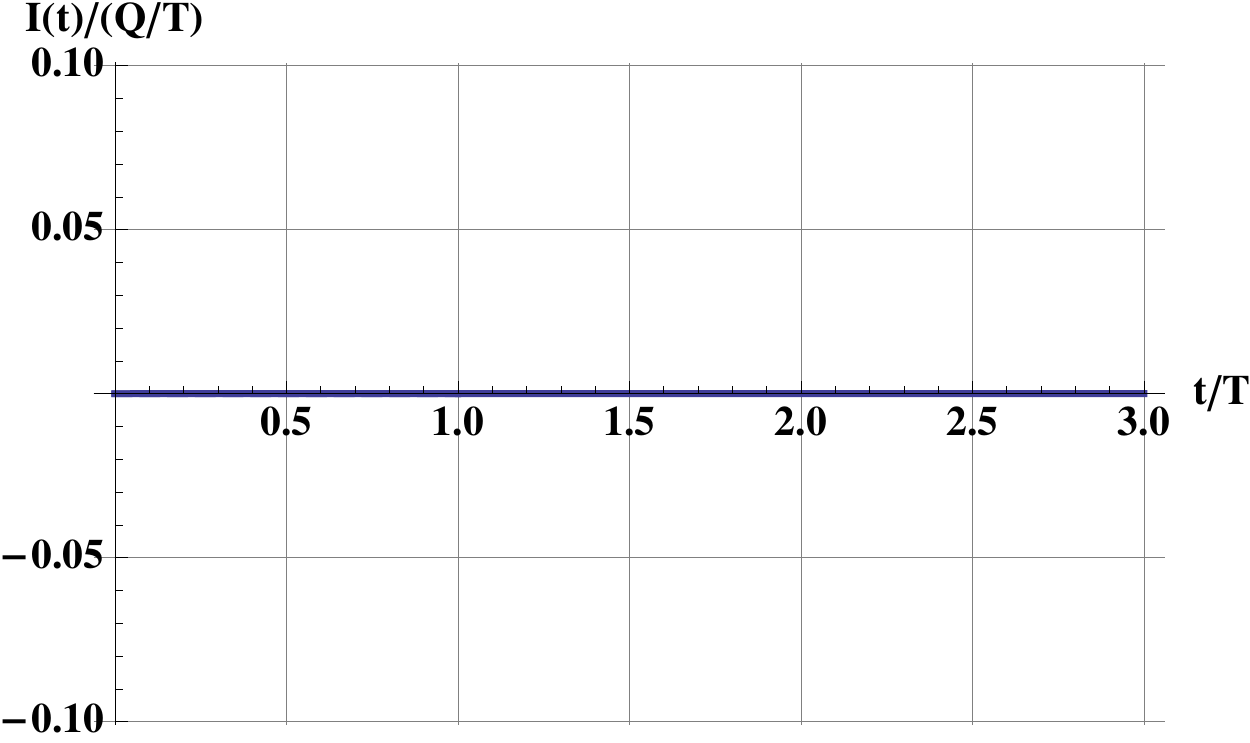}
  \caption{$\vep_r=1, w_x=4g, b=g, T_0=10T$ for $x=0, x=4g,x=8g$}
  \label{surface_signals1}
  \vspace{0.5cm}
  \includegraphics[width=5cm]{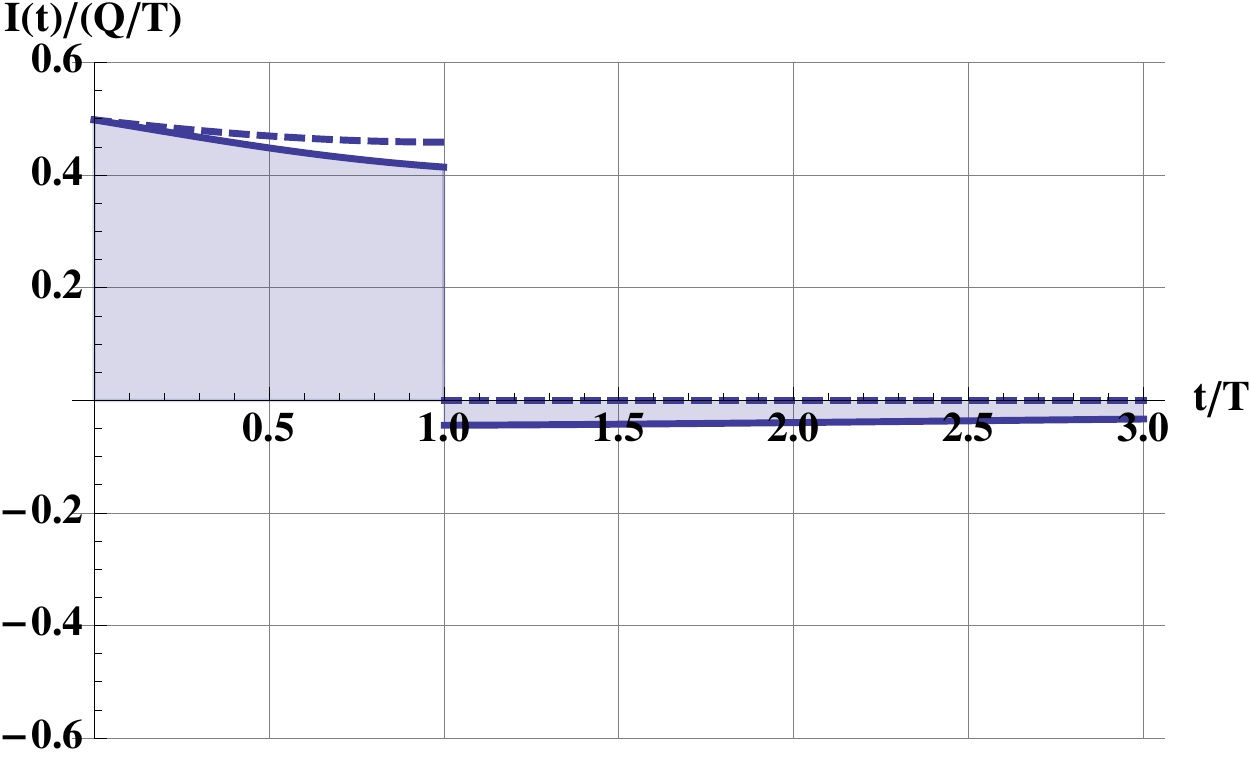}
  \includegraphics[width=5cm]{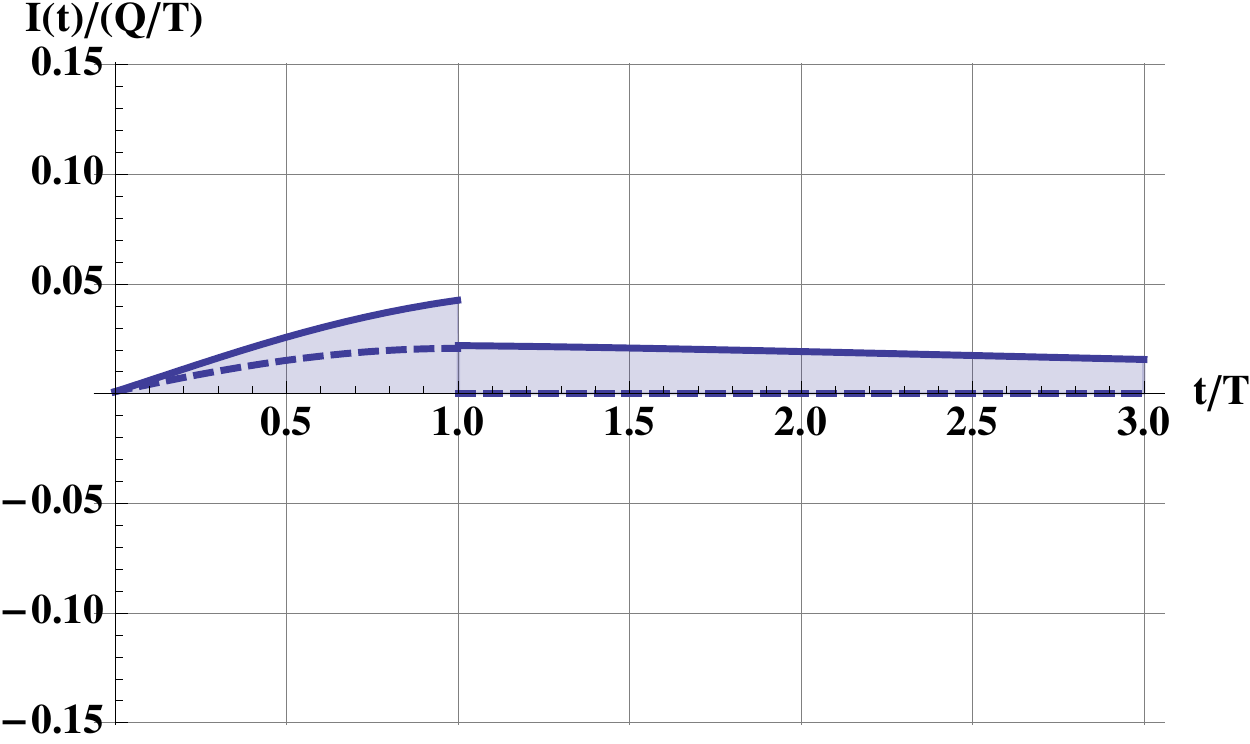}
  \includegraphics[width=5cm]{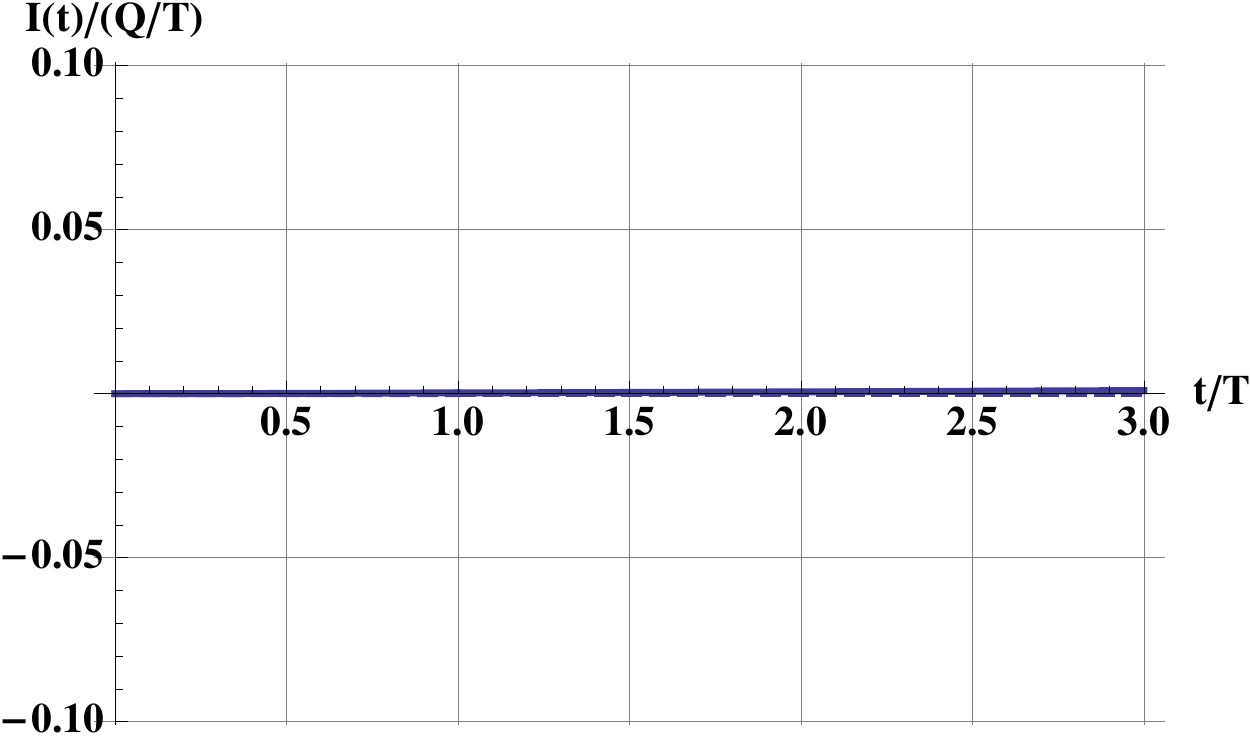}
  \caption{$\vep_r=1, w_x=4g, b=g, T_0=T$ for $x=0, x=4g, x=8g$}
  \label{surface_signals2}
   \vspace{0.5cm}
  \includegraphics[width=5cm]{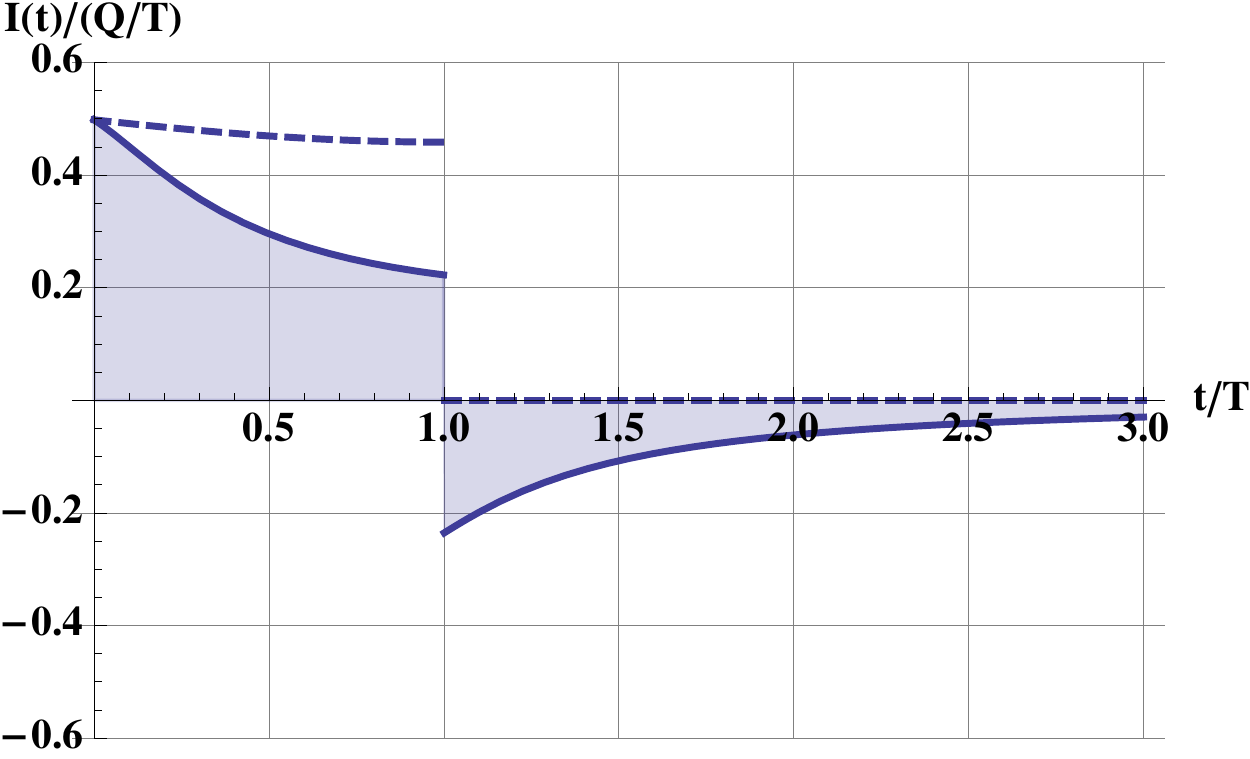}
  \includegraphics[width=5cm]{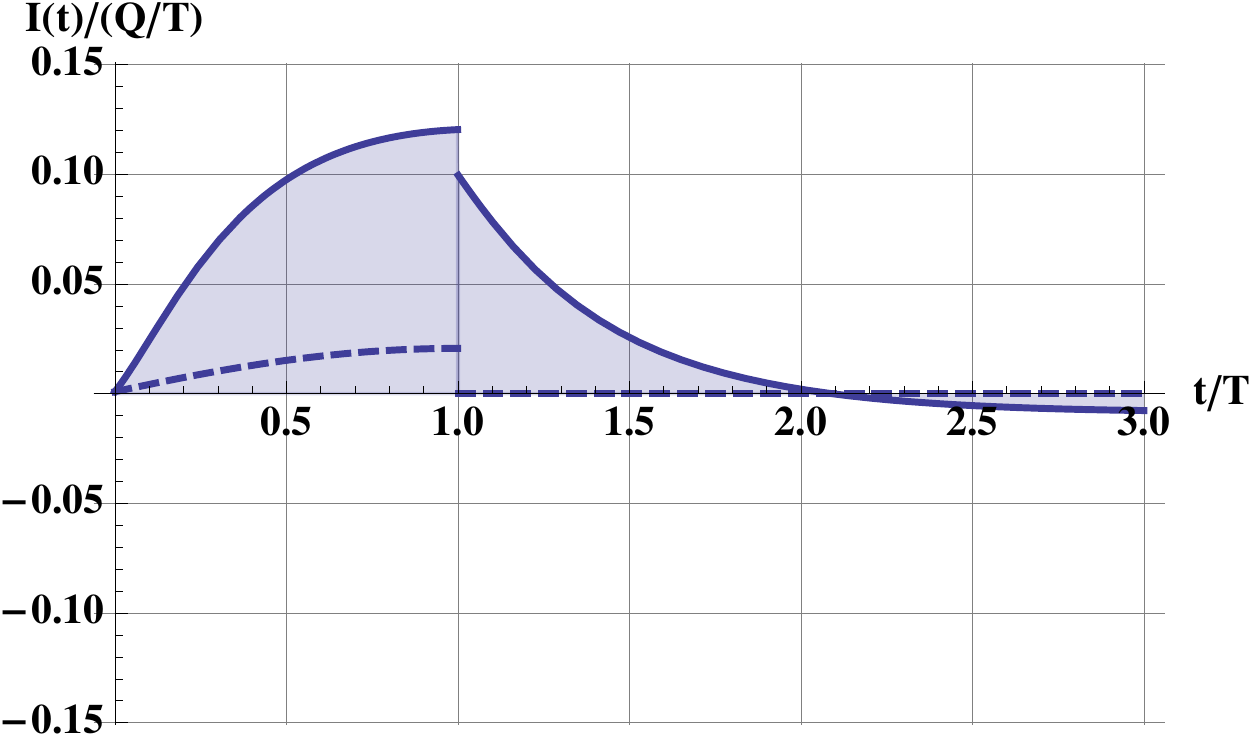}
  \includegraphics[width=5cm]{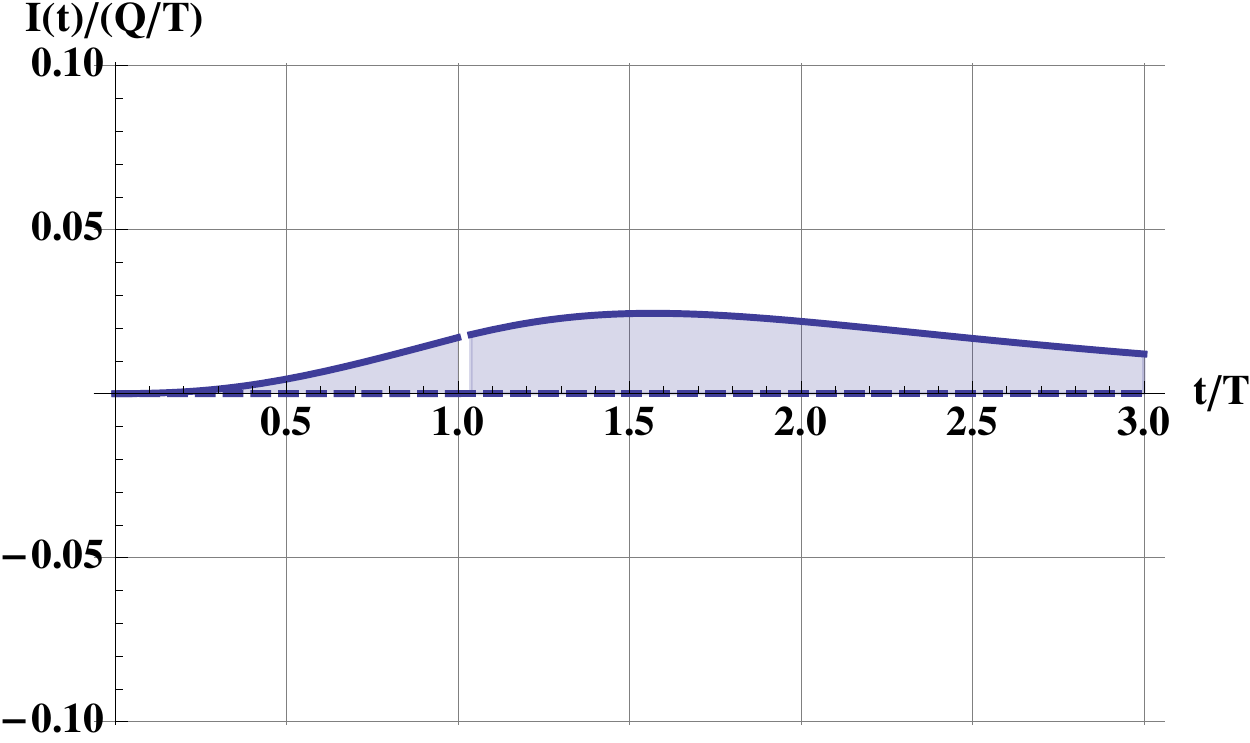}
  \caption{$\vep_r=1, w_x=4g, b=g, T_0=0.1T$ for $x=0, x=4g, x=8g$}
  \label{surface_signals3}
   \vspace{0.5cm}
  \includegraphics[width=5cm]{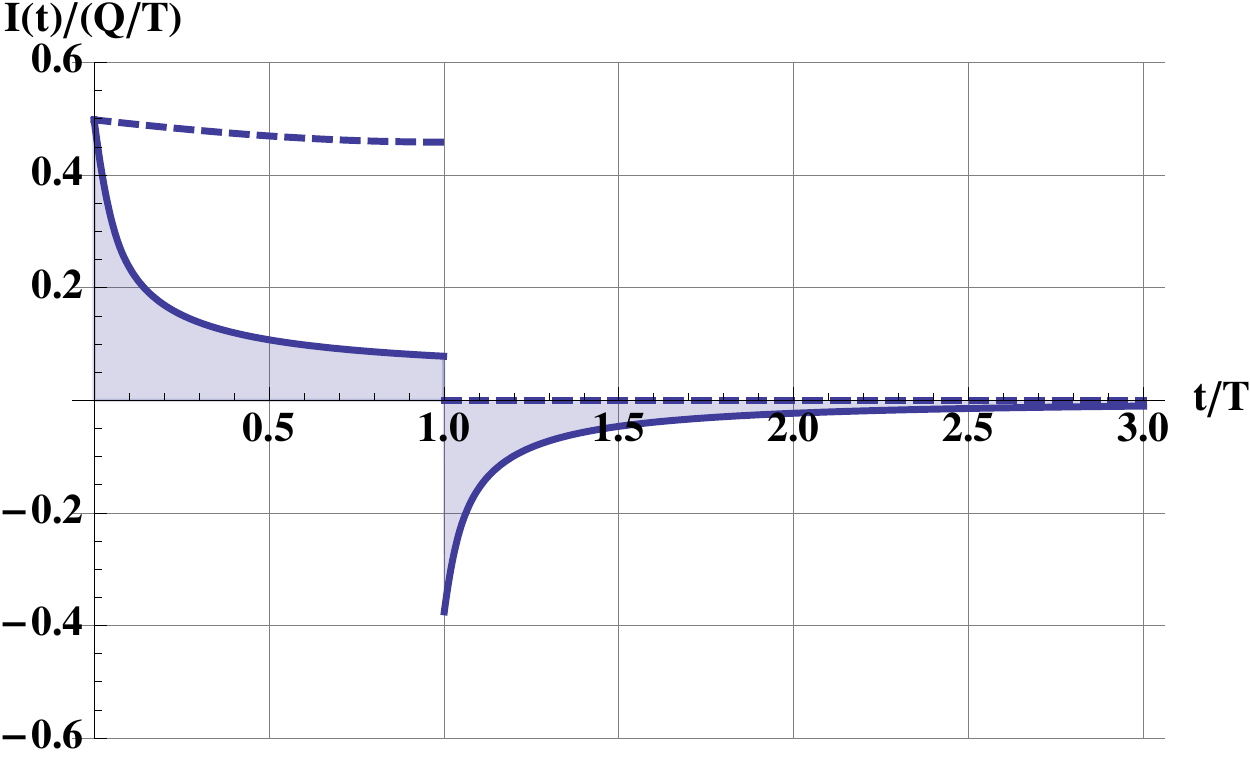}
  \includegraphics[width=5cm]{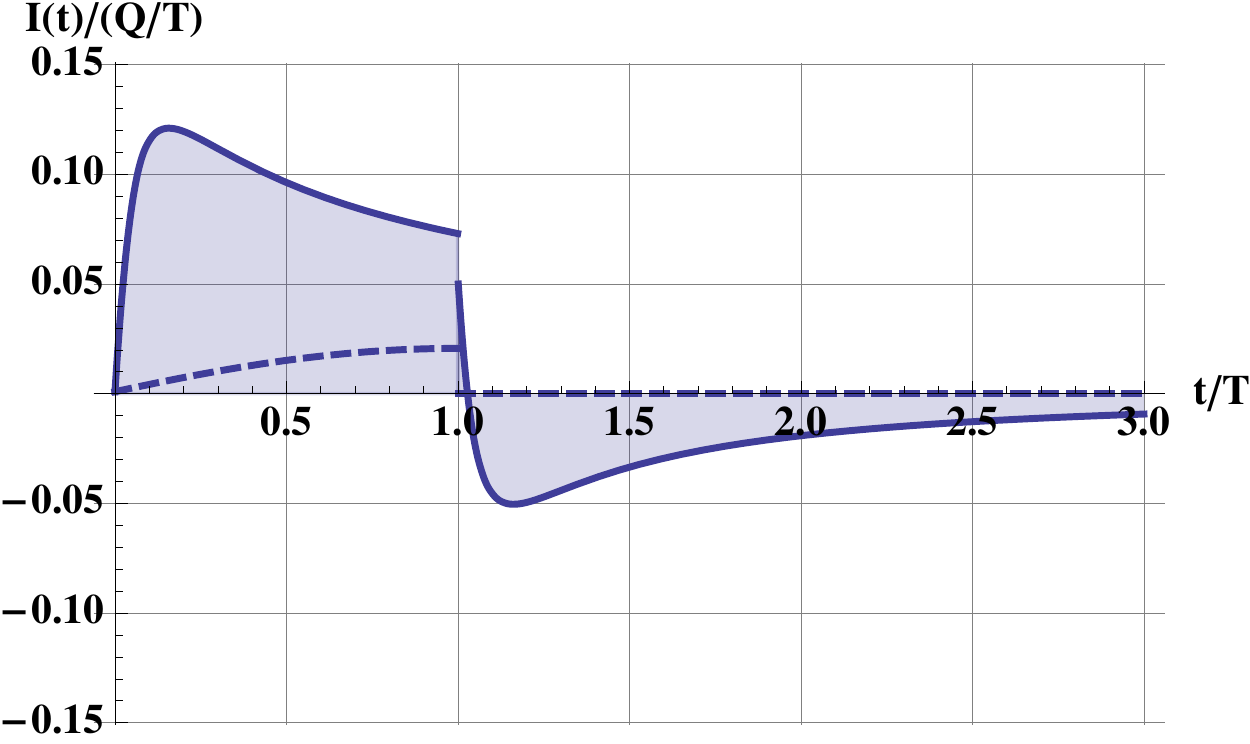}
  \includegraphics[width=5cm]{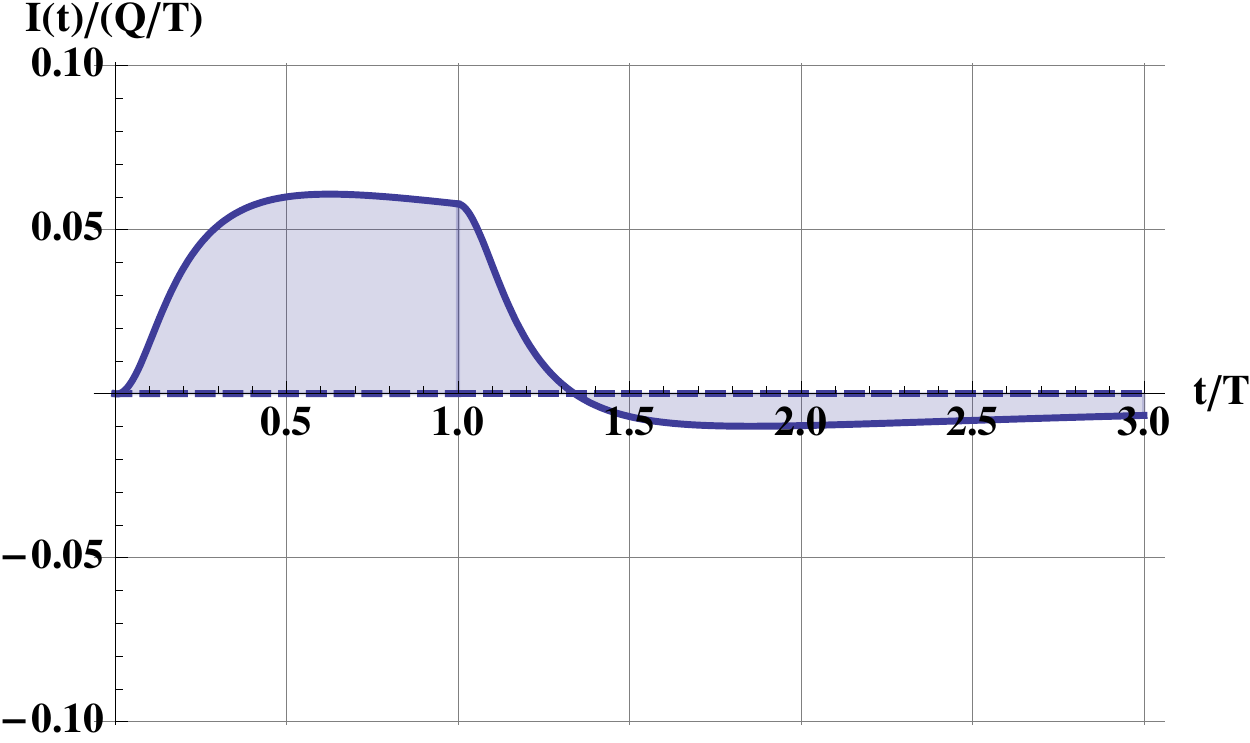}
  \caption{$\vep_r=1, w_x=4g, b=g, T_0=0.01T$ for $x=0, x=4g, x=8g$}
  \label{surface_signals4}
   \vspace{0.5cm}
  \includegraphics[width=5cm]{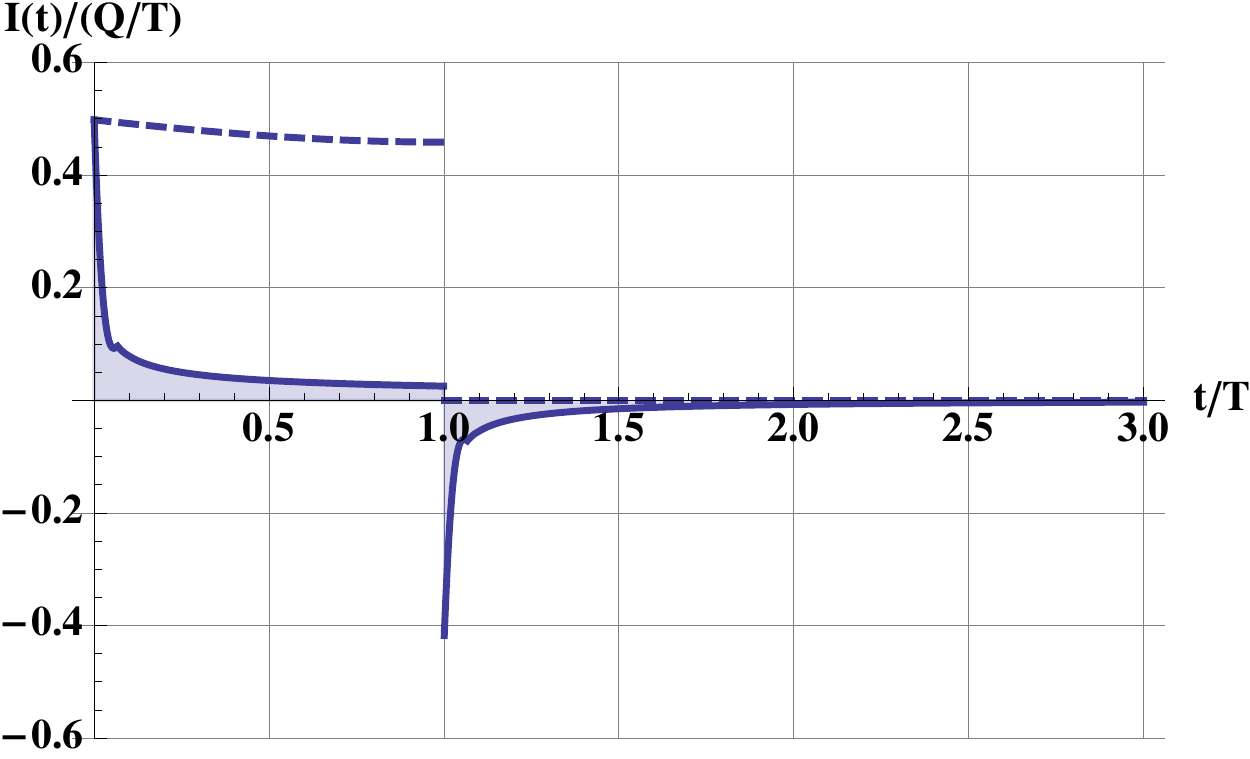}
  \includegraphics[width=5cm]{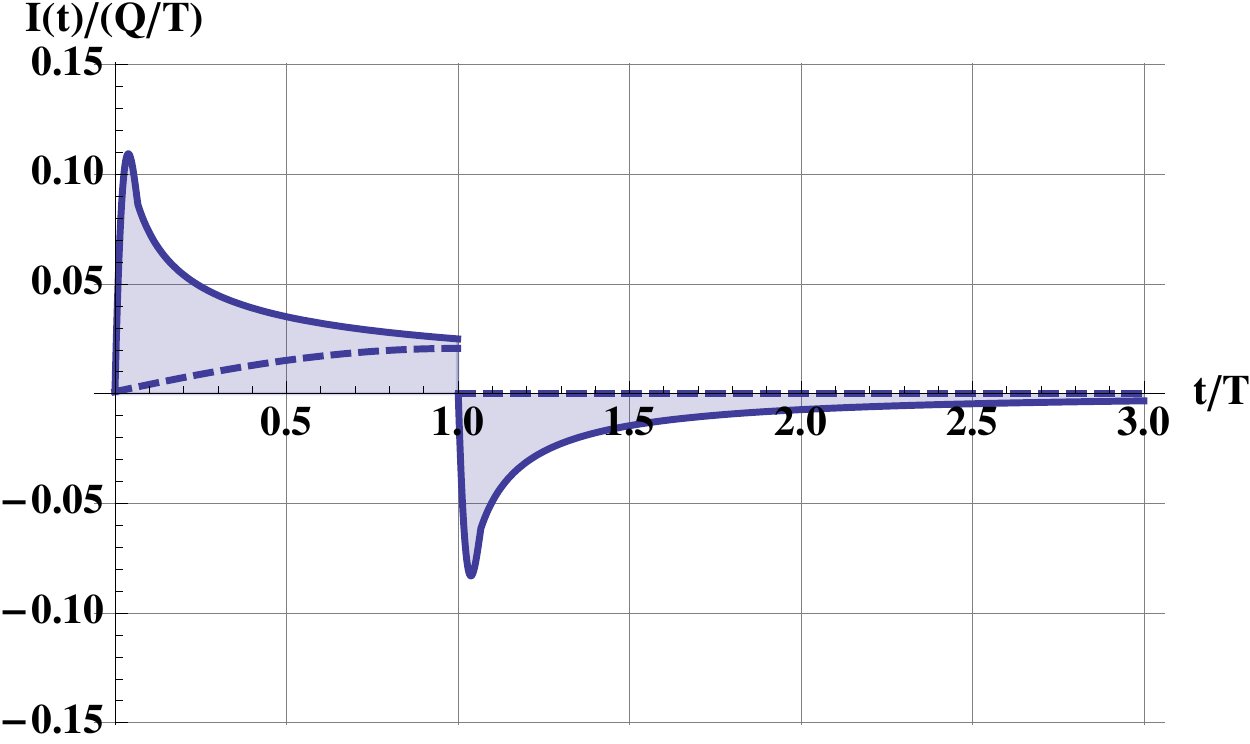}
  \includegraphics[width=5cm]{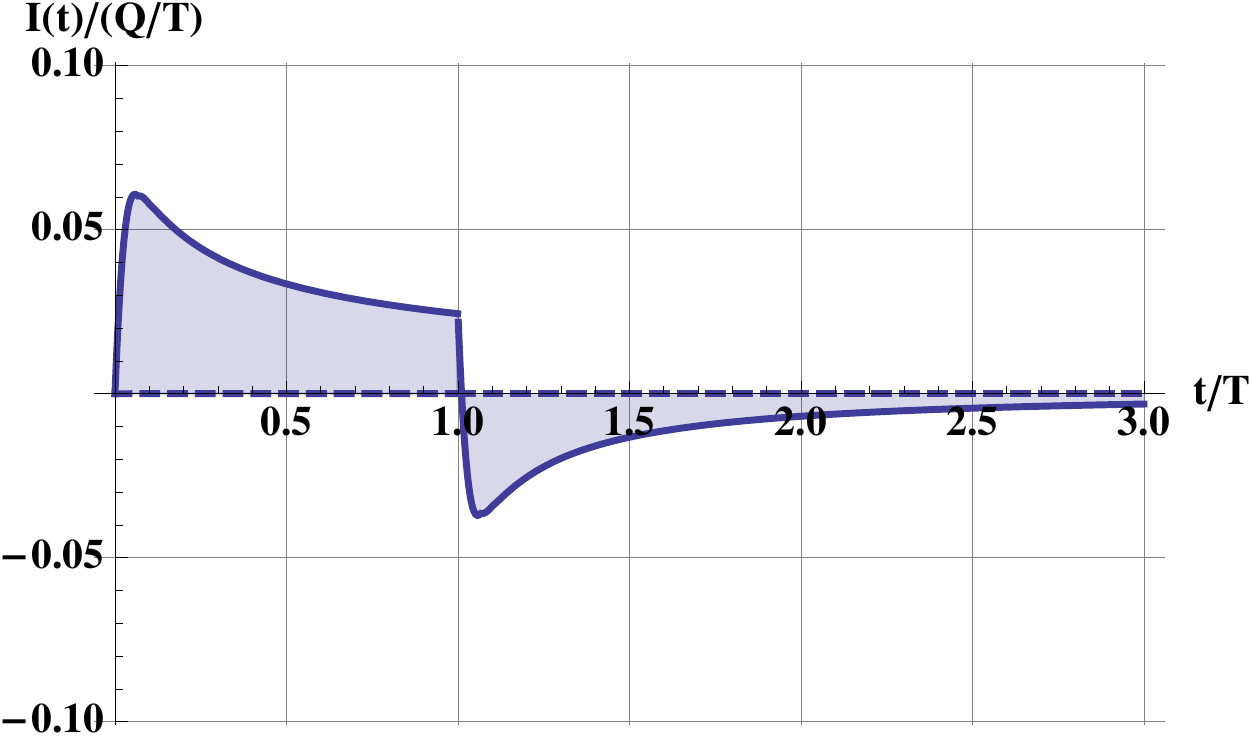}
  \caption{$\vep_r=1, w_x=4g, b=g, T_0=0.001T$ for $x=0, x=4g, x=8g$}
  \label{surface_signals5}
   \vspace{0.5cm}
  \end{center}
\end{figure}

We now turn to the example where there is only a thin layer of surface resistivity $R$ on top of an insulating layer (Fig.\ref{weighting_field_resistive}b,d,f). The expression for $h(k,z)$ is $(0<z<g)$ is
\beq \label{h_surface_resisivity}
  h(k,z,t) = k \cosh(k(1-\frac{z}{g}))
  \left(
  \frac{\vep_r \delta(t)}{D(k)}-\frac{1}{T_0}b_2(k)e^{-\frac{t}{T_0}f_2(k)}
  \right)
\eeq
\beq
  b_2(k) = k\,\frac{\vep_r\sinh(k\frac{b}{g}) \sinh(k)}{D(k)^2} \qquad 
  f_2(k) = k\,\frac{\sinh(k) \sinh(k\frac{b}{g})}{D(k)}
\eeq
where $T_0=\vep_0Rg$ is the 'time constant associated with the resistive layer' in the given geometry. For the case discussed before, where the pair of charges $Q, -Q$ is created at $t=0$ and $Q$ then moves at uniform speed from $z=0$ to $z=g$ during a time $T$, we find 
\bea
    && I(t<T) =\frac{Q}{T} \int_0^\infty \frac{2}{\pi}      \cos (k \frac{x}{g}) \sin (k \frac{w_x}{2g}) \times [ 
     \frac{\vep_r \cosh(k-k\frac{t}{T})}{D(k)} \no \\
     &&  - b_2
     \frac{e^{-\frac{t}{T_0} f_2}(f_2\cosh(k)+\frac{T_0}{T}k\sinh (k))
     -f_2\cosh(k-k\frac{t}{T}) -k \frac{T_0}{T}\sinh(k-k\frac{t}{T})   }
     {k^2\frac{T_0^2}{T^2}-f_2^2}
     ] dk
\eea
\bea
    I(t>T) =\frac{Q}{T} \int_0^\infty \frac{2}{\pi}      \cos (k \frac{x}{g}) \sin (k \frac{w_x}{2g}) 
        b_2\,e^{-\frac{t-T}{T_0}f_1}
     \frac{e^{-\frac{T}{T_0} f_1}(f_1\cosh(k)+k\frac{T_0}{T}\sinh(k))-f_1 }{k^2\frac{T_0^2}{T^2}-f_1^2}
\eea
The limiting case for very high resistivity is equal to the expression from the previous section where there is only an insulating layer. In the limiting case for very small resistance $R$,  $I(t)$ becomes zero since the resistive layer turns into a 'metal plane' that shields the strips from the charges $Q, -Q$. 
\\ \\ 
The signals for a central strip of width $w_x=4g$ as well as the neighbouring strips at $x=4g$ and $x=8g$ as shown in Figures \ref{surface_signals1}-\ref{surface_signals5} for different values of the resistivity $R$ i.e. for different time constants $T_0$.  In case the time constant $T_0$ is large, the effect of the resistivity disappears and the case of $T_0=10T$ in Fig. \ref{surface_signals1} shows signal shapes very close to the on from the previous section for large values of $\tau_0$. For decreasing resistivity, and therefore $T_0$, we see however that the signal on the central strip starts to be 'differentiated' and develops an undershoot and the crosstalk to the other strips increases. 
\\
\\ 
Since for Eq. \ref{h_surface_resisivity} it holds that $\int_0^\infty h(k,z,t)dt=0$, all of the signals are strictly bipolar i.e. $\int_0^\infty I(t)dt=0$. This is due to the fact that the current compensating the point charge $-Q$ is entirely flowing inside the thin resistive layer and no net charge is taken from or is arriving at the strips. 
%
%
%
%
%
%
Finally, the induced current for the movement of the charge from $z=g$ to $z=0$ is 
\bea
     I(t<T) &=&\frac{Q}{T} \int_0^\infty \frac{2}{\pi}      \cos (k \frac{x}{g}) \sin (k \frac{w_x}{2g}) \times \no \\
     && \left[
     \frac{\vep_r \cosh(k\frac{t}{T})}{D(k)} - b_2
     \frac{f_2e^{-\frac{t}{T_0} f_2}-f_2\cosh(k\frac{t}{T}) +k \frac{T0}{T}
     \sinh(k\frac{t}{T})   }{k^2\frac{T_0^2}{T^2}-f_2^2}
     \right] dk
\eea
\bea
     I(t>T)  =  -\,\frac{Q}{T} \int_0^\infty \frac{2}{\pi}    \cos (k \frac{x}{g}) \sin (k \frac{w_x}{2g}) 
        b_2\,e^{-\frac{t-T}{T_0}f_2}
     \frac{f_2e^{-\frac{T}{T_0} f_2}-f_2\cosh(k) +k \frac{T_0}{T}
     \sinh(k)   }{k^2\frac{T_0^2}{T^2}-f_2^2}
    dk
\eea

\clearpage

\section{Integrals with Bessel Functions  \label{integral_app}}

All solutions of the unbounded multilayer problem are expressed in the form 
\beq
  \int_0^\infty J_n(rx)g(x)dx \qquad r>0
\eeq
To evaluate this integral with the method of residues we have to find a proper closed contour in the complex plane. Since $J_n(z)$ is exponentially increasing for large imaginary values, there is no closed contour in the complex plane where the integral vanishes. The Hankel functions $H^{(1)}_n(z)=J_n(z)+iY_n(z)$ are however exponentially decreasing for large positive imaginary values, so we can use a semi-circle in the upper complex half-plane. We use the following trick established in \cite{watson}:
\beq
  \int_{-\infty}^\infty H_n(rx) g(x) dx = \oint _C H_n(rz)g(z)dz = 2\pi i \sum_m \mbox{res}_m = a+bi
\eeq
where $C$ is a closed contour from $x=-R,R$ and a semi-circle of radius $R$ in the upper complex plane over which the integral vanishes for $R\rightarrow \infty$. The values res$_m$ indicate the residuals of the expression $H_n(rz) g(z)$ for all poles of $z_m$ of $g(z)$ in the upper complex plane i.e. for Im$[z]>0$. With the identities 
\beq
J_n(x)=(-1)^nJ_n(-x) \quad
\mbox{Re}[Y_n(-x)]=(-1)^nY_n(x) \quad
\mbox{Im}[Y_n(-x)]=2J_n(-x)
\eeq
we find
\bdi
  \int_{-\infty}^\infty H_n(rx)g(x)dx 
  = \int_0^\infty J_n(rx)[g(x)-(-1)^n g(-x)]dx+i \int_0^\infty Y_n(rx)[g(x)+(-1)^n g(-x)]dx
\edi
so for $g(x)=g(-x)$  with $n$ being uneven, or $g(x)=-g(-x)$ with $n$ being even we have the result
\beq
    \int_0^\infty J_n(rx) g(x) dx = \frac{a}{2}
\eeq
In the following we will make frequent use of the identity
\beq
    K_n(x) = \frac{\pi}{2}i^{n+1} H^{(1)}_n(ix)
\eeq
where $K_n(x)$ denote the modified Bessel functions of second kind \cite{besselK}.

\subsection{Integral 1 \label{integral_1}}

\beq
  \int_0^\infty J_0(rx)\frac{x}{\cosh(x)}dx 
\eeq
The expression has an infinite number of poles in the upper complex half-plane  at $z_m=i\,(2m+1)\pi/2$ with $m=0...\infty$, and the residues are
\bdi
 2\pi i \sum_{m=0}^\infty \lim_{z \rightarrow z_m} \left[(z-z_m)H_0(rz)\frac{z}{\cosh(z)} \right]=  
 \sum_{m=0}^\infty \pi^2 (-1)^m (2m+1)iH_0[ir(2m+1)\pi/2]
 \edi
 \bdi
 =\sum_{m=0}^\infty 2\pi (-1)^m (2m+1)K_0[r(2m+1)\pi /2] = a
\edi
and we have the result
\beq
     \int_0^\infty J_0(rx)\frac{x}{\cosh(x)}dx = 
     \pi \sum_{m=0}^\infty  (-1)^m (2m+1)K_0[r(2m+1)\pi/2]
\eeq
With the exponential behaviour of the $K_0(r)$ the expression is well suited for evaluation for large values of $r$, for small values of $r$ the solution does however converge only slowly and for $r=0$ it even diverges. For small values of $r$ another evaluation is suited:
\beq
  \frac{1}{\cosh(x)}=\frac{2 e^{-x}}{1+e^{-2x}} = 2\sum_{m=0}^\infty (-1)^m\,e^{-(2m+1)x}
\eeq
and using
\beq
   \int_0^\infty x J_0(rx)\,e^{-ax}dx = \frac{a}{(a^2+r^2)^{3/2}}
\eeq
we have
\beq
     \int_0^\infty J_0(rx)\frac{x}{\cosh(x)}dx = 2 \sum_{m=0}^\infty (-1)^m \frac{2m+1}{[(2m+1)^2+r^2]^{3/2}}
\eeq
This expression is very closely related to the method of images discussed in Section \ref{point_charge_N_layer_section}.

\subsection{Integral 2 \label{integral_2}}

\beq
   \int_0^\infty J_0(rx)\,\frac{\sinh(a_1x)\sinh(a_2x)}{\sinh(x)}dx
\eeq
The poles are at $z_m=i\,m\pi$ for $m=1 ... \infty$, so the residuals are given by
\beq
  2\pi i \sum_{m=1}^\infty \lim_{z \rightarrow k_m}
  \left[
  (z-z_m) H_0(rz)\,\frac{\sinh(a_1 z)\sinh(a_2 z)}{\sinh(z)}
  \right]
\eeq
\beq
  =-4\sum_{m=1}^\infty (-1)^m \sin (a_1 m \pi)\sin (a_2 m \pi)   K_0(m\pi r) = a
\eeq

\subsection{Integral 3 \label{integral_3}}

\beq
    \int_0^\infty J_0(rx)\frac{x}{\cosh(x)+ \frac{x}{\beta^2} \sinh(x)}dx 
\eeq
The expression $\cosh(z)+ z\sinh(z)/\beta^2$ has zeroes only on the imaginary axis, so writing $z=iy$ we have to solve
\beq
  \cosh(iy)+\frac{iy}{\beta^2} \sinh(iy)=0 \quad \rightarrow \quad \tan(y)=\frac{\beta^2}{y}
\eeq
Plotting the two functions $\tan(y)$ and $\beta^2/y$ on top of each other it is evident that for $\beta^2>0$  the zeroes $y_n$ satisfy the condition
\beq
   0<y_0<\frac{\pi}{2} \quad  \pi<y_1<\frac{3\pi}{2} \quad 2\pi < y_2 <  \frac{5\pi}{2} \quad \mbox{...} \quad n\pi < y_n < n\pi+\frac{\pi}{2}
\eeq
and that for $\beta^2 \ll 1$ the $y_n$ approach the values
\beq
  y_0=\beta \quad y_n = n\pi \quad \rightarrow \quad z_0=i\beta \quad z_n = in\pi
\eeq
For $\beta \ll 1$ we have 
\beq
  \lim_{z \rightarrow i\beta } \frac{(z-i\beta)}{\cosh(z)+ \frac{z}{\beta^2} \sinh(z)} 
  \approx -\frac{i\beta}{2} 
  \qquad
      \lim_{z \rightarrow in\pi } \frac{(z-in\pi)}{\cosh(z)+ \frac{z}{\beta^2} \sinh(z)} 
    \approx -(-1)^n\frac{i\beta^2}{n\pi}
\eeq
\beq
     2\pi i \sum_{m=0}^\infty \lim_{z\rightarrow z_m }(z-z_m)H_0(rz)\frac{z}{\cosh(z)+\frac{z}{\beta^2}\sinh (z)} \approx 2 \beta^2 \left[
     K_0(\beta r) + 2 \sum_{m=1}^\infty (-1)^m\,K_0(m\pi r)
     \right] = a
\eeq
so we have for $\beta^2 \ll 1$ the result
\beq
 \int_0^\infty J_0(rx)\frac{x}{\cosh(x)+ \frac{x}{\beta^2} \sinh(x)}dx \approx
 \beta^2 \left[ K_0(\beta r) + 2 \sum_{m=1}^\infty (-1)^m\, K_0(m\pi r) \right]
\eeq



\begin{thebibliography}{00}
\bibitem{besselJ} $J_n(x)$ are the Bessel functions of first kind. BesselJ[n,x] in Mathematica 9
\bibitem{besselY} $Y_n(x)$ are the Bessel functions of second kind. BesselY[n,x] in Mathematica 9
\bibitem{besselK} $K_n(x)$ are the Modified Bessel functions of second kind. BesselK[n,x] in Mathematica 9
\end{thebibliography}
\end{document}